\documentclass[10pt,preprint]{aastex}
\usepackage{graphicx,amssymb,natbib}
\usepackage{times,datetime}
\usepackage[toc,page]{appendix}

\shorttitle{HIghMass: optical}
\shortauthors{S. Huang}

\begin{document}

\title{HIghMass -- High HI Mass, HI-rich Galaxies at z$\sim$0 
\\\normalsize Sample Definition, Optical and H$\alpha$ Imaging, and Star Formation Properties
}
\author{Shan Huang\footnotemark[1]$^{~,}$\footnotemark[2],
Martha P. Haynes\footnotemark[2], Riccardo Giovanelli\footnotemark[2], Gregory Hallenbeck\footnotemark[2], Michael G. Jones\footnotemark[2], 
Elizabeth A.K. Adams\footnotemark[2]$^{~,}$\footnotemark[3], 
Jarle Brinchmann\footnotemark[4], 
Jayaram N. Chengalur\footnotemark[5], 
Leslie K. Hunt\footnotemark[6], 
Karen L. Masters\footnotemark[7]$^{~,}$\footnotemark[8], 
Satoki Matsushita\footnotemark[1], 
Amelie Saintonge\footnotemark[9], 
Kristine Spekkens\footnotemark[10]
}
\footnotetext[1]{Institute of Astronomy and Astrophysics, Academia Sinica, 11F of Astronomy-Mathematics Building, National Taiwan University, Taipei 10617, Taiwan, R.O.C.; {\it email:} shan@asiaa.sinica.edu.tw}
\footnotetext[2]{Center for Radiophysics and Space Research, Space Sciences Building, Cornell University, Ithaca, NY 14853}
\footnotetext[3]{ASTRON/Netherlands Institute for Radio Astronomy, Oude Hoogeveensedijk 4, 7991 PD Dwingeloo, The Netherlands}
\footnotetext[4]{Sterrewacht Leiden, Leiden University, NL-2300 RA Leiden, The Netherlands} 
\footnotetext[5]{National Centre for Radio Astrophysics, Tata Institute for Fundamental Research, Pune 411 007, India}
\footnotetext[6]{INAF-Osservatorio Astrofisico di Arcetri, Largo E. Fermi 5, I-50125, Firenze, Italy}
\footnotetext[7]{Institute of Cosmology and Gravitation, Dennis Sciama Building, Burnaby Road, Portsmouth POI 3FX}
\footnotetext[8]{South East Physics Network, www.sepnet.ac.uk}
\footnotetext[9]{Department of Physics and Astronomy, University College London, Gower Place, London, WC1E 6BT, United Kingdom}
\footnotetext[10]{Royal Military College of Canada, Department of Physics, PO Box 17000, Station Forces, Kingston, Ontario, Canada K7K 7B4}


\begin{abstract}
We present first results of the study of a set of exceptional
HI sources identified
in the 40\% ALFALFA extragalactic HI survey catalog $\alpha.40$
as being both
HI massive ($M_{\rm HI} > 10^{10}~M_\odot$) and 
having high gas fractions for their stellar masses:
the HIghMass galaxy sample. We analyze UV- and optical-broadband and
H$\alpha$ images to understand the nature of their
relatively underluminous disks in optical and to test whether their
high gas fractions can be tracked to higher dark matter 
halo spin parameters or late gas accretion.
Estimates of their star formation rates (SFRs) based on SED-fitting
agree within uncertainties with the H$\alpha$ luminosity inferred current massive SFRs. 
The HII region luminosity functions, parameterized as $dN/d\log L \propto L^\alpha$, have standard slopes at the luminous end 
($\alpha \sim -1$).
The global SFRs demonstrate that the HIghMass galaxies exhibit 
active ongoing 
star formation (SF) with moderate SF efficiency, but relative to 
normal spirals, a lower integrated SFR in the past. 
Because the SF activity in these systems is spread throughout their extended disks, 
they have overall lower SFR surface densities and lower surface brightness in the optical bands. 
Relative to normal disk galaxies, 
the majority of HIghMass galaxies have higher H$\alpha$ equivalent widths and are bluer in their 
outer disks, implying an inside-out disk growth scenario. 
Downbending double exponential disks are more frequent than upbending disks among the 
gas-rich galaxies, suggesting that SF thresholds exist in the downbending disks,
probably as a result of concentrated gas distribution. 
\end{abstract}


\section{Introduction}
\label{Int}
While the scenario that galaxies evolve through mergers and gas condensation 
at the center of dark matter (DM) halos is now widely accepted \citep{Springel2006}, 
the detailed processes of gas acquisition and galaxy assembly remain largely unknown. 
The amount of cold gas in a galaxy reflects the complex interplay between processes that either replenish it, 
such as cooling and accretion or mergers with gas-rich galaxies; or deplete it, such as environmental effects, star formation (SF), 
and feedback from massive stars and active galactic nuclei (AGNs). 
In the ``downsizing'' scenario of galaxy evolution \citep{Cowie1996}, massive galaxies are most efficient in consuming 
their gas reservoirs at earlier times. 
Very massive galaxies with a substantial supply of cold gas are thus expected to be rare locally.
However, the recently-completed Arecibo Legacy Fast ALFA (ALFALFA) extragalactic HI survey 
\citep{Giovanelli2005, Haynes2011}, with its significant volume sensitivity, detects thousands of 
massive HI disks with $ M_{\rm HI} > 10^{10}~M_\odot$, comparable to the massive HI disks detected at 
$z\sim0.2$ \citep{Catinella2008}. Understanding the rare, local high HI mass population is important because
they represent the present day counterparts of the populations which are likely to dominate future studies 
of HI in galaxies at higher $z$ with the next generation of radio telescopes, e.g., the Square Kilometre Array (SKA). 
The HIghMass study will establish the local standard of such HI massive 
disks to be used as the $z\sim0$ benchmark in the investigations into 
evolutionary trends with redshift by the future HI surveys. 

Selected examples of very massive but gas-rich galaxies have been studied in recent years, 
with most belonging to the extreme category of low surface brightness (LSB) galaxies 
known as the ``crouching giants'' \citep{Disney1987} or ``Malin 1 cousins''. 
Malin 1 itself is both massive and gas-rich \citep{Lelli2010}, as are other members of the class.
For instance, NGC~765 is shown to have one of the highest HI-to-optical luminosity ratios and one of the largest 
known HI disks, with  $M_{\rm HI}\sim4.7\times10^{10}M_\odot$ \citep{Portas2010}.  
Moreover, UGC~8802 included in the {\it GALEX}-Arecibo-SDSS (GASS) survey of massive
galaxies \citep{Catinella2010} is found to have $M_{\rm HI}\sim2.1\times10^{10}M_\odot$, but a low and remarkably evenly distributed 
star formation rate (SFR) surface density \citep{Moran2010}. The extremely high ratio of current SFR 
to existing stellar mass ($M_*$) surface density in the outer disk implies that all of its stars must have 
formed within the past $\sim$1~Gyr. 

Because it provides the first full census of HI-bearing objects 
over a cosmologically significant volume in the local Universe, the ALFALFA survey is most effective 
in detecting these rare massive {\it and} gas-rich systems 
\citep{Huang2012b}, and enables, for the first time, their {\it systematic study} as an exceptional local population.
Hence, our study focuses on the most HI massive and high HI gas fraction ($f_{\rm HI} \equiv M_{\rm HI}/M_*$)
disks detected in the ALFALFA survey,
hereafter identified as the ``HIghMass'' sample.

The presence of massive HI disks in the local Universe remains a puzzle, 
implying unusual physical conditions of gas depletion and/or supply. 
On the one hand, the higher-than-average HI gas fraction found in the HIghMass galaxies
may result from association with
underluminous stellar disks, i.e., the galaxies may have experienced an arrested stage of evolution in the past
and are thus ``star-poor''. 
Theoretical models predict that galaxies in high spin parameter ($\lambda \equiv J |E|^{1/2} G^{-1} M_{\rm halo}^{-5/2}$, 
characterizing the mass-normalized angular momentum) 
DM halos have unusually extended gaseous disks with large disk scale lengths and low surface densities 
\citep[e.g.][]{Mo1998, Boissier2000, Kravtsov2013}. 
As a result, the gas infall and consumption times are so long that 
a phase of significant SF would never have been reached \citep{Li2012}. 

Although their total HI masses are high, the classical LSB galaxies are indeed found 
to have low HI surface densities. SF is an extremely inefficient process 
in such diffuse disks, e.g., Malin 1 \citep{Bothun1987, Impey1997}, in agreement with the 
expectation of canonical star formation laws (SFLs) and the observed Kennicutt-Schmidt (K-S) relation.
The slowly rising rotation curves found in the Malin-type galaxies imply that they are dominated kinematically by DM,
and their DM halos are less concentrated with larger $\lambda$s \citep{Pickering1997}.
Giant LSB galaxies are observed to have a large amount of angular momentum in their disks, 
in agreement with the formation of LSB galaxies as predicted by recent hydrodynamic simulations \citep{Kim2013}. 
This scenario is also consistent with the small DM perturbations found commonly in cosmological low density regions 
with late formation \citep{Mo1994}. 

In contrast to the explanation of a depressed star formation history, the exceptionally 
high $f_{\rm HI}$ may be attributed to the late accretion of cold gas, leaving
the galaxies with ``too much gas''. 
Rather than hot accretion (post-shock cooling from a quasi-spherical halo), cold accretion (clouds, streams, filaments) 
is theoretically considered as an important aspect in the process of galaxy formation at high $z$, 
and is perhaps still occuring in low-mass galaxies and in low-density regions today \citep{Keres2005, Croton2006}. 
The models that assume a slowly evolving equilibrium between gas inflow, outflow, and SF 
can well reproduce the observed evolution of the SF sequence, the 
Tully-Fisher relation, and other scaling relations \citep[e.g.,][]{Bouche2010, Fraternali2012}, etc.
In the local Universe, it is possible to detect the emission from the diffuse infalling structures with 
sensitive HI observations, e.g., extra-planar HI, HI tails and filaments \citep{Sancisi2008}. 
While in a canonical picture, gas cooling from the virial temperature spends a considerable 
amount of time in the galactic halo so that it has the same specific angular momentum as the underlying DM, 
high-resolution cosmological hydrodynamic simulations suggest that gas accreted in cold mode
enters a galaxy halo along filaments with $\sim70\%$ more specific angular momentum than the DM, 
with the gas well characterized by $\lambda \sim 0.1$ at the time of accretion \citep{Stewart2013}.
The resulting disks of cool gas are dense enough to form HI and can thus help to
explain the frequency of observed extended UV disks \citep{Lemonias2011} and 
extended, warped, or lopsided HI disks \citep{Sancisi2008}. 
Such newly accreted gas may be retarded in producing SF if it is stable against contraction 
and stays in the low surface density outer disk where HI dominates over H$_2$. 
On the other hand, it has been suggested 
that the presence of a bar may induce radial gas flows
which actually trigger SF \citep{Martin1994}. 

To unravel the origin and current evolutionary state of the High HI mass, HI-rich
galaxies at $z\sim0$, we began a campaign in 2009
to gather multi-wavelength data for a volume-limited sample of 34 galaxies selected from ALFALFA by their
high HI mass and high gas-to-stellar mass ratio (for their stellar mass).
The complete HIghMass dataset will eventually enable the understanding of the gaseous, stellar, and DM components
of these exceptional systems.
This paper is the first in a series that presents a summary of the project definition and optical properties of the full sample, 
including the SF as probed by the H$\alpha$ narrowband images and the stellar population as seen in the SDSS images. 
We present the sample, observing status, and basic physical properties of our targets 
relative to other H$\alpha$ or ALFALFA-related surveys in Section \ref{Sam}. 
Details of the H$\alpha$ and {\it R}-band observations, together with the catalogs of optical data are 
given in Section \ref{Obs}. 
SFRs are calculated in multiple ways and are compared against each other in Section \ref{Cal}. 
Section \ref{Opt} introduces optical characteristics of the HIghMass galaxies, e.g., their HII region luminosity functions (LFs). 
Their SF and stellar disk properties relative to the general ALFALFA population, 
as well as the $\lambda$ distribution of host halos inferred from the optical data are discussed in Section \ref{Dis}. 
We summarize the main results and prospective future works in Section \ref{Sum}. 
Image reduction, photometry processing, internal and external data quality checks are described for 
the H$\alpha$ and SDSS bands in Appendix \ref{app:KPNO} and \ref{app:SDSS}, respectively. 

Throughout this paper, we adopt a reduced Hubble constant $h = H_0/(100~{\rm km~s^{-1}~Mpc^{-1}}) = 0.7$ and 
a \citet{Chabrier2003} IMF. 

\section{Sample}
\label{Sam}

\subsection{Sample Selection and Program Overview}
\label{Sel}

The ALFALFA survey achieves its spectral mapping via a drift
scan technique with the Arecibo {\it L}-band feed array (ALFA) on the
Arecibo 305~m antenna \citep{Giovanelli2005}. The final 3-D spectral
cubes yield a median HI centroiding accuracy of 
$\sim$20$^{\prime\prime}$ (dependent on signal-to-noise ratio) and a resolution of $\sim 3.5^{\prime}$.
In addition to properties of the detected HI signals, ALFALFA survey catalogs also include, where
applicable, an assignment of the most probable optical counterpart (OC) to each HI source.
The HIghMass sample galaxies have been selected from the 2011 ALFALFA catalog, $\alpha$.40 \citep{Haynes2011}, 
covering 40\% of the final survey area and including $\sim$16,000 high-quality detections. 
For the area of overlap, the $\alpha$.40 catalog included crossmatches of the OCs to the 
SDSS Data Release 7 \citep{Abazajian2009}. 

\begin{figure*}
\center{
\includegraphics[scale=0.8]{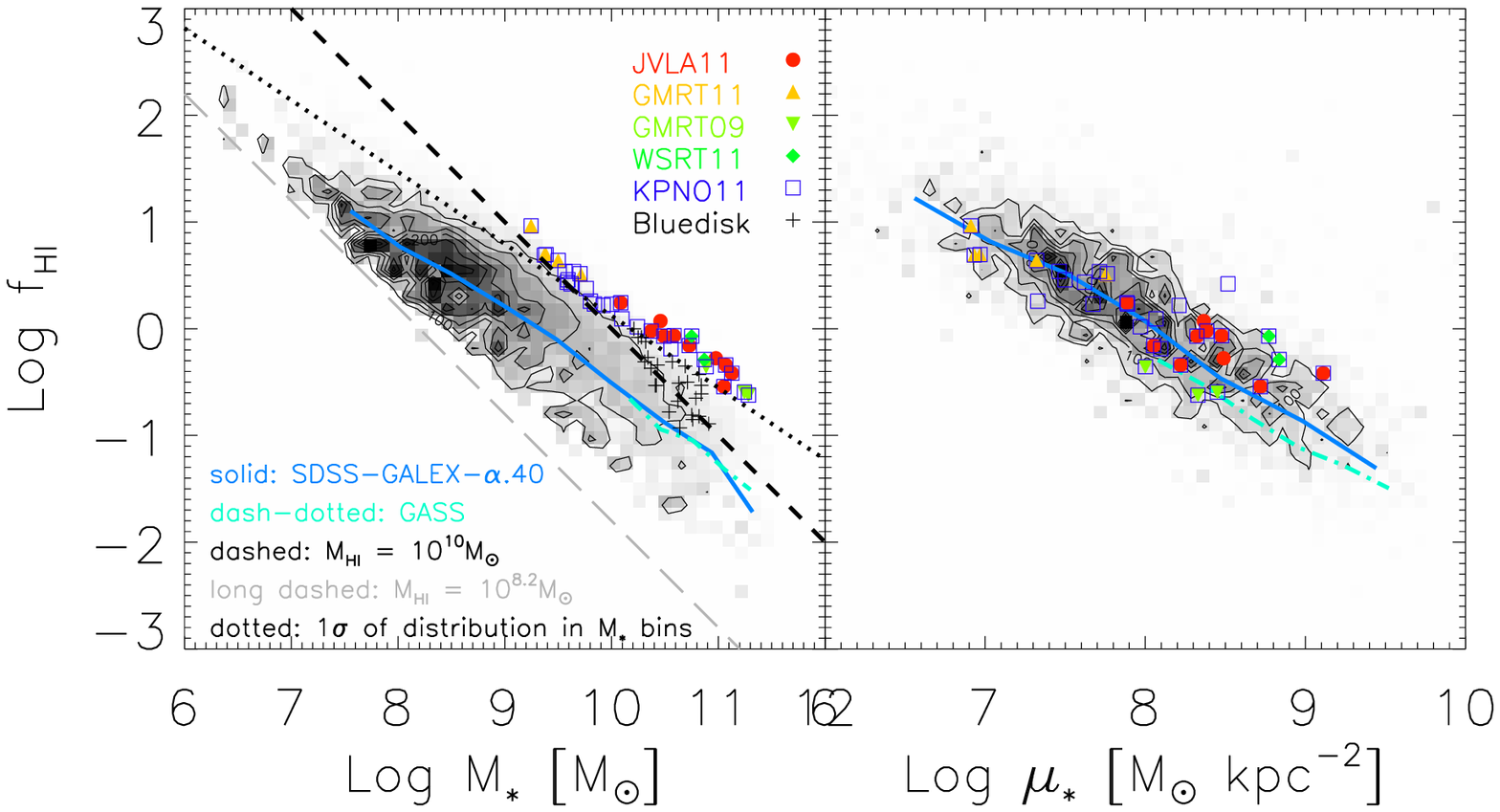}
}
\caption[]{
Left panel: The gas depletion sequence and HIghMass sample selection. Black contours and grayscales represent the 
SDSS-{\it GALEX}-$\alpha$.40 common sample, weighted by the $V/V_{\rm max}$ values as presented by \citet{Huang2012b} . 
The gray long dashed line shows the approximate lower limit of $M_{\rm HI} = 10^{8.2}~M_\odot$. 
The solid blue curve illustrates the weighted running average, in agreement with the GASS result \citep{Catinella2013},
shown as the cyan dash-dotted line. Colored symbols are overlaid on the diagram for the 34 HIghMass galaxies. 
They are selected to have $M_{\rm HI} > 10^{10}~M_\odot$ (above the dashed line) and have HI fraction more than 
1$\sigma$ above the running average (above the dotted line). Targets of the HI synthesis mapping and H$\alpha$ 
imaging programs are denoted by filled and open symbols respectively. The massive ``Bluedisk'' galaxies \citep{Wang2013} are 
represented by the black crosses lying below the HIghMass distribution. 
Right panel: Similar diagram showing the variation in f$_{\rm HI}$ with the stellar mass surface density $\mu_*$.
The HIghMass galaxies follow the general trend between the HI mass fraction and the stellar mass surface density. 
}
\label{fig:fHI}
\end{figure*}

Because of the high sensitivity of the Arecibo telescope,
the $\alpha$.40 catalog includes over 2800 galaxies with $M_{\rm HI} > 10^{10}~M_\odot$. 
Among these, the HIghMass
sample selects the most gas-rich, relative to their stellar mass. 
To derive $f_{\rm HI}$, we estimate $M_*$ by spectral energy distribution (SED) 
fitting to seven {\it GALEX}-SDSS bands \citep{Huang2012a, Huang2012b}. 
Similar to Figure 2(c) of \citet{Huang2012b},
the left panel of Fig.~\ref{fig:fHI} shows the $f_{\rm HI}$ {\it vs.}~$M_*$ diagram
for the SDSS-{\it GALEX}-$\alpha$.40 sample, 
but weighted by the $V/V_{\rm max}$ values, 
where $V$ is the overlapped survey sky volume and $V_{\rm max}$ is the maximum distance 
at which an HI source can be detected by ALFALFA. 
Relatively HI-poor galaxies are observable in a smaller $V_{\rm max}$ than 
the total survey volume, so that they are assigned a higher weight above unity. 
This approach corrects for the ALFALFA survey incompleteness \citep{Haynes2011} 
in order to mimic the scaling relations derived from a volume limited sample \citep{Huang2012b}. 
It is similar to what was applied in \citet{Baldry2004}, 
accounting for the fact galaxies of a given absolute magnitude can be observed only within 
a certain redshift range. 
Black contours and grayscales represent the SDSS-{\it GALEX}-$\alpha$.40 common sample. 
The sharp lower edge of the general distribution 
arises because only the galaxies 
with weights below 60, or equivalently with $M_{\rm HI} \gtrsim 10^{8.2}~M_\odot$, are included. 
The gray long dashed line in Fig.~\ref{fig:fHI} shows the approximate lower limit of $M_{\rm HI} = 10^{8.2}~M_\odot$. 
The solid blue curve illustrates the weighted running average defined by the SDSS-{\it GALEX}-$\alpha$.40 
galaxies, in agreement with the result derived from the GASS survey at $M_* > 10^{10}~M_\odot$ 
\citep[][final data release, cyan dash-dotted line]{Catinella2013}. 
The decreasing trend of $f_{\rm HI}$ with $M_*$ confirms the expectation that massive galaxies are mostly gas poor. 
The hint of a $f_{\rm HI}$ bimodal distribution as driven by the higher portion of gas-poor galaxies 
at the high stellar mass end will be more prominent if the ALFALFA detections with higher weights 
(i.e., the HI poor galaxies with $M_{\rm HI} \lesssim 10^{8.2}~M_\odot$ detected only nearby) are included in this plot. 
The $f_{\rm HI}$ bimodal distribution is consistent with the bimodal distribution in an optical color-magnitude diagram, 
as well as the fact that the relative number of red galaxies increases among massive galaxies with $M_* \gtrsim 10^{10} M_\odot$. 

As an HI-selected population, galaxies detected by ALFALFA are strongly biased
towards gas-rich systems with the HIghMass galaxies being the extreme cases.
The black lines in Fig.~\ref{fig:fHI} corresponds to the selection criteria of HIghMass galaxies: 
({\it i}) has an $M_{\rm HI} > 10^{10}~M_\odot$, lying above the dashed line; 
and ({\it ii}) has an $f_{\rm HI}$ more than than 1$\sigma$ above the running average in a given $M_*$ bin, lying above the dotted line. 
These selection criteria are partly motivated by the semi-analytic model prediction in \citet{Boissier2000} that 
those galaxies which reside in high-$\lambda$ DM halos have preferentially higher gas fractions at fixed circular velocity. 
To avoid unreliable measures, we visually inspected these galaxies to eliminate cases in which the HI 
fluxes suffer from confusion in the ALFA beam or those in which the SDSS magnitudes suffer from significant shredding or blending.
{\it The final HIghMass program sample consists of 34 galaxies, listed in Table \ref{tab:HI}.}
Different subsets are observed in our comprehensive multiwavelength observing programs, 
accounting for the technical feasibility. 
For instance, targets close to bright stars were avoided in our {\it GALEX} program (GI-6; Haynes PI); 
targets close to strong continuum sources or with low HI fluxes  were avoided in the 
HI synthesis mapping conducted with the Jansky Very Large Array (JVLA),
the Giant Meterwave Radio Telescope (GMRT) and the Westerbork Synthesis Radio Telescope (WSRT). 

The 34 HIghMass galaxies are shown superposed on the ALFALFA population as colored symbols in Fig.~\ref{fig:fHI} 
with separate symbols identifying the galaxies included in our HI 
synthesis program (20 filled symbols) and those with narrowband H$\alpha$ imaging (31 open squares: KPNO11). 
Most of these objects are also included in other multiwavelength programs with other facilities ({\it GALEX}, 
Isaac Newton Telescope, warm-{\it Spitzer}, {\it Herschel}, IRAM 30-m, CARMA, SMA). 
The $f_{\rm HI}$ ranges between 0.24 and 9.2, with 20/34 having $f_{\rm HI} > 1$.
Because ALFALFA does not provide the angular resolution needed to resolve the HI disks,
we have obtained HI synthesis maps of 20 objects which could be feasibly mapped with
existing HI synthesis instruments, 
including eight observed with the GMRT  
(in 2009; Chengalur PI; downward triangles in Fig.~\ref{fig:fHI}; 
and in 2011; Adams PI; upward triangles), two with the WSRT (in 2011; Haynes PI; diamonds), 
and last ten with the JVLA (in 2011; Haynes PI; circles). By mapping the distribution and velocity
of the HI gas, these observations will yield clues on the origin 
and nature of the HIghMass galaxies via the constraints on their DM halos placed by gas dynamics 
\citep{Hallenbeck2014}. In addition, recent accretion of intergalactic medium (IGM) or minor mergers of 
faint companions may leave signatures in the HI morphology or velocity fields. 

The HIghMass sample was also targeted for medium-depth observations (1200 sec) with the {\it GALEX} satellite,
but only NUV observations were obtained because of the early failure of the {\it GALEX} FUV detector.
However, existing {\it GALEX} archival data give insight into the intriguing behavior in some of the HIghMass galaxies. 
{\it GALEX} images reveal that SF in more gas-rich galaxies often extends much farther in radius than had 
previously been appreciated \citep{Thilker2007}. Whereas only two faint  loose spiral arms are seen in optical 
images of UGC~9234 in the HIghMass sample, an extended UV disk hosting irregular and patchy SF is clearly evident,
indicative of recent gas inflow and disk regrowth. 
It is known that the extent of the UV disk (normalized to the optical size) is strongly correlated to the 
integrated $f_{\rm HI}$, as expected if that the amount of HI regulates the growth of SF disk in the outskirt of galaxies 
\citep{Cortese2012}. 
In combination with the resolved HI column density maps, dust maps available from {\it Herschel} observations,
insights into the underlying stellar population from {\it Spitzer}, and CO fluxes from the IRAM 30-m telescope for
significant subsets of the HIghMass sample,
future work with the resolved SF measures derived from 
both UV photometry and H$\alpha$ imaging will address the questions concerning the gas-SF interplay. 
We will study the HI-to-H$_2$ transition and the empirical K-S relation 
in more detail to explore how such massive HI disks can exist without having converted the bulk of gas into stars yet.

The right panel of Fig.~\ref{fig:fHI} illustrates the trend of HI mass fraction as a function of 
stellar mass surface density, $\mu_*$
for the ALFALFA population (contours and grayscale) with the HIghMass galaxies identified individually 
as in the left panel.
\citet{Catinella2010} found $\mu_*$ to be one of the optical-derived quantities that can be 
used to predict accurately the $f_{\rm HI}$ in galaxies, in addition to the NUV$-r$ color, or the specific SFR defined as SSFR~$\equiv$~SFR/$M_*$. 
We follow \citet{Catinella2013} to define $\mu_* \equiv M_*/(2\pi r^2_{50, z})$ for the parent SDSS-{\it GALEX}-$\alpha$.40 sample, 
where $r_{50, z}$ is the radius containing 50 percent of the Petrosian flux in {\it z}-band measured by the SDSS pipeline. 
No inclination correction is applied to obtain face-on values because the Petrosian flux is determined within 
circular apertures by the standard procedure. 
However, we have reprocessed the photometry on the SDSS images with elliptical apertures (see Appendix \ref{app:SDSS}), 
so that $\mu_* \equiv M_*/(2\pi r^2_{50, z}~q)$ for the HIghMass galaxies, where $q$ is the minor-to-major axial ratio. 
The running average as determined by our SDSS-{\it GALEX}-$\alpha$.40 sample (blue solid line) lies slightly above 
the GASS result \citep[cyan dash-dotted line,][]{Catinella2013}. 
This small discrepency can be easily explained by the fact that the ALFALFA survey extends to a lower $M_*$ range while 
the $f_{\rm HI}$ at fixed $\mu_*$ is overall higher for less massive galaxies \citep{Catinella2010}. 
In comparison with the parent ALFALFA sample, 
the HIghMass galaxies have extraordinarily high $f_{\rm HI}$ at given $M_*$, 
but overall lower $\mu_*$ (see Section \ref{Dis}). 
In an $f_{\rm HI}$ {\it vs.}~$\mu_*$ diagram, because the $f_{\rm HI}$ value generally increases with decreasing $\mu_*$ 
for the overall population, 
the HIghMass galaxies are not outliers in the right panel of  Fig.~\ref{fig:fHI}, 
i.e., they follow the global scaling relation between $f_{\rm HI}$ and $\mu_*$. 

\subsection{Basic Physical Properties of the HIghMass galaxies}

Table \ref{tab:HI} presents the existing data for selected observations and 
basic properties 
for the full sample of 34 HIghMass galaxies. Columns are as follows:
\begin{itemize}
\item Col(1): ALFALFA catalog identifier (also known as the AGC number). 
\item Col(2): Other name resolution by NED; same as AGC numbers in the UGC cases. 
\item Col(3) and (4): J2000 position of the OC assigned to the HI source, in degrees. 
\item Col(5): Galaxy morphology classification according to NED. The presence of a strong bar is denoted by `B' 
	based on the the data release for Galaxy Zoo 2 \citep{Willett2013}. 
\item Col(6): The HI line width $W_{50}$, in km~s$^{-1}$, taken from the $\alpha$.40 catalog \citep{Haynes2011}. 
\item Col(7): The HI disk rotational velocity $V_{\rm rot}$, in km~s$^{-1}$,  taken to be the deprojected $W_{50}$ 
	using the axial ratio of the optical disk to correct for inclination (see Section \ref{Dis}).
\item Col(8): The systemic heliocentric recessional velocity $cz$, in km~s$^{-1}$. 
\item Col(9): The adopted distance in Mpc, taken from the $\alpha$.40 catalog \citep{Haynes2011}. 
\item Col(10): The logarithm of the $M_{\rm HI}$ and its error, in $M_\odot$, taken from the $\alpha$.40 catalog \citep{Haynes2011}. 
\item Col(11): Type of archival {\it GALEX} images used in Section \ref{Cal}, 
	AIS = all sky imaging survey; MIS = medium imaging survey; DIS = deep imaging survey; GII = guest invited investigation; null if outside of footprint. 
\item Col(12): Source of H$\alpha$ images, 
	K = our KPNO 2011 run; K* = KPNO 2012 run by Angie Van Sistine; G = GOLDmine data \citep{Gavazzi2003}; null if too far away.
\item Col(13): Code for our HI synthesis mapping programs as discussed in Section \ref{Sel}. 
\end{itemize}

\begin{figure*}
\center{
\includegraphics[scale=0.9]{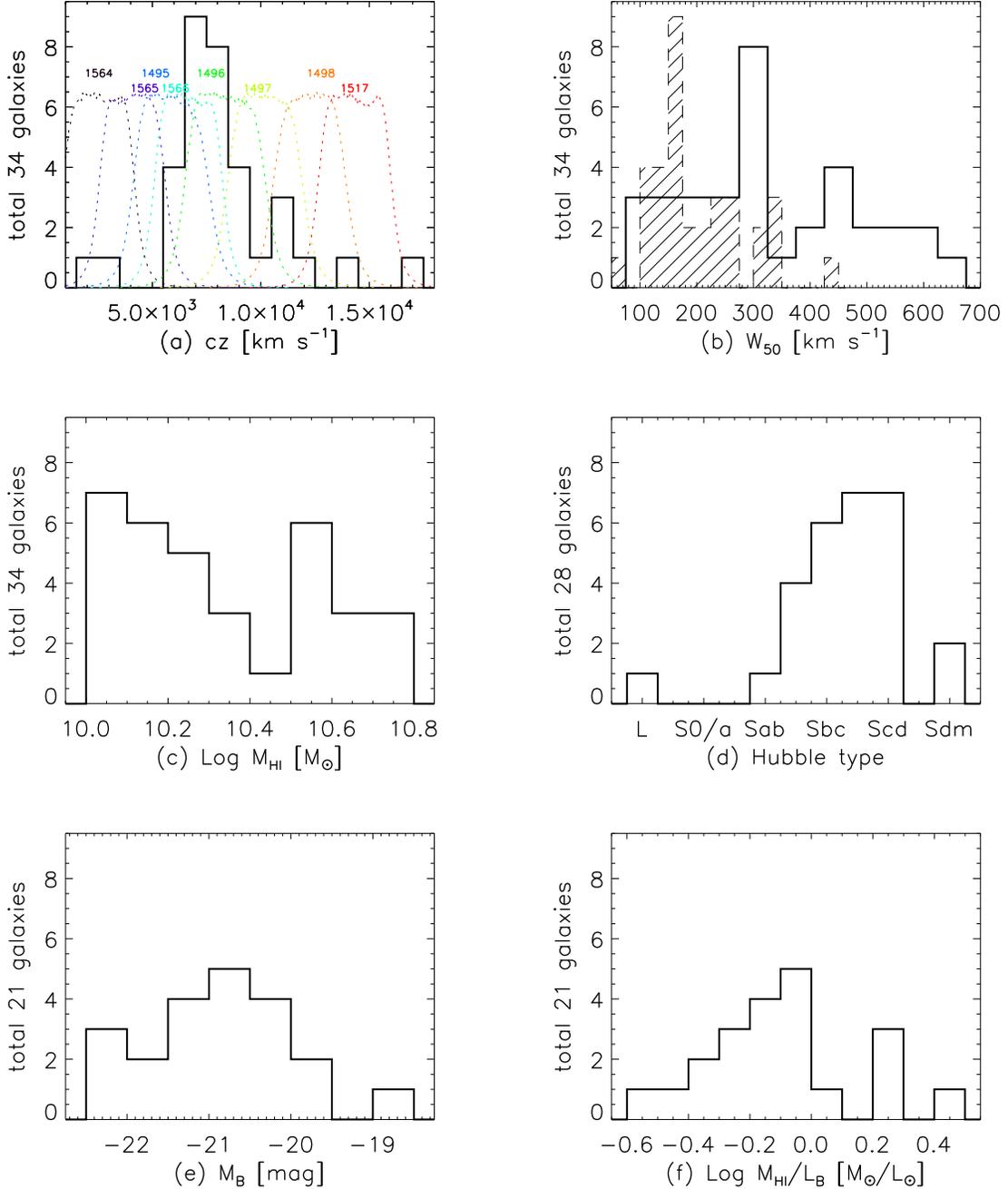}
}
\caption[]{
Black solid lines in all panels show distributions of the HIghMass galaxies: 
(a) HI systemic velocity; (b) observed HI line width; (c) logarithm of HI mass; 
(d) morphological type; (e) {\it B}-band absolute magnitudes; (f) logarithm of the HI mass 
to {\it B}-band luminosity ratio. 
Colored dashed lines in panel (a) illustrate the transmission curves of the H$\alpha$ 
filters used in our KPNO run, together with the corresponding filter names. 
The filled dashed histogram in panel (b) indicates the distribution of inclination-corrected HI rotational velocities.
}
\label{fig:basic}
\end{figure*}

Histograms in panel (a)-(d) of Fig.~\ref{fig:basic} show, respectively, the distributions of the heliocentric recessional velocity, 
observed HI line width, logarithm of the $M_{\rm HI}$, 
and NED morphology classification in the Third Reference Catalog (RC3) system \citep[][available for 28 galaxies]{deV1991}. 
Adopting an intrinsic axial ratio of $q_0 = 0.2$, the inclination $\cos i$ is inferred by $\sqrt{(q^2-q_0^2)/(1-q_0^2)}$. 
The rotational velocity of the HI disk, $V_{\rm rot}$, is thus estimated by $(W_{50}/2) /\sin i$. 
The resulting $V_{\rm rot}$ values, overlaid in Panel (b) by the filled histogram, are large, 
consistent with their status as HI massive disk galaxies (see Panel c). 
The only early type galaxy in the sample, UGC~4599, is a face on S0 galaxy with an outer ring
where active star forming regions are found. 
The total asymptotic magnitudes in the {\it B} band are available from RC3 for 21 galaxies, 
ranging from 15.73 mag to 12.60 mag, with a median of 14.35 mag. 
The final two panels in Fig.~\ref{fig:basic} show the RC3 {\it B}-band absolute magnitude and 
logarithm of the $M_{\rm HI}/L_{B}$, assuming $B = 5.47$ for the sun. 
These {\it B}-band magnitudes are systematically brighter than our reprocessed 
SDSS magnitudes (measured inside elliptical Petrosian apertures, see Appendix \ref{app:SDSS}), 
but the differences are within the uncertainty. 
The final set of 34 HIghMass galaxies span a range of colors, morphologies, luminosities, 
$M_*$, and SFRs (see Section \ref{Cal}). 

\begin{figure*}
\center{
\includegraphics[scale=0.9]{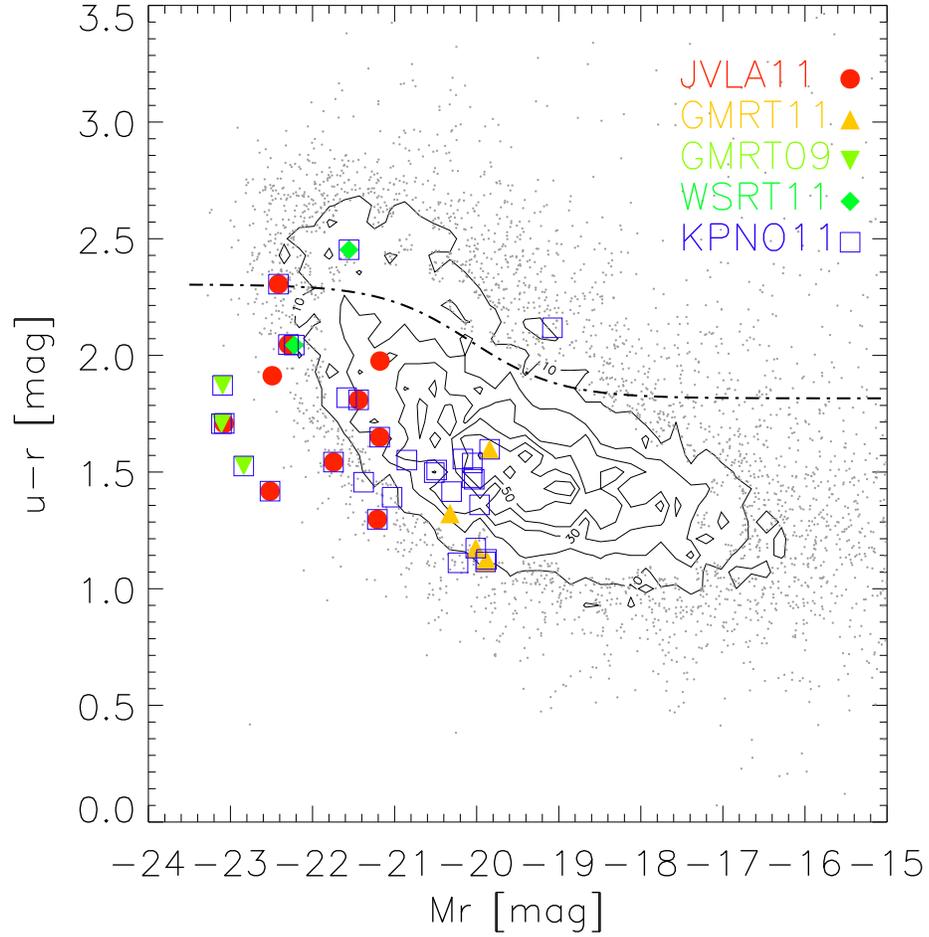}
}
\caption[]{Optical color-magnitude diagram. 
Definition of the colored symbols is the same as that in Fig.~\ref{fig:fHI}. 
Contours and points represent the $\alpha$.40-SDSS(DR8) sample 
in high and low number density regions, respectively. 
The approximate dividing line which separates the ``red sequence'' from the 
``blue cloud'' as presented by \citet{Baldry2004} is shown as the dash-dotted curve.
}
\label{fig:CMD}
\end{figure*}

Fig.~\ref{fig:CMD} illustrates the unweighted distribution of the 
parent $\alpha$.40-SDSS sample (contours and points) on an optical color-magnitude diagram;
the HIghMass galaxies are superposed with the same colored symbols as in Fig.~\ref{fig:fHI}. 
The approximate dividing line which separates the ``red sequence'' from the ``blue cloud'' as presented by 
\citet{Baldry2004} is shown as the dash-dotted curve. 
In this plot, the SDSS magnitudes of the parent sample are the DR8 pipeline values, but those of the HIghMass 
galaxies are our reprocessed measurements (see Appendix \ref{app:SDSS}). 
The vast majority of the galaxies in the $\alpha$.40-SDSS population are found in the blue cloud region. 
Similarly, most of the HIghMass galaxies, especially the GMRT09 and JVLA11 targets, are exceptionally blue 
given their high luminosities. 
Only three HIghMass galaxies lie in the red sequence above the division, UGC~6066, UGC~9234, and UGC~4599. 
UGC~6066 is an edge-on red spiral and UGC~4599 is an S0 galaxy; they have the earliest  morphological
types in the HIghMass sample. 
Although the stellar light in both UGC~9234 and UGC~4599 is dominated by red central regions, 
the first galaxy has two loose arms and the second one has an outer ring, both being blue and LSB features along which 
multiple star-forming regions are identified in our continuum-subtracted H$\alpha$ images (see Appendix \ref{app:KPNO}). 
The huge HI reservoirs in both of these red galaxies are apparently involved in the current re-growth of the outer features. 
Future work will investigate the star formation histories of these systems exploiting other multiwavelength 
datasets.

\subsection{HIghMass in the Context of other Surveys}

In recent years, a number of other surveys have amassed a significant body of H$\alpha$ imaging data on 
selected populations of star-forming galaxies, 
including the GOLDmine survey of the Virgo cluster and Coma superclusters \citep{Gavazzi2003}. 
Out of the HIghMass galaxies, 
UGC~7686 happens to be a background galaxy in the Virgo direction so that is imaged by the GOLDmine survey. 
However, the ALFALFA galaxies are biased against cluster environments \citep{Martin2012}. 
Another ALFALFA-related H$\alpha$ program, H$\alpha$3 \citep{Gavazzi2012}, 
is in fact lacking in objects with $\log M_{\rm HI} > 9.75~M_\odot$, 
because of the well-known cluster HI deficiency and the relatively small volume sampled. 

The galaxies in the HIghMass program can be 
most easily distinguished from very local surveys, e.g, the survey of 52 dwarf-dominated 
galaxies in the local volume \citep{Karachentsev2010}, the surveys of 140 local irregulars by Hunter and collaborators \citep{Hunter2004}, 
or the 11~Mpc H$\alpha$ UV Galaxy Survey \citep[11HUGS;][]{Kennicutt2008}, by its inclusion of massive and luminous
galaxies which are too rare to be included in surveys of such limited volume.
Note that \citet{Hunter2004} have used 74 spiral galaxies spanning the range of morphologies from Sab to Sd 
as compiled by \citet{Kennicutt1983} for comparison. However, the $\log M_{\rm HI}/L_B$ distribution 
of that Sab-Sd sample peaks around $-1$, which is significantly HI poorer than our galaxies 
with similar Hubble types (see Fig. \ref{fig:basic}f). Instead, the less massive blue compact dwarfs (BCDs) in \citet{Hunter2004} have comparable 
$\log M_{\rm HI}/L_B$ values to our galaxies (see Fig.~4 in their paper). This can be understood 
as a result of the general increasing trend of $f_{\rm HI}$ with decreasing $M_*$. 

%

Most H$\alpha$ surveys covering the complete morphological spectrum have also been 
limited to distances of less than 30~Mpc, including the SIRTF Nearby Galaxies Survey \citep[SINGS;][]{Kennicutt2003}, 
the H$\alpha$ Galaxy Survey sampling 334 galaxies \citep[H$\alpha$GS;][]{James2004}, 
and the JCMT Nearby Galaxies Legacy Survey of 156 nearby galaxies \citep[NGLS;][]{SG2012}. 
Similar to ours, the sample of the last survey has been HI flux selected 
in order to avoid an SFR-driven selection while ensuring that the galaxies have a rich ISM. 
The Hubble-type distributions of spirals from these surveys are similar, 
peaking around Sc; in fact, the HIghMass sample has a larger fraction of galaxies 
with later types, consistent with the well-known correlation between 
high $f_{\rm HI}$ and late-type morphology. However, even a volume out to 30 Mpc contains 
few truly massive HI disks, with only a handful of galaxies of $M_{\rm HI} > 10^{10}~M_\odot$
found in those surveys.

An exception is the Survey for Ionization in Neutral-Gas Galaxies \citep[SINGG;][]{Meurer2006}, 
consisting of 468 galaxies selected from the HI Parkes All-Sky Survey \citep[HIPASS;][]{Meyer2004}. 
All of their targets are detected in H$\alpha$, confirming that non star-forming galaxies 
with $M_{\rm HI}\gtrsim 10^7~M_\odot$ are very rare. 
SINGG includes galaxies as distant as 80~Mpc, although the majority still have $cz < 3000$~km~s$^{-1}$. 
By comparison, only one of the HIghMass galaxies, UGC~4599 has $cz < 3000$~km~s$^{-1}$, 
whereas our median $cz =$~7660~km~s$^{-1}$ ($D \sim$100~Mpc). 
While SINGG Release 1 \citep{Meurer2006} includes 13 HIPASS sources with 
$10 < \log M_{\rm HI} < 10.3$ and 3 more with $10.3 < \log M_{\rm HI} < 10.6$ (none with $\log M_{\rm HI} > 10.6$), 
multiple galaxies luminous in H$\alpha$ are found to be associated with a single HI source 
for 9/16 of them.  In contrast, we have dropped targets with massive companions to 
minimize such kind of confusion. 
As a result, the HIghMass galaxies, being extremely HI massive, are underrepresented by even the SINGG sample.

In addition, there are several H$\alpha$ surveys of particular galaxy types. 
Analysis of Interstellar Medium of Isolated Galaxies \citep[AMIGA;][]{VM2005} 
compiles 206 galaxies from the Catalog of Isolated Galaxies \citep[CIG;][]{Karachentseva1973}. 
Only UGC~5711 in our sample is included in the CIG. 
\citet{Schombert2011} present a recent H$\alpha$ imaging survey for a large sample of LSB galaxies. 
Although the optical color of their galaxies are blue, being comparable to the 
dwarf galaxies and gas-rich irregulars, their SSFRs are a factor of ten less 
than other galaxies of the same baryonic mass \citep{Schombert2011}. 
In contrast, the HIghMass galaxies exhibit healthy ongoing SF globally (see Section \ref{Dis}), 
neither do they belong exclusively to the isolated population. 




The $f_{\rm HI}$ {\it vs.}~$M_*$ correlation obtained by the GASS survey \citep{Catinella2010} is illustrated along with the ALFALFA population 
in Fig.~\ref{fig:fHI}. The parent sample of GASS is defined by the intersection of the footprints of the SDSS primary spectroscopic
survey, the projected {\it GALEX} medium imaging survey (MIS) and the ALFALFA region. 
Further selection criteria include a stellar mass cut ($10 < \log M_*/M_\odot < 11.5$) and a redshift cut ($0.025 < z < 0.05$). 
The final GASS targets are randomly drawn from this parent sample in a manner which balances the distribution across $M_*$, 
but any sources already detected by ALFALFA are not re-observed. 
By design, the less massive objects 
with $\log M_* < 10$ but high gas fractions are not probed by the GASS. 
Above this $M_*$ cutoff, HIghMass galaxies fall within the extreme tail of the $f_{\rm HI}$ distribution in each $M_*$ bin, 
which are likely to be excluded statistically from the GASS target list. 
Given the $f_{\rm HI}$ and $M_*$ measurements, 
there is only one galaxy, UGC~8802, lying above the dotted line in our Fig.~\ref{fig:fHI}, 
out of the 250 galaxies in total from the GASS final data release \citep{Catinella2013}. 
This galaxy meets all our selection criteria. 
Lying north of Dec.$\sim 35 ^\circ$, it is outside of the footprint of the $\alpha$.40 catalog, 
but has comparable physical properties to the HIghMass galaxies, e.g., 
low and evenly distributed SFR surface density, strong color gradient, and extremely high SSFR in the outer disk \citep{Moran2010}. 
 

Finally, motivated by the study of GASS galaxies, a project entitled ``Bluedisk'' has been devoted to obtaining HI maps 
of 25 nearby galaxies predicted to have unusually high $f_{\rm HI}$ \citep{Wang2013}. 
The sample is selected from the SDSS DR7 MPA/JHU catalogue, using the follow criteria: 
$10 < \log M_*/M_\odot < 11$, $0.01 < z < 0.03$, Dec.~$>30^\circ$, 
with high signal-to-noise ratio (S/N) NUV detection in the {\it GALEX} imaging survey, 
and have high $f_{\rm HI}$ relative to the average trend as predicted from measurements of $\mu_*$, 
NUV$-r$ color, and $g-i$ color gradient of the galaxy. 
Given their final measurements of $M_*$ and $M_{\rm HI}$, however, only one target with the lowest $M_*$ 
among their ``gas-rich'' sample (see black crosses in Fig.~\ref{fig:fHI}) sits above the dotted line, 
denoting one of the selection criteria of the HIghMass galaxies. Again, the HIghMass sample is more extreme
than the ``Bluedisk'' galaxies.

In summary, because of their rarity, the HIghMass galaxies form a unique sample of exceptionally
gas-rich massive HI disks and their study will benefit future HI surveys
of their higher redshift counterparts which are likely to dominate the planned deep field surveys 
by the SKA and its pathfinders. 
The expected number of resolved detections peaks in the HI mass bin of $\sim 10^{10}~M_\odot$ \citep{Duffy2012}, 
for both the ASKAP WALLABY survey (out to $z=0.26$, with a mean redshift of 0.05 at S/N = 6) 
and the DINGO DEEP survey (out to $z=0.26$, with a mean redshift of 0.12 at S/N = 6). 
Therefore, the HighMass galaxies are representative of this class of object, in terms of HI mass. 
Meanwhile, there is an ongoing study of 53 SDSS-selected galaxies at 
$0.16 < z < 0.26$, using $\sim$400 hr Arecibo telescope time (Catinella \& Cortese, in prep). 
Those HI bearers at $z\sim0.2$ are found to be HI-rich, blue, and of low stellar mass surface densities, 
sharing similar properties with the HIghMass galaxies (see below). 
Preliminary analysis shows that the stellar massive HIghMass galaxies align with those $z\sim0.2$ HI detections 
in the $f_{\rm HI}$ {\it vs.}~$M_*$ plot (Fig.~\ref{fig:fHI}), and can be their low redshift analogues (private communication). 
The HIghMass galaxies will be studied along with the $z\sim0.2$ HI bearers to further illustrate this (Catinella \& Cortese, in prep).

\section{Observations and Optical Data}
\label{Obs}

\subsection{H$\alpha$ and {\it R}-band Observations}

All of the H$\alpha$ and {\it R} imaging reported hereafter was obtained in Feb--Mar 2011 over 5 nights 
(2 other nights were lost to bad weather), using the T2KB detector on the 2.1-m telescope at KPNO. 
The CCD has a pixel scale of 0.43$^{\prime\prime}$~pixel$^{-1}$ and a chip size of 2048~$\times$~2048, 
so that all our targets can be easily imaged by a single pointing. 
The detector was used in the 3.1 $e^{-}$~ADU$^{-1}$ gain mode. 
The standard {\it R} filter is used to provide broadband imaging along with continuum subtraction for the narrowband images. 
Due to the wide spread of $cz$, a series of eight H$\alpha$ filters are used, in ascending central wavelength, 
kp1564, kp1565, kp1495, kp1566, kp1496, kp1497, kp1498, and kp1517. 
Their bandpasses are superimposed on the velocity histograms in Fig.~\ref{fig:basic}(a) 
by dotted lines in colors, with peak transmission $\sim75\%$. 
The H$\alpha$ filter for each source is selected to be the one whose central wavelength 
best matches the $cz$ of HI line; in all cases, the FWHM of the filters ($\sim75~\rm \AA$) are sufficient 
to cover the velocity width of the HI line. 
However, in most cases, the [NII]$\lambda\lambda$6548 and [NII]$\lambda\lambda$6584 are also contained
within the filter bandpass. For the sake of clarity, while we will correct for this contamination when we calculate 
the SFR from the H$\alpha$ luminosity in Section \ref{Cal}, we generically refer to H$\alpha$ as the total detected
H$\alpha$+[NII] flux in this section. Because of the lack of an appropriate 
H$\alpha$ filter, the most distant HIghMass galaxy UGC~8797 could not be observed.
Two other HIghMass galaxies UGC~12506 (the lone ALFALFA fall sky target, an image of which has been kindly provided
to us by A. van Sistine and J. Salzer) and UGC~7686 
(which had an extant H$\alpha$ image from the GOLDmine) were also not observed in this run.

We utilized an observing mode consisting of
three H$\alpha$ exposures (15~min each) and two or three {\it R} band exposures (3~min each). 
Calibration frames included afternoon bias and dome flat fields. 
Observations of galaxies were bracketed by standard star exposures of both 
spectrophotometric standards HD19445, HD84937, and BD +26 2606 \citep{Oke1983}, 
and \citet{Landolt1992} standards (in both {\it R} and {\it I} bands yielding the color term), 
at regular intervals throughout the night to calibrate the flux zero points. 
Galaxies requiring the same H$\alpha$ filter were grouped into observations on the
same nights to minimize standard star exposures. 
There were three clear nights under photometric conditions; on these the standard stars were imaged
and then galaxy frames taken on the other non-photometric nights were calibrated by bootstrapping to 
the photometric frames. 
The smoothed images (see Appendix \ref{app:KPNO}) have an average seeing of $\sim1.^{\prime\prime}5$. 
Unfortunately, none of the images of UGC~4599 were taken on photometric nights and
a very bright star near UGC~190277 causes serious bleeding on the CCD. 
As a result, the final H$\alpha$ measurements to be presented are for 29 HIghMass galaxies. 
Unsurprisingly, H$\alpha$ emission is detected in all of our targets as HI-selected galaxies. 

\subsection{H$\alpha$ and {\it R}-band Photometry Catalog}

Details of the image reduction, continuum subtraction, surface photometry, and external checks of the H$\alpha$ data quality can be found in 
Appendix \ref{app:KPNO}. 
Here we present the H$\alpha$ and {\it R}-band measurements from our KPNO images of 29 HIghMass galaxies 
in Table~\ref{tab:KPNO}. 
The [NII] and extinction corrections are applied only when calculating SFRs in this work; 
hence any H$\alpha$ EW values and line fluxes here are in fact for H$\alpha$+[NII], uncorrected for internal or Galactic extinction, 
but corrected for continuum over-subtraction (see Appendix \ref{app:KPNO}). 
However, we have checked that none of these corrections affect the qualitative results to be addressed. 
Columns are as follows:
\begin{itemize}
\item Col(1): The ALFALFA catalog identifier (also known as the AGC number). 
\item Col(2): KPNO H$\alpha$ filter used in our observing run. 
\item Col(3): $\theta$, the position angle (measured eastward from North) 
      and its associated error, measured on the {\it R}-band image, in units of degrees.
\item Col(4): $\epsilon$, the ellipticity and its associated error measured on the {\it R}-band image, defined as $1-b/a$. 
\item Col(5): $\mu_0$, the disk surface brightness extrapolated to the galaxy center 
        as measured on the {\it R}-band image, in units of mag arcsec$^{-2}$, and 
	not corrected for inclination (see Section \ref{Opt}). 
\item Col(6): $r_{\rm d}$, the disk scale length measured on the {\it R}-band image, in units of arcsec. 
\item Col(7): $r_{\rm d, out}$, the disk scale length of the outer disk (see Section \ref{Opt}); the value is null if only
        a single disk is fit. 
\item Col(8): $r_{\rm petro, 50}$, the semi-major axis of the ellipse that encompasses 50\% of the 
        {\it R}-band Petrosian flux, in units of arcsec. 
\item Col(9): $r_{\rm petro, 90}$, the semi-major axis of the ellipse that encompassess 90\% of the 
        {\it R}-band Petrosian flux, in units of arcsec. 
\item Col(10): $d_{25}$, the major axis of the ellipse defined by the {\it R}-band isophote of 25 mag arcsec$^{-2}$, in units of arcsec. 
\item Col(11): $R$, the {\it R}-band magnitude and its associated
	error in the Johnson-Cousins system; with or without superscription if the magnitudes are measured as
        mag$_8$ (the magnitude extrapolated to 8$r_{\rm d}$) or the Petrosian magnitude. 
        See Appendix \ref{app:KPNO} for the definitions of the two types of global magnitudes and validation of their consistency. 
        These magnitudes are not corrected for internal or Galactic extinction. 
\item Col(12): Logarithm of the line flux of H$\alpha$+[NII] and its associated error, 
        in units of erg s$^{-1}$ cm$^{-2}$; with (without) superscription if measured as mag$_8$ (Petrosian magnitude). 
\item Col(13): Equivalent width (EW) of the H$\alpha$+[NII] lines and its associated error, in units of {\rm \AA}. See Appendix \ref{app:KPNO} for
        details of its derivation. 
\item Col(14): Logarithm of the SFR derived from the H$\alpha$ luminosity (see Section \ref{Cal}), in units of $M_\odot$~yr$^{-1}$, 
       with all corrections applied. 
\end{itemize}


\subsection{SDSS photometry Catalog}

In addition to our own broadband {\it R} magnitudes, we use public SDSS images to provide magnitudes
at its filter bands. However, because of their relatively large size, low surface brightness and patchy nature, we 
use our own photometric analysis of the SDSS magnitudes rather than relying on the standard SDSS pipeline which is
not optimized for systems of these characteristics.
Details of the photometry reprocessing, an internal comparison with the KPNO measurements, and external checks of data quality 
(including a comparison with the SDSS pipeline derived values) can be found in Appendix \ref{app:SDSS}. 
Here we present our SDSS measurements for all 34 HIghMass galaxies, 
along with the SED fitting derived quantities to these five bands in Table~\ref{tab:SDSS}. All magnitudes are in the AB system. 
Columns are as follows:
\begin{itemize}
\item Col(1): The ALFALFA catalog identifier (also known as the AGC number). 
\item Col(2)-Col(6): Magnitudes and their associated errors in the five SDSS bands, respectively; 
        with (without) superscription if measured as mag$_8$ (Petrosian magnitude) and
	not corrected for internal or Galactic extinction. 
\item Col(7): $\cos i$, where $i$ is the inclination of the disk. A $\cos i$ value close to zero corresponds to a large 
	correction term for surface brightness deprojection (see Section \ref{Opt}), 
	whereas a $\cos i$ value close to unity corresponds to a large uncertainty in the $V_{\rm rot}$ estimate (see Section \ref{Dis}). 
\item Col(8): $\mu_e(r)$, the {\it r}-band face-on surface brightness at the half light radius, in units of mag arcsec$^{-2}$ (see Section \ref{Opt}). 
\item Col(9): $A_r$, the {\it r}-band internal extinction and its associated error, derived by 
SED fitting using the five SDSS bands \citep[prior $A_r$ distribution applied;][]{Huang2012b}, in units of mag. 
\item Col(10): $M_r$, the final {\it r}-band absolute magnitude after applying all corrections. 
We adopt the $E(B-V)$ measurements by DIRBE \citep{Schlegel1998} to account for foreground reddening,
and the internal extinction is corrected given the $A_r$. 
\item Col(11): Logarithm of the stellar mass and its associated error, derived by SED fitting, in units of $M_\odot$ 
	(see Section \ref{Cal}). 
\item Col(12): Logarithm of the stellar mass surface density in units of $M_\odot$~kpc$^{-2}$ (see Section \ref{Sel}); 
	the value is null if the radius containing 50 percent of the Petrosian flux in {\it z}-band is undetermined. 
\item Col(13): Logarithm of SFR and its associated error, derived by SED fitting, in units of $M_\odot$~yr$^{-1}$ with
     all corrections applied (see Section \ref{Cal}). 
\item Col(14): Logarithm of the birth-parameter, $b \equiv {\rm SFR}/\langle {\rm SFR}\rangle$, and its associated error, 
	derived by SED fitting (see Section \ref{Opt} and \ref{Dis}). 
\end{itemize}

\section{Calculation of SFRs}
\label{Cal}

\subsection{SFRs from SED fitting, SFR(SED)}

Through a coincidence of dust and stellar population evolution physics, the dust-age-metallicity degeneracy actually helps 
in the estimate of $M_*/L$ from optical colors \citep{Taylor2011}. 
While H$\alpha$ is recognized as an instantaneous indicator sensitive to SF activity over the timescale of 
$\sim$10~Myr, images obtained in the SDSS {\it u}-band respond to variations in the SFR over 100-500~Myr. 
Our future study of the {\it Spitzer} IRAC imaging will trace the low-mass star population which dominates the mass 
budget in galaxies, so as to better constrain the $M_*$. 
In this section, we make use of the photometry derived in Appendix \ref{app:SDSS} to perform SED fitting to the five SDSS bands
in order to obtain the $M_*$ and SFR estimates. 

Further details of the method and fitting quality for the $\alpha$.40--SDSS (DR7) sample are found in \citet{Salim2007} and 
\citet{Huang2012a}. 
We use the \citet{Bruzual2003} stellar population synthesis code to construct the model SEDs, adopting a \citet{Chabrier2003} IMF. 
Random bursts are allowed to be superimposed on a continuous star formation history (SFH). 
We correct the observed magnitudes for Galactic reddening and implicitly apply $K$-corrections by convolving the 
redshifted model SEDs with SDSS bandpasses in the rest frame. 
The full likelihood distributions of parameters are derived following a Bayesian approach, 
so that the errors due to model degeneracies are properly characterized by the width of the resulting probability density functions. 
In addition, the Gaussian prior distribution of the effective optical depth in the {\it V} band, $\tau_V$, is applied, 
accounting for the fact that the extinction depends not only on the inclination 
but also on the luminosity of galaxy \citep{Giovanelli1997}. 
As a result, the fitting-derived internal extinction is improved \citep{Huang2012b}, in terms of both systematic 
and random uncertainties. 
Although dust has little effect on the $M_*$ estimates, the uncertainty of the extinction correction can be the dominant term in the 
estimate of the SFR from optical or UV SFR indicators, e.g., H$\alpha$. 
Approximately half of the starlight emitted in the optical and UV is absorbed by interstellar dust and re-radiated at 
infrared wavelengths \citep{Kennicutt2012}, so that it is important to assume a prior $\tau_V$ distribution 
to better constrain the extinction. 
Our SED fitting-derived measures of $M_*$, SFR(SED), and $A_r$ are given in Table~\ref{tab:SDSS}. 
We also follow the same fitting procedure as for the $\alpha$.40--SDSS (DR8) sample, yielding physical quantities of the 
general population to be presented together with the HIghMass galaxies in Section~\ref{Dis}.  

\subsection{SFRs from $L_{\rm H\alpha}$, SFR(H$\alpha$)}

H$\alpha$ emission traces stars with masses greater than $\sim15~M_\odot$ and hence provides a measure of very recent massive SF. 
In addition to the need to assume an IMF to extrapolate the number of high mass stars (relevant for most SFR indicators), 
H$\alpha$-based SFRs need large and uncertain extinction corrections (like UV and other blue indicators, e.g., [OII] and H$\beta$). 
Central AGN may also contribute to the observed H$\alpha$ line emission (like [OII], [OIII], radio, and IR indicators).  
The specific drawbacks of H$\alpha$-based SFRs is 
contamination in most of the filters used in the observations by the [NII]$\lambda\lambda$6548 and 6584~$\rm \AA$ lines. 
In general, the largest systematic errors are dust attenuation and sensitivity to the population of the upper 
IMF in regions with low absolute SFRs \citep{Kennicutt2012}. 
Given their integrated SFRs, the incomplete IMF sampling has no impact on the HIghMass galaxies, 
and, the uncertainties in the internal extinction corrections to the H$\alpha$ fluxes dominate over 
photometric errors as the main error in the derived SFRs. 

We derive the SFR from the H$\alpha$ luminosity adopting a recent H$\alpha$ calibration in \citet{Kennicutt2012}, 
\begin{equation}
\label{eqa:sfr}
	\log {\rm SFR}~[M_\odot~{\rm yr}^{-1}] = \log {L_{\rm H\alpha}}~[{\rm erg~s^{-1}}] - 41.27 + 0.06. 
\end{equation}
Here, the last term converts a Kroupa IMF used in their work to a \citet{Chabrier2003} IMF used by us \citep{Bell2003}. 
We note that the correction for stellar absorption underlying the H$\alpha$ emission is already taken care of by the 
continuum subtraction \citep{Kennicutt2008}. 
In addition to the (1) continuum over-subtraction and (2) Galactic extinction corrections explained in Appendix \ref{app:KPNO}, 
the following corrections are applied on the observed $L_{\rm H\alpha}$: 


(3) [NII] contamination: Although SDSS spectra are available for 33/34 of the HIghMass galaxies, 
they are limited by the small aperture (3 arcsec) of the fibers, covering only the nuclear regions in the majority. 
Additionally, according to the BPT diagram \citep{Brinchmann2004}, 
there are five AGNs, 
two low S/N LINERs, 
and two composite galaxies 
with both AGN and SF contributions among our sample. 
The nuclear [NII]/H$\alpha$ ratios are most likely to deviate from the overall values. 
Alternatively, we adopt the scaling relation between [NII]/H$\alpha$ and $M_B$ given by \citet{Kennicutt2008}, 
\begin{eqnarray*}
	\log ({\rm [NII]\lambda\lambda6548,6584/H\alpha}) = 
	\left\{
	\begin{array}{lr}
		-0.173 M_B - 3.903,~{\rm if }~M_B > -21,\\
		\log 0.54,~~{\rm if }~M_B \leq -21,
	\end{array}
	\right.
\end{eqnarray*}
where $M_B$ comes from the combined SDSS magnitudes, assuming $h=0.75$ as in \citet{Kennicutt2008}. 
Despite a large scatter, we have confirmed this relation to be systematically consistent with the SDSS flux measurements, 
excluding the AGNs. 
We further adopt a line ratio [NII]$\lambda\lambda$6584/[NII]$\lambda\lambda6548 = 3$ to exclude the 
[NII]$\lambda\lambda$6584 contribution in five galaxies because it falls outside of the H$\alpha$ filter. 

(4) Dust extinction correction: We make use of the continuum extinction given by the SED fitting and 
assume a constant ratio between the nebular line and stellar continuum extinctions at the same wavelength, 
$A_r = 0.44 A_{\rm H\alpha}$ \citep{Calzetti2000}. 
Due to the same concern raised by the small fiber aperture, we choose here not to rely on the Balmer decrement derived from 
the SDSS spectra in order to infer $A_{\rm H\alpha}$. 
Use of the Balmer decrement may lead to an underestimate of the extinction because lines of sight with low extinction 
are more heavily weighted within the beam \citep{Kennicutt2012}, 
or an overestimate of the extinction induced by the fact that the disk central regions are 
less transparent \citep{Giovanelli1994}. 
In the future, we plan to explore this issue further with long-slit optical spectroscopy and {\it Herschel} data. 

The final SFR(H$\alpha$) values with these corrections applied are presented in the last column of Table~\ref{tab:KPNO}. 
The SFR(H$\alpha$)s for the HIghMass galaxies
range from 0.34 to 21~$M_\odot~{\rm yr^{-1}}$, with a median value of 2.5~$M_\odot~{\rm yr^{-1}}$, 
significantly higher than the median SFR found by the other H$\alpha$ surveys, e.g., 11HUGS, H$\alpha$GS, or SINGG. 
The difference cannot be attributed to the different corrections we applied because a comparison of the 
$L_{\rm H\alpha}$ distributions (only continuum over-subtraction corrected and Galactic extinction corrected) 
show as well that the HIghMass galaxies tend towards high $L_{\rm H\alpha}$. 
This result is in contrast with previous understanding that HI-selected galaxies have somewhat 
lower $L_{\rm H\alpha}$ and SFRs \citep{SG2012}, but agree with the finding in \citet{Huang2012b} 
that the HI-rich galaxies have, on average, higher SFRs at fixed $M_*$. 

\subsection{SFR(SED) vs. SFR(H$\alpha$)}

\begin{figure*}
\center{
\includegraphics[scale=0.85]{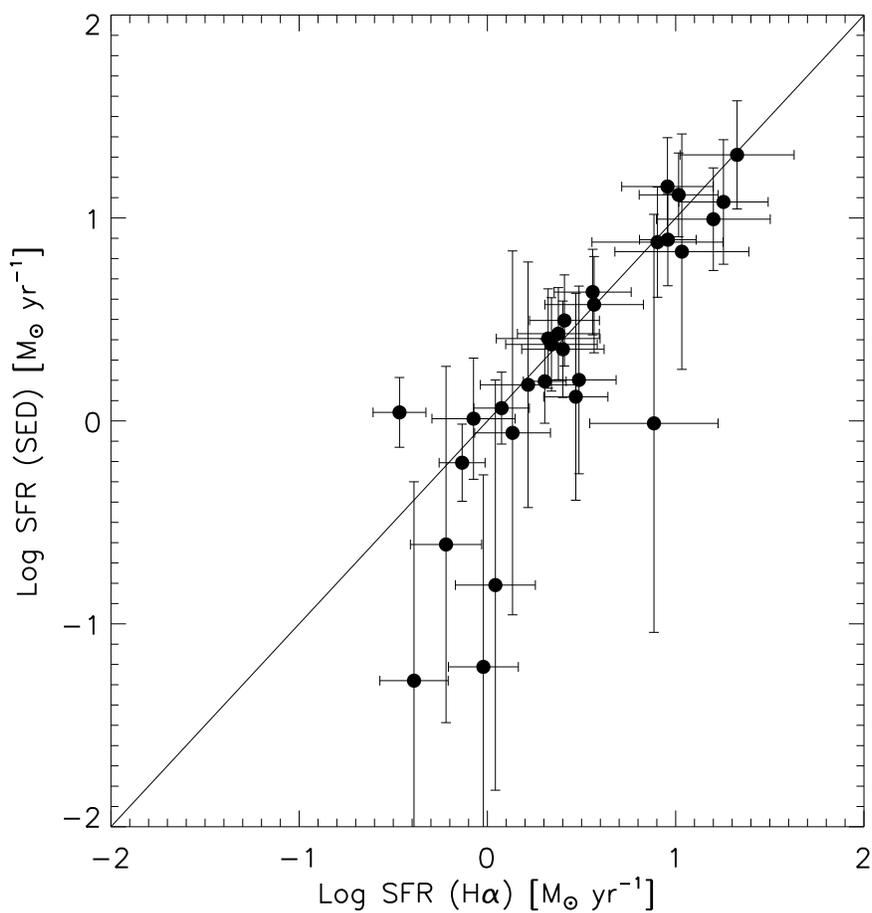}
}
\caption[]{
A comparison between the SFR(SED)s and SFR(H$\alpha$)s with all corrections applied. 
The two estimates generally agree within the uncertainty, whereas the SFR(H$\alpha$)s are preferred in cases of disagreement. 
}
\label{fig:SFRcom}
\end{figure*}

Fig.~\ref{fig:SFRcom} shows the comparison of the values of the SFR derived from the SEDs,
SFR(SED), and from the H$\alpha$ luminosity, SFR(H$\alpha$)s, for the HIghMass galaxies. For the majority,
the two quantities are in rough agreement within the uncertainties. 
Therefore, the standard calibration of the H$\alpha$ SFR indicator, assuming a \citet{Chabrier2003} IMF, is applicable to our galaxies. 
However, for a few sources, the SFRs derived from SED-fitting result in much lower values by almost a factor of ten. 
In such cases, huge error bars on the SFR(SED)s result from broad probability density functions of the SFR estimates, 
suggesting that the SFR(H$\alpha$)  values represent the more realistic result. 
We further assess this conclusion by referring to the {\it GALEX} archive for NUV measurements; note that FUV measures
which would also be useful are available for a smaller subset.
With a poorer resolution ($\sim4$~arcsec FWHM) and a larger impact from dust extinction, the NUV images are less 
valuable for the purpose of tracing SFRs than our high quality H$\alpha$ data, but we can use them 
here to judge the relative reliability of SFR(SED) and SFR(H$\alpha$). 
Excluding the shredded pipeline NUV magnitudes, we apply corrections for Galactic extinction, internal extinction 
(self-consistent values given by the SED fitting), and $K$-corrections. The end result is that
the inferred SFRs from the final NUV luminosity, adopting a \citet{Kennicutt2012} calibration, 
reach better agreement with the higher values of SFR(H$\alpha$) than with the SFR(SED)s. 

In summary, we have double checked the robustness of the SFR(SED)s and SFR(H$\alpha$)s, adopting a 
conservative error analysis to conclude that the H$\alpha$-derived SFRs provide more reliable measures of the
SFR in those cases where the two approaches differ. We conclude that there is no evidence that the exceptional gas 
fractions of the HIghMass galaxies result from abnormal levels of SF at least among the massive stars.
We can also rule out a significant decaying or rising SF history in the last $\sim$10~Myr relative to the last 100-500~Myr. 

\section{Optical Characteristics of the HIghMass galaxies}
\label{Opt}

In this section, we examine the characteristics of the HIghMass galaxies as a class by 
studying their detailed optical properties relative to other samples of galaxies. 
We first take a look at the global EW$\rm_{H\alpha+[NII]}$ values and then inspect various radial profiles. 
The HII region LFs of six HIghMass galaxies are also presented. 
Intriguing features indicative of unusual behavior in their gas accumulation and SF,
are identified in some systems as discussed below.

\subsection{H$\alpha$ Properties}

\subsubsection{The Birth Parameter b}

In order to understand the EW${\rm_{H\alpha+[NII]}}$ distribution of the HIghMass galaxies, we have to account for the potential 
variation of EW with luminosity, considering the different luminosity ranges probed by other H$\alpha$ surveys. 
A number of works have shown controversial results regarding this issue: 
a weakly declining trend of EW with brighter $M_R$ was seen in \citet{Jansen2000b} and \citet{Lee2007}, 
in contrast to the conclusions of others who found no significant trend between the two quantities \citep{James2004, SG2012}. 
We investigated this issue in the HIghMass galaxies 
and found no significant correlation between the two. Thus we can directly compare the EW${\rm_{H\alpha+[NII]}}$ distribution 
with the other surveys. 


The H$\alpha$ EW is related to the ratio of current SF to the past-averaged level, i.e., 
the ``birth parameter'' (see also Section \ref{Dis}). 
For instance, the EW distribution of galaxies included in the SINGG is centered and peaked at $\sim 24~{\rm \AA}$, 
indicating a $b \sim 0.2$ \citep{Meurer2006}. 
In the luminosity range of $-22 < M_B < -19$, \citet{Lee2007} obtained $\langle \log {\rm EW} \rangle = 1.17$ ($\approx$ 15 \AA); 
$\sigma = 0.40$. 
The EWs of the NGLS galaxies, also being HI selected, range from 1 to 880~${\rm \AA}$ with a median value of 27~${\rm \AA}$. 
For the later-type spirals (Sc-Sm), \citet{Kennicutt1983} and \citet{James2004} obtain slightly higher average EWs: 29~$\rm \AA$ 
and 35~$\rm \AA$, respectively. 
By comparison, the HIghMass galaxies have $\langle \log {\rm EW} \rangle = 1.41$ ($\sigma = 0.26$) and the average is 30~$\rm \AA$. 
In fact, our average is nominally higher than most of the others \citep[except for][]{James2004}, 
although not significant given the dispersion. 
The galaxy with the lowest $M_*$ in our sample, AGC~203522, has the highest EW (90.66~${\rm \AA}$), 
so that can be classified as a starburst galaxy. 
This rules out the hypothesis that the high $f_{\rm HI}$ values in the HIghMass galaxies result from gas being inhibited from 
converting into stars {\it at the present epoch.}
Either past SF has been suppressed, perhaps due to the high $\lambda$ value of the DM halo, 
or the galaxy has undergone recent gas accretion, giving it a fresh supply of gas and enhancing the star formation
rate. Based on the initial examination of the HI synthesis mapping data for two representative HIghMass galaxies,
UGC~9037 and UGC~12506,
\citet{Hallenbeck2014} find that both explanations may be applicable.

\subsubsection{The Radial Distribution of SFR Surface Density}

As discussed in Appendix \ref{app:KPNO}, we use the continuum-subtracted H$\alpha$ images to 
construct azimuthally-averaged radial profiles of the surface brightness $\Sigma_{\rm H\alpha+[NII]}(r)$ for each galaxy. 
These $\Sigma_{\rm H\alpha+[NII]}$ radial profiles are expected to be less smooth than the starlight since they 
represent a shorter-lived evolutionary stage.  
Here we convert them to $\Sigma_{\rm SFR}$ for our future studies of the SF threshold and associated K-S relation. 
In addition to the global corrections applied to the total $L_{\rm H\alpha}$ discussed in Section \ref{Cal}, 
including Galactic extinction, 
and [NII] contamination corrections, we follow \citet{Leroy2008} to deproject 
the $\Sigma_{\rm H\alpha}$ by a factor of $\cos i$, where $i$ is the inclination of the disk. 
We do not apply a spatially-resolved internal extinction correction in order to be able to compare with other
authors who use H$\alpha$ ({\it vs.}~IR) emission to characterize unobscured ({\it vs.}~obscured) SF, separately. 
Deriving the radial profile of internal extinction is beyond the scope of this work but will be part
of our future work using  
Balmer decrements from long-slit optical spectroscopy. 

We present the $\Sigma_{\rm SFR}$ radial profiles of 29 HIghMass galaxies in Fig.~\ref{fig:spf}, in order of ascending 
integrated SFR(H$\alpha$), with galactocentric radius normalized by $r_{25}$. 
The vertical dash-dotted lines mark the inner and outer edges of the assumed disk regions (see Appendix \ref{app:KPNO}). 
Previous works show that
the typical value of the 
SFR$_{25}$ defined as the integrated SFR normalized by the area $\pi r_{25}^2$, 
is 10$^{-3}~M_\odot$~yr$^{-1}$ among irregular galaxies (Im) \citep{Hunter2004}. 
Typical spirals reach this $\Sigma_{\rm SFR}$ between 0.35$r_{25}$ and 0.81$r_{25}$, 
the average being 0.5$r_{25}$ \citep{Kennicutt1989c}. In terms of $\Sigma_{\rm SFR}$, the Im 
galaxies more closely resemble the outer parts of spirals. 
The HIghMass galaxies have overall a {\it lower} 
SFR$_{25}$ relative to the Sab-Sd galaxies in \citet{Kennicutt1989c}, 
but, marginally, a higher 
SFR$_{25}$ relative to the Im galaxies in \citet{Hunter2004}. 
This result is consistent with the earlier finding that the galaxies with higher ratios of 
gas relative to their luminosity or total baryonic mass have lower $\Sigma_{\rm SFR}$ \citep{Hunter2004}.
In fact, the $\Sigma_{\rm SFR}$ drops below 10$^{-3}~M_\odot$~yr$^{-1}$ within 0.35$r_{25}$ in nine of the
29 HIghMass galaxies (see Fig.~\ref{fig:spf}). Compared to a subset of the SINGS galaxies that are dominated by HI over H$_2$ 
\citep{Bigiel2008, Leroy2008}, most being late-type spirals or dwarf irregulars, the HIghMass galaxies have, on average,
slightly higher $\Sigma_{\rm SFR}$. However, many of the massive spirals in the SINGS with H$_2$ dominated centers 
have significantly higher $\Sigma_{\rm SFR}$ than those found here. Furthermore, although
it was found that most of the SF activity takes place in the Im galaxies within 3$r_{\rm d}$ \citep{Hunter2004}, 
H$\alpha$ emission is traceable beyond that radius in most of the HIghMass galaxies, and in some, to as far as 6$r_{\rm d}$,
e.g., UGC~6043, UGC~6967 (LSB galaxy), UGC~5648, and AGC~248881. 

Although the HIghMass galaxies have high integrated $f_{\rm HI}$s, $L_{\rm H\alpha}$s 
(absolute value), and EWs (vigorous current SF relative to the past), the SF activity in them 
is spread throughout the disks and, conversely, centrally concentrated intense SF is uncommon. 

\subsubsection{EW${\rm_{H\alpha+[NII]}}$ Radial Profiles}

We use the same set of tilted rings to measure the $\Sigma_{\rm H\alpha}$ and $\Sigma_{R}$ radial profiles, 
yielding the EW${\rm_{H\alpha+[NII]}}$ radial profiles as presented for all of the 29 KPNO targets in 
Appendix \ref{app:KPNO}. 
Comparable EW profiles are shown in \citet{James2004}, among which the centrally-concentrated 
SF, characterized by a central peak in the EW curve and a decline with radius, is quite common. 
Similarly, the EWs of BCDs are most often found to drop steeply with radius, implying that the SF has 
migrated to the center within the last Gyr \citep{Hunter2004}. 
Only two HIghMass galaxies exhibit prominent central EW peaks: 
UGC~9023 and UGC~9037, both likely members of the poor cluster Zw1400+0949 \citep{Giovanelli1985}. 
According to the BPT diagram derived from their SDSS spectra, AGN activity makes little contribution to the 
line emission in their nuclear regions. The cluster environment may play a role in the inwards 
gas driven and subsequently the enhanced nuclear SF. Our JVLA maps of UGC~9037 show strong
evidence of inward streaming motions and may suggest
the inflow of recently accreted gas \citep{Hallenbeck2014}.

On the other hand, most HIghMass galaxies have higher EWs in the outer regions of their disks, 
implying younger outer disks relative to the older central bulges dominated by the optical continuum emission. 
Such strong variation of the EWs prompts a caveat of a widely-used method to correct for the aperture effect 
when using nuclear emission line strengths to infer global SFRs, i.e., scaling the SDSS nuclear SFRs by the 
ratio of the nuclear and the overall broadband luminosities \citep[e.g.,][]{Hopkins2003}. 
In particular, the EW profiles rise almost monotonically in AGC~203522, 
AGC~726428, AGC~721391, and UGC~6692; 
the stellar population becomes increasingly dominated by young stars with radius. A
similar behavior of the EW is found in UGC~8802 \citep{Moran2010}, consistent with a scenario of 
inside-out disk growth as such the SF activity migrates outwards. 

\citet{James2004} suggest a relationship between moderately enhanced SFRs and the presence of bars; 
the five galaxies with the highest SFRs in their sample are all barred. 
A characteristic structure of the radial EW profile is always found in barred galaxies: 
a strong central peak, followed by a broad dip at intermediate radius (a SF ``desert'' in the region swept out by the bar), 
and a gentle outermost rise to the plateau level at 15-30~$\rm \AA$ (substantial SF in HII regions scattered around the disk). 
However, \citet{Masters2012} have shown that the bar fraction is significantly lower among gas-rich 
disk galaxies than gas-poor ones. 
According to the morphology classifications of Galaxy Zoo 2 \citep{Willett2013}, 
only two HIghMass galaxies have strong bars (UGC~8089 and UGC~8408), 
and additional 10 are very likely to have weak bars. 
In fact, UGC~8089 and UGC~8408 both have relatively low SFRs and only weak central peaks 
are seen in their EW profiles. It appears then that bars have not left clear imprints 
in the SF activity of the HIghMass galaxies.

%

\subsubsection{HII Region Luminosity Function}

Other useful measures of the state of massive star formation in galaxies consider the properties of the luminosity
distribution of their HII regions: the brightest HII region or the HII region LF. 
The brightest HII regions have $L_{\rm H\alpha} \sim 10^{(39-41)}~{\rm erg~s^{-1}}$ 
and are brighter on average in galaxies with later Hubble types, 
indicating a physical change in the HII region populations \citep{Kennicutt1988}. 
Notably, \citet{Kennicutt1989a} analyzed 30 galaxies with 
Hubble type ranging from Sb to Irr and found that the 
differential HII LFs can be parameterized as $dN/dL_{\rm H\alpha} \propto L_{\rm H\alpha}^{-2 \pm 0.5}$ for 
luminosities $L_{\rm H\alpha} \gtrsim 10^{37}~{\rm erg~s^{-1}}$. 
Both the normalization and shape of the LF change systematically with Hubble type \citep{Caldwell1991, Helmboldt2005}. 

\citet{Kennicutt1989a} also found that a subsample ($\sim20\%$ of the total) of the galaxies is better described by a double 
power-law LF with a break in the slope at $L_{\rm H\alpha} \sim 10^{38.7}~{\rm erg~s^{-1}}$. 
\citet{Oey1998} suggest that the break at an H$\alpha$ luminosity of $L_{\rm br} \sim 10^{(38.5-38.7)}~{\rm erg~s^{-1}}$ 
occurs around the luminosity contributed by a single star at the upper mass limit of the Salpeter IMF 
(100~$M_\odot$, $38.0<\log l_{\rm up}<38.5$), 
and they refer to the regions ionized by rich clusters with good stellar statistics as ``saturated'' ($L_{\rm H\alpha} > L_{\rm br}$). 
The LF is predicted to fall off more steeply beyond $L_{\rm br}$. 
Therefore, to characterize the shapes of the LFs tracing the distribution of 
masses for rich clusters full sampling IMF, \citet{Helmboldt2005} fit a power law only to 
$L_{\rm H\alpha} > L_{\rm br}$. 
However, the hypothesis that the break in slope has a physical nature rather than being an observational artifact has been 
under suspicion for a long time \citep{Liu2013}. 
Blending due to limited spatial resolution can induce catalog incompleteness in crowded environments. 
As a result, the observed LF at the faint end is artificially flattened 
and the turnover point of the LF is shifted to higher $L_{\rm H\alpha}$ \citep{Thilker2000, Helmboldt2005}. 
Specifically, experiments of degrading the resolution to 200--400~pc 
causes a significant turnover of the LF at the faint end but the shape of the {\it upper} LF is still preserved. 
Beyond a resolution of 300--500~pc, however, the blending affects the entire LF and causes a spurious increase 
in the luminosities of the first-ranked HII regions \citep{Kennicutt1989a}. 

Previous studies of the HII LF are mostly restricted to galaxies 
within 40~Mpc, and care here must be taken to account for blending in crowded regions. 
Therefore, we will fit a power law only to the upper LF, which is less vulnerable to blending, 
in hope to characterize the regions ionized by rich clusters in a ``saturated'' regime, following \citet{Helmboldt2005}. 
Most of the HIghMass galaxies lie sufficiently distant that limits in resolution prohibit the determination of a robust HII region LF. 
We examined the HII regions of the six HIghMass galaxies with the best linear resolutions. 
At distances of 88.5~Mpc to 110.3~Mpc, the final resolutions
(corresponding to the PSF FWHM of the frame with the worst seeing; see Appendix \ref{app:KPNO}) amount to
534--630~pc at their distances. Although some star forming complexes are measured to have sizes of 
820--1080~pc \citep{Caldwell1991}, 
the common star forming complexes in nearby galaxies (with diameters of $\sim$100--200~pc) 
will be unresolved by our dataset, and the trend at the faint end of the LF will not be 
accessible. Furthermore, the luminosities of the first-ranked HII regions will have to be interpreted with care.

In order to extract the HII regions from these six galaxies, 
we make use of the IDL program ``HIIphot'' \citep{Thilker2000}. 
Three combined images per galaxy are supplied to the program so that the noise level is estimated: 
a net H$\alpha$ image after continuum subtraction, an {\it R}-band image as `OFF' frame, and an H$\alpha$ 
image before continuum subtraction as `ON' frame. 
HIIphot has a robust and automatic algorithm to identify HII region seeds and then grow the HII regions from these seeds 
until it meets another HII region or reaches a given emission measure (EM) gradient. 
This approach provides the most substantial benefit during analysis of highly resolved systems, 
but is sufficiently general to work well also for distant galaxies \citep{Thilker2000}. 
The diffuse ionized gas contribution is determined and subtracted as a background contribution by the program, 
and both luminosities and S/Ns of the identified regions are measured. 
The emission line surface brightness is given in units of rayleighs, defined as 
1$R = 10^6/4\pi~{\rm photons~cm^{-2}~s^{-1}~sr^{-1}} = 5.67\times10^{-18}~{\rm erg~cm^{-2}~s^{-1}~arcsec^{-2}}$, 
and EM = 2.78$R~{\rm pc~cm^{-6}}$ for an assumed electron temperature $T_e = 10^4~$K \citep{Meurer2006}. 
We incorporated the corrections into our calibration for Galactic extinction, continuum oversubtraction, 
and [NII] contamination (calculated in Section \ref{Cal}) but not for internal dust extinction, 
to be consistent with other similar HII LF studies. 

HII LFs of the six HIghMass galaxies are presented in Fig.~\ref{fig:LF}. 
We adopt the convention to bin the number counts in luminosity logarithmically instead of linearly, 
i.e., $dN/d\log L \propto L^\alpha$, or equivalently $dN/dL \propto L^{\alpha-1}$. 
Only the HII regions detected with S/N $>$ 5 are plotted, following the detection limit as recommended by \citet{Thilker2000}. 
The vertical dotted line in each panel shows the corresponding luminosity to the 5$\sigma$ detection limit 
derived from a fit to the well-defined luminosity {\it vs.}~S/N correlation. 
The 5$\sigma$ detection limits range from $\log L_{\rm H\alpha} = $~37.92 to 38.20, 
lying well below the slope break luminosity of $\log L_{\rm br} \sim 38.5-38.7$, 
reaffirming that we have the sensitivity to trace the shape of the {\it upper} LF. 
The dashed line in each panel is the best-fit power law, 
with the $\alpha$ value listed in the upper right corner. 

The linear resolutions of the six H$\alpha$ images used here are already in a regime for which 
\citet{Kennicutt1989a} conclude that blending affects the entire LF and
spuriously increases the luminosities of the first-ranked HII regions. 
No attempt is made to address the existence of the slope break or the behavior at the faint end. 
In fact, the LFs drop significantly at the faint end as expected for the case of 
moderate resolution \citep{Kennicutt1989a, Thilker2000}, 
also being consistent with the adopted 5$\sigma$ detection limit. 
The fit to a power law is performed only to points above the apparent turnover point (which itself 
is likely to be an observational artifact) toward the bright end. 
Perhaps surprisingly, our $\alpha$ values are still in excellent agreement with 
the result of \citet{Kennicutt1989a}: 
$\alpha=-1 \pm 0.5$ in that work (converted to logarithmic binning) relative 
to an average of $-$0.96 and a median of $-$1 in this work. 
Errors in the $\alpha$ values are given in the upper right corners of all panels in Fig.~\ref{fig:LF}, 
ranging from 0.08 to 0.33. 
We confirm the statement in \citet{Thilker2000} and \citet{Kennicutt1989a} that the slope of 
the LF above the low-luminosity turnover was
rather insensitive to the ``upward contamination'' potentially brought about by the blending. 
The somewhat unexpected agreement of LF slope indicates that the HII regions in these 
galaxies are not particularly crowded but rather, are sparsely distributed throughout the disks
so that blending is alleviated at least partly. It is also consistent with our previous suggestion
that the HIghMass galaxies exhibit lower overall $\Sigma_{\rm SFR}$. 

For the majority, a power law fit can easily characterize the bright end of the HII LF. 
The only exception, UGC~9037, is unusual among the HIghMass galaxies in that
it has centrally concentrated SF and, as demonstrated
by \citet{Hallenbeck2014}, shows evidence from the HI synthesis mapping of inwards streaming motion of 
gas in its central regions.  The blending of HII regions in its nucleus is likely the cause of its 
peculiar LF shape with multiple gaps (see Fig.~\ref{fig:LF}), 
so that the fitting for $\alpha$ value fails. Alternatively, 
its distinctive HII LF may be due to intrinsically different physical properties 
of the HII regions situated in its galactic nucleus, in the circumnuclear region and
its more extended disk \citep{Kennicutt1989b}.

In contrast to the  more robust values of $\alpha$ determined by detailed studies
which include hundreds of HII regions per galaxy, we hesitate to draw here any strong
conclusion about the luminosity of the first-ranked HII regions. We note only that the six
HIghMass galaxies studied here show $\log L_{\rm H\alpha}$ values of the first-ranked HII regions 
from 39.2 to 40.1, within the range derived in \citet{Kennicutt1989a}. 
In contrast to the LFs for Sa galaxies which are very steep with few or no regions having 
$L_{\rm H\alpha} > 10^{39}~{\rm erg~s^{-1}}$ \citep{Caldwell1991}, 
most of our LFs extend well beyond that. The one with the faintest first-ranked HII region, UGC~9023, 
also has the lowest normalization as well as the steepest bright end slope of LF. It happens to be the only galaxy 
among these six which is included in our list of LSB galaxies (see below), 
suggesting a possible correlation between the surface brightness of the host galaxy 
and the luminosity of the first-ranked HII region \citep{Helmboldt2005}.
Most importantly, the bright end shape of the HII LF, as characterized by $\alpha$, demonstrates 
no evidence of any abnormal luminosity distribution of the star forming regions in 
this ``saturated'' regime, as might have been invoked to explain their higher than average 
gas fraction. 

\subsection{Broadband Properties}

\subsubsection{LSB Galaxies in the HIghMass sample}

The correlation between gas richness, LSB disks, and high $\lambda$ halos are predicted 
by both semi-analytical models and hydrodynamic simulations of galaxy formation 
\citep[e.g.,][]{Mo1998, Boissier2000, Fu2010, Kim2013, Kravtsov2013}. 
A first question that arises is whether the HIghMass galaxies are just newly-identified
members of the class of giant LSB galaxies typified by Malin 1. Indeed, with log $M_{\rm HI} = 10.82$
and $f_{\rm HI} \sim 0.87$ \citep{Lelli2010}, Malin 1 meets the selection critera of the HIghMass sample, but its 
distance places it outside the volume sampled by ALFALFA, i.e., its HI emission is redshifted
below the frequency range covered by the ALFALFA bandpass. Among the HIghMass galaxies,
three - UGC~6536, AGC~190796 (Malin-like), and UGC~6967 - have been part of previous studies of 
LSB galaxies \citep{Bothun1985b, Schombert1992, Sprayberry1995}. 
The overall lower $\Sigma_{\rm SFR}$ of the HIghMass galaxies is already indicative of the low surface brightness in optical. 
Here we use the broadband images
to estimate two different widely-used measures of surface brightness to evaluate 
which of the HIghMass galaxies may belong to the LSB class. 



The first test uses the effective surface brightness, $\mu_e$, defined as the 
face-on surface brightness at the half light radius, $r_{\rm petro,50}$. 
We correct the {\it R}-band $\mu_e$ for Galactic extinction and convert it to the AB magnitude system, 
to be consistent with the SINGG calculation. First, the cumulative histogram of $\mu_e(R)$ demonstrates that 
the HIghMass galaxies have, on average, lower surface brightness than the SINGG galaxies as shown in
Fig.~15 of \citet{Meurer2006}.  We also use the SDSS photometry to calculate $\mu_e(r)$, shown 
as the distribution of points along the $x$-axis in Fig.~\ref{fig:SB}. Using a low redshift sample
of galaxies selected from the SDSS spectroscopic sample which is itself biased against LSB galaxies 
at low luminosity, \citet{Blanton2005} identify LSB galaxies as those with $\mu_e(r)$ fainter than 23.5 
~mag~arcsec$^{-2}$ (the vertical dashed line in Fig.~\ref{fig:SB}). By that criterion,
11/34 HIghMass galaxies would be classified as LSB galaxies (see Table \ref{tab:SDSS}):
AGC~188749 (marginally), AGC~190796, AGC~190277, AGC~203522, UGC~6043, AGC~213964, UGC~7220 (marginally), 
UGC~7686, UGC~8089, UGC~9023, and UGC~12506 whereas UGC~6536 and UGC~6967, identified as LSB earlier,
would not be. The classification of LSB by this single measure seems insufficient.
 
Since the $\mu_e$ values can be easily affected by the bulge component, a fairer estimate uses a
property of the disk itself, specifically 
the disk surface brightness interpolated to the center, $\mu_0$.
Using $\mu_0$ as the measure of surface brightness, LSB galaxies are those which have central 
disk surface brightnesses substantially below the canonical Freeman value for normal disks. 
As for $\mu_e$, an inclination correction is applied to give a face-on value of $\mu_0$. It should be
noted that the two
quantities $\mu_e$ and $\mu_0$ measured in the same band assuming a perfect exponential light profile 
should be related by $\mu_0 = \mu_e - 1.822$. For comparison with the study of LSB galaxies 
in \citet{Schombert2011}, we use the combined SDSS magnitudes to characterize 
$\mu_0(V)$, $V = g - 0.5784(g - r) - 0.0038;  \sigma = 0.0054$, plotted on the $y$-axis in Fig.~\ref{fig:SB}. 
Compared to the central surface brightness $\mu_c(V)$ of the LSB galaxies in \citet{Schombert2011}, 
the $\mu_0(V)$ values of the HIghMass galaxies are brighter in general. 
The majority of HIghMass galaxies (28/34) have $\mu_0(B)$ fainter than the Freeman value 
\citep[21.65~mag~arcsec$^{-2}$;][]{Freeman1970}. 
Adopting a standard cutoff of $\mu_0(V) > $23~mag~arcsec$^{-2}$,
four HIghMass galaxies fall in the category of extreme LSB galaxies: 
UGC~190277, UGC~6066, UGC~9234, and UGC~12506, all of which fall below the horizontal 
dashed line in Fig.~\ref{fig:SB}. Of these, UGC~6066 and UGC~9234 do not appear
among the LSB systems identified by the $\mu_e(r)$ criteron above. Due to the presence of 
both prominent bulges and extended outer disks in these galaxies, 
their $\mu_e$ is significantly brighter than the $\mu_0 + 1.822$ in the same band. 

Perhaps not surprisingly given the general association of ALFALFA and star-forming populations
\citep{Huang2012b}, the HIghMass galaxies have on average lower surface brightness 
relative to the galaxies included in optically-selected samples, but they do not in general meet the surface brightness
criteria of extreme LSB galaxies. As a class, the HIghMass galaxies are not ``crouching giants''.

\subsubsection{Broken Exponential Disks}

The structural properties of the outer disks must be intimately linked to the mechanisms 
involved in the growing and shaping of the host galaxies. 
For example, in a study of nearby late-type spirals selected from the SDSS,
\citet{Pohlen2006} found that $\sim 90\%$ of their surface brightness profiles 
are better described as broken exponential rather than single exponential disks. 
About 60\% of these galaxies show a break in the exponential profile between $\sim$~1.5 -- 4.5$r_{\rm d}$ 
followed by a downbending (steeper slope) in the outermost regions. 
Another $\sim30\%$ show a clear break between $\sim$~4 -- 6$r_{\rm d}$, followed by an upbending (shallower slope) outer region. 
Quite intriguingly, the shape of the profiles correlates with Hubble type. 
Downbending breaks are more frequent in later Hubble types, while the fraction of upbending breaks rises towards earlier types. 

In fact, among the relatively HI poor BCDs, most of the broken exponential disks show upbending \citep{Huang2012a}. 
In contrast, the HIghMass galaxies are selected 
to be the most HI-rich sources among the massive galaxies, most of which have late Hubble types. 
As discussed in the appendices, 
if a significant change in slope exists in the light profile, we mark the inner and outer disk regions and 
fit two exponential functions to each portion individually. 
Among the 29 HIghMass galaxies 
in Table~\ref{tab:KPNO}, seven show broken exponential disks in the {\it R}-band. 
As expected, the majority of them have steeper outer disks:
AGC~203522, AGC~721391, UGC~6692, UGC~7220, UGC~7899, and UGC~9023, with the only upbending optical disk occurring in  
UGC~6043. In agreement with the results in \citet{Pohlen2006}, the downward breaks occure at 2 -- 3$r_{\rm d}$ (2.5$r_{\rm d}$ on average) 
in the six downbending disks, in contrast to UGC~6043 at $\sim$~5$r_{\rm d}$. 
In addition, double disks are evident in the SDSS images of two HIghMass galaxies UGC~8797 and UGC~7686 
(not included as KPNO targets), both being downbending. 
We find an interesting coincidence of three overlapping sources in this list of downbending disks and 
the list of four disks with radially rising H$\alpha$ EW profiles identified earlier: 
AGC~203522, 
AGC~721391, and UGC~6692. 
\citet{Hunter2010} suggest that 
the upbending profiles may be explained as a result of shrinking of the size of the actively star-forming disks in BCDs, 
perhaps due to gas removal in the outer disks. 
The shallower outer profiles trace the underlying old stellar 
population, while the steeper inner profiles are dominated by the centrally concentrated and intense 
regions of recent SF. 
Our results altogether 
suggest a correlation between the gas-richness, a downbending outer profile, 
and active outer disk growth. 

Although we have confirmed that the downbending breaks are more frequent in late-type disks with high $f_{\rm HI}$ 
in which the SF is migrating outwards, the origin of this downbending remains a puzzle, and even seems to be 
contrary to the presence of an actively forming outer disk. 
Two downbending features were differentiated in \citet{MN2012}: 
an innermost `transition' radius at $\sim0.77\pm0.06r_{25}$ and a second characteristic radius, 
or `truncation' radius, close to the outermost optical extent $\sim1.09\pm0.05r_{25}$.  
Those authors propose that such a `transition' might be related to a threshold in the SF, 
while `truncations' more likely reflect a real drop in the stellar mass density of the disk associated with the maximum 
angular momentum of the stars \citep{MN2012}. 
Similarly, \citet{Roskar2008} claimed that the transition corresponds to a rapid dropoff in the SFR associated with a drop 
in the cool gas surface density. 
An inspection of the six downbending HIghMass galaxies shows that the changes in slope all happen within 
$r_{25}$, with an average of $\sim0.7r_{25}$, implying a threshold in SF rather than a truncation of the stellar disk. 
In fact, most of the HIghMass galaxies have H$\alpha$ emission traceable to $\sim r_{25}$ and beyond (Fig.~\ref{fig:spf}), 
except for UGC~721391, UGC~6692, UGC~7899, and UGC~9023. 
This matches very well with the list of downbending disks. 
Therefore, we conclude that the inner transition is related to a threshold in the SF as proposed by \citet{MN2012}. 
It will be interesting to see if these H$\alpha$ truncations are also visible in the FUV, 
because stochasticity may lead H$\alpha$ to show signs of knees and turnoffs while the 
FUV remains smooth \citep{Boissier2007}. 

Overall, 
there are a handful of downbending double exponential disks in the HIghMass sample. 
Although their disks are growing inside-out at the current epoch, SF thresholds also exist. 
The SF threshold is nearly always explained as the result of a gas surface density dropoff or 
of global dynamic stability \citep{Leroy2008}. On that basis, we would predict, for example, that UGC~7220 
has a relatively compact distribution of its HI gas; we will test this with our HI synthesis mapping data.

\subsubsection{Disk Color Gradients}

If galaxies indeed grow from the inside out, stars should be younger on average 
in the outer parts, leading to radial color gradients. Such trends have been
previously reported, for example for the SINGS galaxies \citep{Munoz-Mateos2011}, as well 
as in the HI-rich massive GASS galaxy, UGC~8802 \citep{Moran2010}. 

The $u-r$ surface brightness profiles of the 29 KPNO targets are presented 
in Appendix \ref{app:KPNO}. 
Indeed, most of them exhibit strong color gradients. 
While many still have relatively red centers, typical for their luminosities,
the colors become continuously bluer throughout the disks (e.g., UGC~8475), 
in a way that does not simply reflect the color differences between the bulge and disk components.
Relative to the $B-V$ color maps of the LSB galaxies presented in \citet{Schombert2011}, 
the HIghMass galaxies demonstrate systematic trends becoming clearly bluer in their outer disks. 


Measuring age gradients in disks from color profiles is not straightforward, since the radial decrease 
in the internal extinction and metallicity also conspire to yield bluer colors at larger radius \citep{Munoz-Mateos2011}. 
In the case of our observations, 
because the H$\alpha$ lines lie within the wavelength range of the {\it R} filter, differential reddening 
should have less impact on the H$\alpha$ EW gradients than on the optical color gradients. 
We present the H$\alpha$ EW profiles in 
Appendix \ref{app:KPNO}. 
A comparison between the EW and color profiles of the HIghMass galaxies 
shows evidence of an anti-correlation between the EWs and the colors. 
Globally, bluer galaxies have overall higher EWs (e.g., AGC~203522) and vice versa. 
Within a galaxy, the reddening trend of color as a function of radius is always associated with the 
declining EW (e.g., AGC~188749) and vice versa (e.g., AGC~726428). 
The overall trend in most of the HIghMass galaxies of lower central H$\alpha$ EWs 
supports the inference of older central regions
relative to the disks, in agreement with the scenario of inside-out disk formation. 

Nevertheless, reverse reddening trends in the outermost regions are observed in a number of galaxies, 
leading to the `U'-shape color profiles in, e.g., AGC~714145, UGC~6168, UGC~5711, UGC~5543, etc. 
Their EW profiles appear to have `$\Lambda$' shapes correspondingly. 
Simulations show that such features may result from a combination of a drop in the SFR 
(seeded by warps in the gaseous disk, the radial distribution of angular momentum, misalignment 
between the rotation of the inflowing gas and the disk, etc.) and radial stellar migration, which would 
populate the outskirts of disk with old stars formed further in \citet{Munoz-Mateos2011}. 

\section{HIghMass Galaxies in the Context of the ALFALFA Parent Sample}
\label{Dis}
As a population, HI-selected galaxies are among the least-clustered population known \citep{Martin2012},
and, as illustrated in Fig. \ref{fig:CMD}, in the great majority, they are star-forming,
blue-cloud galaxies. Overall, ALFALFA-detected galaxies
are less evolved and have overall higher SFR and SSFR at a given $M_*$
but lower star formation efficiency (SFE) and extinction, relative to an optically-selected sample 
\citep{Huang2012b}. The HIghMass galaxies are identified within the highest HI mass subset
of the ALFALFA population
because of their exceptional gas richness for their stellar mass. In this section,
we examine whether and in what ways besides their high gas fraction the HIghMass
galaxies differ from the overall HI-selected population detected by the ALFALFA survey.

\subsection{HIghMass Galaxies on the SF Sequence}

On an SFR {\it vs.}~$M_*$ diagram, star forming galaxies form a narrow band, the so-called 
``main sequence'' of SF; an even tighter correlation has been identified on a similar SSFR {\it vs.}~$M_*$ plot \citep[e.g.][]{Salim2007}. 
$M_*$ appears to be the crucial quantity governing the SF along this sequence. 
In the absence of mergers or other events that trigger a starburst, it has been suggested that
blue galaxies on the SF sequence evolve towards higher $M_*$ and and lower SSFRs. 
In this scenario, the tightness of the SF sequence, with an rms scatter of less than 0.3 dex, indicates 
that the SF of ``main sequence'' galaxies is not driven mainly by merger-induced 
starbursts but rather by a continuous mass-dependent process that is gradually declining with time, 
e.g., smooth gas accretion \citep{Bouche2010}. 
\citet{Huang2012b} have previously presented the results for the ALFALFA parent population, confirming 
the expectations that SFRs generally increase but SSFRs decrease with increasing stellar mass.

Whereas the SF sequence reflects directly the relationship
between the accretion rate and the halo mass \citep{Bouche2010}, the DM halo should play a 
more fundamental role than the stellar mass in the regulation of disk growth. 
For this reason, we examine the SF behavior of the ALFALFA sample overall,
and the HIghMass galaxies in particular, as a function of the disk rotational 
velocity, $V_{\rm rot}$,  taken to be the deprojected HI line width using the axial ratio of the
optical disk to correct for inclination. Within some scaling factor that accounts
for the (unknown) extent of the HI disk, we expect $V_{\rm rot}$ to provide a measure of
the dynamical mass of the DM halo, more reliable for example than $M_*$ alone, 
in order to characterize the overall gravitational effect leading to material infall and thus SF. 
This expectation is particularly justified here since 
$f_{\rm HI} > 1$ in 20/34 of the HIghMass galaxies. We could also derive the
dynamical masses ($M_{\rm dyn}$) from the values of $V_{\rm rot}$ and the optical disk size,
but because the axial ratios are less vulnerable to shredding by the SDSS DR8 pipeline 
relative to the disk size measurements (see Appendix \ref{app:SDSS}), we choose to present 
the results simply using $V_{\rm rot}$ rather than $M_{\rm dyn}$. 
We note however that the qualitative results to be drawn from this analysis will 
not be altered if any of $M_*$, $M_{\rm dyn}$, or $M_*+M_{\rm HI}$ is used instead of $V_{\rm rot}$. 

Fig.~\ref{fig:SFS} shows the parameter space occupied by the HIghMass galaxies (colored symbols) 
as compared to that by the  general ALFALFA population (black contours and gray dots). 
The $x$-axis is $V_{\rm rot}$ in all panels with $y$-axes showing SFR, SSFR, and SFE, respectively 
from top to bottom. 
In order to make the plots here, we adopt SFR(H$\alpha$)s for the HIghMass sample if available,
but use SFR(SED)s otherwise. Note that our parent sample here
is defined by galaxies included in that $\alpha$.40 catalog and with SDSS photometry from 
the SDSS/DR8. 

In the SFR {\it vs.}~$V_{\rm rot}$ diagram (top panel), the HIghMass galaxies in general lie slightly 
above the running average as defined by the ALFALFA galaxies. 
The only extreme outlier beneath the SF sequence is the only early type galaxy in our sample, UGC~4599, 
having an SFR of 0.01~$M_\odot$~yr$^{-1}$ in the center region (the outer ring is excluded from the Petrosian aperture 
because of the existence of the central bulge, see Appendix). 
This happens to be the only KPNO target without standard star frames 
bracketed, so that the estimate of the SFR comes from SED fitting. 
However, there are SF regions situating along the outer ring of this galaxy as evident in the 
continuum-subtracted H$\alpha$ image (uncalibrated). 
Recent work by \citet{Finkelman2011} derived the SFR within the entire ring to be 
0.04~$M_\odot~{\rm yr}^{-1}$ from {\it GALEX} UV flux. 
Those authors concluded that both a merger between two HI-rich galaxies and the cold accretion 
of gas from the IGM can account for the observed properties of this galaxy, as well as the huge amount of HI detected. 
It is very possible that the
SFR(SED) value obtained in the current work is an underestimate in this relatively red galaxy (see discussion in Section \ref{Cal}). 
Except for UGC~4599, the current SFRs range from 0.34 (UGC~8089, a LSB galaxy with low $M_*$) 
to 21 (UGC~8475, a well-developed spiral), and have a median of 2.5~$M_\odot$~yr$^{-1}$, in rough agreement with the 
range derived from Sc galaxies but higher than that of the Sa galaxies in \citet{Caldwell1991}. 
Again, we emphasize that the majority of the HIghMass galaxies are not crouching giants. 
Instead of being quiescent objects like Malin 1, most have blue outer disks and exhibit healthy onging SF, 
as gauged by their $V_{\rm rot}$ via this standard scaling relation. 
Although they have slightly lower surface brightnesses, they are different from the extreme LSB galaxies 
which typically have lower SSFRs by a factor of ten than other galaxies of the same baryonic mass \citep{Schombert2011}. 

In the middle panel, the upward deviation of the HIghMass galaxies from the general trend is even more evident. 
The HIghMass galaxies have only moderate to high current SFRs but significantly higher SSFRs at fixed $V_{\rm rot}$. 
Given the fact that $M_*$ represents the integrated past SF, this offset can either due to 
a higher ratio of current to past-averaged SFR and/or 
the HIghMass galaxies have formed the bulk of their stars more recently relative to the typical ALFALFA galaxy, 
and thus have a shorter integrated SFH. 

In order to test these scenarios, we characterize the shape of the SFH and estimate the formation time, 
again by SED fitting. The relevant parameters are defined as follows: 
$T_{\rm form}$ is the time since the galaxy first started forming stars; 
$T_{\rm wm}$ is the mass-weighted age; 
$\gamma$ is the timescale of the SFH so that the continuous component of SFH is given as $e^{\gamma t}$; and
$b$-, the birthrate-parameter, is the present SFR (averaged over the last 100~Myr) divided by the past average SFR, 
accounting for the bursts in SFH. 
Due to model degeneracies, these four parameters are poorly constrained by the SED fitting of 
five optical bands, with typical uncertainties being 60\%. 
Despite this caveat, it is indicative that the HIghMass galaxies in general started to form stars more recently,
with a median $T_{\rm form}$ of 4.16~Gyr, in comparison with a value of 4.51~Gyr for the ALFALFA galaxies.
This also suggests that the HIghMass galaxies  
have overall younger stellar population (median $T_{\rm wm} =$~2.28 {\it vs.}~2.54~Gyr). 
The median timescales of SF are comparable between the two populations, with a median of $\gamma =$~0.41 for both,
seemingly in contradiction with the prediction that the galaxies in high $\lambda$ halos have longer timescales of SF, 
or equivalently, that their SFHs decline more slowly. 
However, taking into account the possibility of bursts, the median $b$-parameter is higher among the HIghMass galaxies, 
being 0.38 ({\it vs.}~0.31 for the full ALFALFA sample). It is interesting to note that the
two HIghMass galaxies with the highest $b$-parameters (AGC~203522 and AGC~248881) also
have the highest EW${\rm_{H\alpha+[NII]}}$ values among our KPNO targets, i.e., they are undergoing a
burst of star formation at the present time.


The HIghMass galaxies in fact follow the standard calibration of the $f_{\rm HI}$ fundamental plane 
\citep[e.g., the $f_{\rm HI}$--SSFR--$\mu_*$ correlation in ][]{Huang2012b}, despite the higher than average 
$f_{\rm HI}$ at fixed $M_*$, $M_{\rm dyn}$, $M_*+M_{\rm HI}$, or $V_{\rm rot}$. 
This consistency suggests that the more gas-rich galaxies with higher $f_{\rm HI}$s tend to 
have higher SSFRs and lower surface densities by a corresponding amount, so that they follow the 
same $f_{\rm HI}$ fundamental plane as defined by their gas-poorer counterparts. 
Hence, it would be impractical to identify the gas-rich galaxies in high $\lambda$ halos by selecting outliers from the 
$f_{\rm HI}$ fundamental plane, because of the seemingly self-regulated process between gas supply and SF. 

The upper panel of Fig.~\ref{fig:muext} confirms the finding that the HIghMass galaxies have overall 
lower stellar mass surface densities at fixed $V_{\rm rot}$, relative to the parent $\alpha$.40--SDSS (DR8) sample. 
The concentration index in the {\it r}-band, $r_{\rm petro,90}/r_{\rm petro,50}$, 
is plotted as a function of $V_{\rm rot}$ in the lower panel of Fig.~\ref{fig:muext}. 
Higher values of $r_{\rm petro,90}/r_{\rm petro,50}$ correspond to bulge-dominated systems 
($r_{\rm petro,90}/r_{\rm petro,50} = 3.35$ for a de Vaucouleurs model; 2.29 for an exponential disk). 
Except for UGC~4599, all of the HIghMass galaxies can be categorized as disk systems. 
However, in comparison with the parent sample, the overall lower concentration indexes of the HIghMass galaxies 
at fixed $V_{\rm rot}$ is insignificant but rather results from the very large scatter in the 
concentration index {\it vs.}~$V_{\rm rot}$ relation. 

In summary, the moderate to high current SFRs of the HIghMass galaxies, together with their 
overall higher EWs, SSFRs, and $b$-parameters 
support the idea that SF in them has been suppressed in the past, but 
that they are undergoing an active period of disk growth at the current epoch. This evolutionary
state is predicted to be the typical behavior of galaxies residing in high $\lambda$ DM halos 
by semi-analytical models \citep[e.g., ][]{Boissier2000}.  And/Or, the HIghMass galaxies
appear to have been forming the bulk of their stars at later times, possibly as triggered by 
gas infall events. Future work will explore evidence for gas infall in the HI synthesis maps 
and better determine the SFH via NIR photometry from warm-{\it Spitzer}.

\subsection{HI-based SFEs in the HIghMass Galaxies}

Among the various forms of applicable SFLs, the most commonly studied is the empirical K-S 
relation which attempts to describe how the SFR surface density
$\Sigma_{\rm SFR}$ is regulated by the gas surface density, e.g., 
$\Sigma_{\rm SFR} \propto \Sigma_{\rm HI+H_2}^\alpha$, where the exponent is determined globally to be 
$\alpha = 1.4 \pm 0.15$ in \citet{Kennicutt1998}. 
By definition, K-S relation gives the surface density of SF efficiency, $\Sigma_{\rm SFE} \equiv 
\Sigma_{\rm SFR} / \Sigma_{\rm gas} 
\propto \Sigma_{\rm gas}^{\alpha-1}$, and the reciprocal of the SFE is the SF timescale.  
Within the optical disk where H$_2$ dominates, it has been established that 
the $\Sigma_{\rm SFR}$ correlates better with 
the $\Sigma_{\rm H_2}$ than with the total gas density 
\citep{Leroy2008, Bigiel2008}, indicating that SF occurs exclusively in the molecular phase of the ISM. 
In spirals, the $\Sigma_{\rm SFE}$ of H$_2$ alone is nearly constant with an H$_2$ depletion time of $\sim$2~Gyr 
\citep{Leroy2008, Bigiel2011, Schruba2011}, i.e., $\alpha = 1$ at a resolution of $\sim$kpc. 
Controversial results are found in low-metallicity dwarf galaxies and in the outer parts of large spirals, 
where the ISM is mostly atomic.
The HI-based $\Sigma_{\rm SFE}$ is much smaller, with a SF timescale of $\sim$100~Gyr and, in systems
where HI dominates, the K-S relation is likely to have a distinct form, 
e.g., $\Sigma_{\rm SFR}$ also begins to correlate with the $\Sigma_{\rm HI}$ \citep{Bigiel2010, Krumholz2013}. 

In the current work, we only inspect the global HI-based SFE ($\equiv {\rm SFR}/M_{\rm HI}$) and 
the corresponding gas depletion timescale $t_R \equiv {\rm SFE^{-1}}$. 
The distribution of $t_R$ is known to be broad, ranging from 0.3~Gyr (on the scale of starbursts) 
to 100~Gyr (many times the Hubble time, $t_H = 13.6$~Gyr). 
\citet{Huang2012b} have already
shown that as an HI-selected sample, the ALFALFA galaxies have overall lower SFEs, or equivalently longer 
$\langle t_R \rangle = 8.9$~Gyr, relative to an optically-selected sample.
The bottom panel of Fig.~\ref{fig:SFS} shows the distributions of HIghMass galaxies and the ALFALFA parent sample 
on an SFE {\it vs.}~$V_{\rm rot}$ diagram. 
The horizontal dash-dotted line in green corresponds to $t_R = t_H$; the cyan dashed line on top marks the 
average SFE derived from the GASS sample \citep[$t_R = 3$~Gyr,][]{Schiminovich2010}. 

Contrary to the expectation that the high $f_{\rm HI}$ results from a low efficiency of SF in the HIghMass galaxies, 
their HI-based SFEs are moderate relative to the parent ALFALFA sample (see Fig.~\ref{fig:SFS}). 
The median SFE is $10^{-9.83}$~yr$^{-1}$ ($t_R = 6.8$~Gyr), which is slightly higher than the average of 
the ALFALFA population overall, 
but still lower than that of the stellar mass-selected GASS galaxies. 
The $t_R$ distribution is confirmed to be broad. 
However, except for the early-type UGC~4599 which has a $t_R$ comparable to the 
$\Sigma_{\rm SFE}$ obtained in spiral outer disks (100~Gyr), the global HI depletion timescales are not extremely long 
in comparison with the H$_2$ depletion time (2~Gyr), with 74\% of the sample having $t_R < t_H$. 
The majority of the HIghMass galaxies with $t_R > t_H$ are in our list of LSB galaxies, including 
AGC~188749, AGC~190796, AGC~190277, UGC~8089, and UGC~9234, in support of the 
correlation between lower surface density and lower SFE.  

The HIghMass galaxies have overall lower $\Sigma_{\rm SFR}$ and stellar surface densities, 
suggesting lower $\Sigma_{\rm gas}$ given the K-S relation. 
One may naively expect the HIghMass galaxies to have lower SFEs based on our understanding of the K-S relation. 
However, such a reasoning implicitly relies on several questionable assumptions.  
In the inner disks with higher levels of $\Sigma_{\rm SFR}$, H$_2$ can still be the dominant component of the ISM locally. 
The lower $\Sigma_{\rm SFE}$ caused by the lower $\Sigma_{\rm gas}$ is possible if the K-S relation is significantly 
super-linear with $\alpha > 1$. 
This slope is highly debated due to the different background subtraction or the treatment of diffuse emission unrelated to SF. 
However, a number of authors obtained $\alpha$ value close to 1 \citep{Leroy2008, Bigiel2011, Schruba2011}, 
at least in the inner disk.  
The exact form of the K-S relation is less clear towards the outskirts, but the 
total $\Sigma_{\rm gas}$ there is not necessarily low for the HIghMass galaxies relative to the more normal spiral
population. Despite the lower central surface density, disks formed in higher $\lambda$ halos have longer scale lengths 
so that the gas may be distributed in extended disks. 
In fact, except for a few cases with truncated $\Sigma_{\rm SFR}$ radial profiles, the star forming disks are 
traceable to large radii in most of the HIghMass galaxies. 
Finally, the inference of global properties such as the SFE according to the results derived from resolved samples 
which yield the K-S relation is subject
to spatial averaging over a range of $\Sigma_{\rm gas}$, timescales and conditions within the ISM. 

Although models predict that a lower SFE should lead to a higher gas fraction and vice versa \citep[e.g.,][]{Bouche2010}, 
the HIghMass galaxies with high $f_{\rm HI}$s do not have particularly low HI-based SFEs relative to the other galaxies. 
Given that the $M_{\rm H_2}$-to-$M_{\rm HI}$ ratios are lower than average, 
the H$_2$-based SFEs should be even higher than normal spirals. 
However, the global SFEs are still low in comparison with the $\Sigma_{\rm SFE}$ 
with the former set by the inefficient conversion of atomic to molecular gas in the far outer galaxy \citep{Kennicutt2012}. 
Future work will attempt to compare the properties of both HI and molecular distributions to probe further
the K-S relation applicable in the HIghMass galaxies.


\subsection{Spin Parameter Distribution}

Both semi-analytical models and hydrodynamic simulations predict that the stellar disks in high $\lambda$ halos 
are formed to be less concentrated and more gas-rich \citep[e.g.,][]{Mo1998, Boissier2000, Kim2013, Kravtsov2013}. 
In support of this, the stellar disks of the HIghMass galaxies are confirmed to be extended with overall lower surface densities 
at fixed $M_*$. 
A quantitative angular momentum statistic, the ``spin parameter'' $\lambda$, can be expressed on terms of observable quantities as:
\begin{eqnarray}
	\lambda = \frac{\sqrt{2} V_{\rm rot}^2 r_{\rm d}}{G M_{\rm halo}}, 
\end{eqnarray}
where $V_{\rm rot}$ is the rotational velocity as derived above, $r_{\rm d}$ is the
scale length of the optical disk and $M_{\rm halo}$ is the mass of the associated DM halo \citep{Hernandez2006}.
We follow the same approach as 
in \citet{Huang2012b} to derive the empirical $\lambda$ distribution of host halos,
adopting the $M_{\rm halo}$ derived from 
the $V_{\rm halo}- V_{\rm rot}$ relation given in \citet{Papastergis2011}. 
Underlying assumptions include an exponential surface density profile of the optical disk, 
a flat disk rotation curve at $V_{\rm rot}$, 
an isothermal density profile of the halo (virialized), 
and that the potential energy of the galaxy is dominated by that of the halo. 
The last two assumptions are known to be unrealistic or highly uncertain, but are adopted for simplicity. 
Therefore, the relative $\lambda$ distributions are more indicative than the absolute values. 


Although \citet{Hernandez2007} find that 
the $\lambda$ distribution of a disk galaxy sample selected from the SDSS is consistent
with values derived from simulations,
\citet{Huang2012b} have shown that the ALFALFA population overall shows a $\lambda$ distribution 
that is no longer lognormal and has a mean value well in excess of that predicted, 
implying that the HI-selected galaxies preferentially occupy DM halos of higher than average
$\lambda$. In  Fig.~\ref{fig:spin}, we examine the statistical measures of $\lambda$
derived as above for the full HIghMass galaxies (solid line) superposed on a normalized distribution for the
ALFALFA population overall (dashed line). 
The shaded area traces the subsets included
in two of our HI synthesis imaging programs. All of the JVLA11 targets (shaded in red) are stellar massive disks 
($M_* > 10^{10}~M_\odot$), while the GMRT11 targets (shaded in cyan) all have $M_* < 10^{10}~M_\odot$. 

There is evidence that the HIghMass galaxies do
indeed reside in DM halos characterized by higher than average values of $\lambda$, 
and the average $\lambda$ of the less stellar massive GMRT11 targets is similar to that of the JVLA11 ones. 
However, the significance of a departure from the rest of the ALFALFA population and possible
dependence on other paramters such as $M_*$ or color is presently unclear, given the 
statistical uncertainties and small sample size. The HI synthesis maps will
permit us to constrain the halo density profile. In fact, \citet{Hallenbeck2014}
present UGC~12506 as a first case of a HIghMass galaxy whose spin parameter, derived from
the fitting the HI velocity field, is truly high. Future work will use the halo density
profiles derived from the HI maps to constrain the spin parameters of the HIghMass
galaxies more directly.


\subsection{The Local Environment of HIghMass Galaxies}

It is well-known that the HI mass of disk galaxies can be depleted by interactions with environments of galaxies, especially
via tidal or ram pressure stripping. 
Here, we examine whether or not the exceptionally gas-rich massive
HI disks included in HIghMass are different from the rest of the HI-selected population. 
To quantify the environment we used a galaxy's $n$th nearest neighbors as a metric. 
Using as the source catalog the SDSS DR7 spectroscopic catalog galaxies \citep{Abazajian2009}, the ten nearest 
neighbors of all of the high S/N ratio detections in the
$\alpha$.40 sample were found. The catalogs were cut in R.A. and Dec., as well as redshift, 
such that the ALFALFA sample was volume-limited in the range 60-140 Mpc 
(within which 
21 of the HIghMass galaxies fall), and the SDSS sample was volume-limited out to 
165 Mpc, or equivalently an absolute {\it r}-band petrosian magnitude of $-$18.6. 
The search for neighbors then proceeded 
within an expanding spherical bubble (in the plane of the sky and in redshift distances), out to a limit of 45~Mpc. 
Finally a minimum separation of 10 arcsec was applied because such close neighbors are unlikely to be indicative 
of the surrounding environment and would not be distinguishable from the parent galaxy in ALFALFA survey anyway. 

Measures of local galaxy density were estimated by averaging the logarithm of the 1st, 2nd, and 3rd nearest neighbor densities 
\citep[similar to][]{Baldry2006}, and are plotted for the ALFALFA sample, the ALFALFA sample with stellar mass 
above $10^{10}~M_{\odot}$, and the HIghMass galaxies in Fig.~\ref{fig:env}. 
K-S tests comparing 
the distributions suggest that the high stellar mass sample is distinct from its parent ALFALFA sample 
(with a P-value of $1.1 \times 10^{-16}$), with the high stellar mass galaxies typically avoiding regions of very low local density. 
The K-S tests for the HIghMass galaxies indicate that their environment is more similar to that 
of the high stellar mass galaxies (P-value of 0.71), than that of the average ALFALFA galaxy (P-value of 0.013). 
ALFALFA is naturally biased to blue galaxies which tend to be less clustered; \citet{Martin2012} have shown that the gas-rich galaxies
detected by ALFALFA is the most weakly clustered population known. 
Thus, it may be surprising to conclude that the local environment of a HIghMass galaxy is not especially under dense, 
when compared with that of similarly stellar massive galaxies in ALFALFA. Much the same result is found when they are compared 
with other high HI mass galaxies within ALFALFA.



\section{Summary and Future Work}
\label{Sum}

The main goal of this paper has been to describe the optical properties of
the 34 HIghMass program galaxies, a set of exceptionally gas-rich, massive HI disks
identified among the 15000 HI detections presented in the $\alpha$.40 catalog
\citep{Haynes2011}. In addition to summarizing the characteristics of the
HIghMass galaxies as a subset of the overall HI-selected population, we 
present a dataset of high-quality H$\alpha$ imaging for 29 HIghMass galaxies. 
The higher-than-average (for their stellar mass) gas fractions may result
from suppressed SF in the past or may be attributed to late cold gas accretion. 
Understanding the past history and current evolutionary state of
such extremely HI-rich galaxies as a population in the local Universe will provide insight
into the nature of the massive and gas rich systems which are likely to dominate 
the planned deep field surveys of HI in galaxies at higher redshift with the SKA and 
its pathfinders \citep[e.g.][]{Meyer2009, Duffy2012}. 

This work is the first systematic H$\alpha$ study of galaxies in this $M_{\rm HI}$ regime. 
Relative to the existing large H$\alpha$ surveys of field galaxies, the HIghMass galaxies 
lie at further distances and are thus more massive in general. 
We report the details of our observation and the procedures of image reduction, continuum subtraction, 
as well as H$\alpha$ and {\it R}-band surface photometry in the Appendix. 
KPNO measurements are presented in Table \ref{tab:KPNO}. 
Because of the shortcomings of the standard SDSS photometry pipeline, e.g., 
shredding and sky background over-subtraction, 
we follow a similar approach to obtain supplementary measurements from the SDSS images 
which are presented in Table \ref{tab:SDSS}. 
Multiple internal and external checks ensure the consistency of the KPNO and SDSS results 
and the quality of absolute flux calibration. 
Current SFRs of the HIghMass galaxies are therefore derived in two ways: 
SED fitting to the SDSS bands and converting from the $L_{\rm H\alpha}$. 
Albeit the poorer constraint of SFR(SED) for some galaxies, 
optical SED- and $L_{\rm H\alpha}$-derived values agree within the uncertainty ranges for most of the sources; 
there is no evidence that the high $f_{\rm HI}$ values are due to 
an excess or deficit in the formation of massive stars. 

The 34 galaxies that comprise the HIghMass sample span a range of luminosities, $M_*$, and SFRs
and all but three, UGC~6066, UGC~9234, and UGC~4599, clearly reside in the blue cloud region of the 
color-magnitude diagram. Several, e.g., AGC~190796, AGC~190277, and UGC~12506 meet the criteria
of low surface brightness characteristics of other samples, but the majority are not
''crouching giants'' like Malin 1. In comparison with the SFRs derived in other H$\alpha$ surveys, 
the HIghMass galaxies have overall higher integrated SFRs. 
This characteristic rules out the hypothesis that the high HI fractions in the HIghMass galaxies 
result from a simple inhibition of gas conversion into stars {\it at the present time}. 
Relative to the overall $\alpha$.40--SDSS parent sample, the HIghMass galaxies exhibit healthy ongoing SF 
and (HI-based) SFEs despite the evidence
by their overall high H$\alpha$ EWs and $b$-parameters that they may have
have been relatively inactive in forming stars {\it in the past}. 
The derived high SSFRs at fixed HI rotational velocities also support a picture of more recent 
formation times. However, the SFR surface density profiles show that the SF activity is 
spread throughout their extended disks so that they have lower SFR surface densities overall. 
Similarly, the HIghMass galaxies have, on average, lower surface brightness in the {\it R}-band or 
the stellar mass surface density. Albeit limited by resolution, 
the slopes of the HII region LFs in the luminous part agree with the results derived from 
samples of normal spirals; again, there is no sign of abnormal behavior in the formation of massive stars. 

Only two of the HIghMass galaxies have prominent central EW peaks: UGC~9023 and UGC~9037. 
In fact, \citet{Hallenbeck2014} find evidence
from the HI velocity field that the gas in UGC~9037 is inflowing and suggest
that it may be on the verge of a starburst phase. 
All the other HIghMass galaxies have higher EWs in the outer disk regions, implying inside-out disk growth. 
In particular, the EW profiles rise almost monotonically in AGC~203522, AGC~726428, AGC~721391, and UGC~6692, 
i.e., the SF activity is migrating outwards. 
Shallower outer stellar disks (upbending profiles) are seldom seen in gas-rich galaxies but there are a handful of 
downbending double exponential disks in the HIghMass sample. 
Although their disks are growing inside-out at the current epoch, SF thresholds exist 
in the downbending disks, probably as a result of a concentrated gas distribution. 
The majority of HIghMass galaxies have strong color gradients, being redder in the center, 
in agreement with the scenario of inside-out disk formation. 
The color gradients reverse at the very outermost regions in some disks, 
which may result from a combination of a drop in the SFR and radial stellar migration. 

In the future, we will use gas dynamics derived from the HI maps to constrain the properties of DM halos, 
as well as to search for the imprints of recent accretion left on the HI morphology or velocity fields. 
In addition to the atomic gas content, the mass of cold molecular gas will be given by IRAM 30-m observations
of the $^{12}$CO($J=1-0$) line emission, and,
for a few galaxies, resolved CARMA maps will trace its distribution. In combination with the 
H$\alpha$ images, these additional data will allow us to study in detail the 
HI-to-H$_2$ conversion, K-S relation, and SF threshold in these massive gas-rich disks. 
{\it Herschel} images will be used to probe the obscured SF, cold dust emission, 
and to identify the sources or regions with extremely low dust-to-gas ratios 
which may be responsible for the low efficiency of H$_2$ formation. 
The SFHs and old stellar populations will be further constrained by {\it Spitzer} observations. 
Optical spectra will help determine the CO-to-H$_2$ conversion factors and yield metal enrichment histories throughout the disks. 
A synthesis of these multi-wavelength data will allow us to study all the key processes 
involved in gas consumption and star formation in these exceptional HIghMass galaxies.

\acknowledgements
The authors acknowledge the work of the entire ALFALFA collaboration team 
in observing, flagging, and extracting the catalog of galaxies used in this work. 
The ALFALFA team at Cornell is supported by NSF grants AST-0607007 and AST-1107390 to RG and MPH and 
by grants from the Brinson Foundation. 
SM is supported by the National Science Council (NSC) of Taiwan, NSC 100-2112-M-001-006-MY3. 
We thank Yiming Li for useful discussions and
Angela van Sistine and John Salzer for supplying the H$\alpha$ image of UGC~12506. 

{\it GALEX} is a NASA Small Explorer, launched in 2003 
April. We gratefully acknowledge NASA's support for construction, operation and science 
analysis for the {\it GALEX} mission, developed in cooperation with the Centre National 
d'Etudes Spatiales of France and the Korean Ministry of Science and Technology. 
SH, RG and MPH acknowledge support for this work 
from the {\it GALEX} Guest Investigator program under NASA grants NNX07AJ12G, NNX07AJ41G,
NNX08AL67G, NNX09AF79G and NNX10AI03G.

Kitt Peak National Observatory, National Optical Astronomy Observatory, 
is operated by the Association of Universities for Research in Astronomy (AURA) under 
cooperative agreement with the National Science Foundation.

Funding for the SDSS and SDSS-II has been provided by the Alfred P. Sloan Foundation, 
the participating institutions, the National Science Foundation, the US Department 
of Energy, the NASA, the Japanese Monbukagakusho, the Max Planck Society and 
the Higher Education Funding Council for England. The SDSS Web Site is 
http://www.sdss.org/.
The SDSS is managed by the Astrophysical Research Consortium for the 
participating institutions. The participating institutions are the American 
Museum of Natural History, Astrophysical Institute Potsdam, University of Basel, 
University of Cambridge, Case Western Reserve University, University of Chicago, 
Drexel University, Fermilab, the Institute for Advanced Study, the Japan 
Participation Group, Johns Hopkins University, the Joint Institute for Nuclear 
Astrophysics, the Kavli Institute for Particle Astrophysics and Cosmology, the
 Korean Scientist Group, the Chinese Academy of Sciences (LAMOST), Los Alamos 
National Laboratory, the Max Planck Institute for Astronomy, the MPA, New Mexico 
State University, Ohio State University, University of Pittsburgh, University 
of Portsmouth, Princeton University, the United States Naval Observatory and 
the University of Washington.

Funding for SDSS-III has been provided by the Alfred P. Sloan Foundation, the Participating Institutions, 
the National Science Foundation, and the U.S. Department of Energy Office of Science. The SDSS-III web site is http://www.sdss3.org/.
SDSS-III is managed by the Astrophysical Research Consortium for the Participating Institutions of the SDSS-III Collaboration including the University of Arizona, the Brazilian Participation Group, Brookhaven National Laboratory, Carnegie Mellon University, University of Florida, the French Participation Group, the German Participation Group, Harvard University, the Instituto de Astrofisica de Canarias, the Michigan State/Notre Dame/JINA Participation Group, Johns Hopkins University, Lawrence Berkeley National Laboratory, Max Planck Institute for Astrophysics, Max Planck Institute for Extraterrestrial Physics, New Mexico State University, New York University, Ohio State University, Pennsylvania State University, University of Portsmouth, Princeton University, the Spanish Participation Group, University of Tokyo, University of Utah, Vanderbilt University, University of Virginia, University of Washington, and Yale University.

\newpage
\begin{appendices}

\section{H$\alpha$ and {\it R}-band Data Reduction}
\label{app:KPNO}

\subsection{Image Processing and Continuum Subtraction}

The preliminary H$\alpha$ and {\it R}-band image reduction follows standard procedures in IRAF: 
fixing header keywords, 
pedestal removing, 
trimming, bias subtracting, 
flat fielding, 
interpolating across the columns of bad pixels, 
and cosmic ray cleaning. 
All the frames have noticeable curvature at the edges; however, all the sample galaxies occupy 
only the inner 10\% of the chip or less. 

The processed narrowband images contain contributions from both H$\alpha$ and underlying stellar continuum, 
and the accuracy of the continuum scaling and subtraction can be the dominant error source for galaxies 
with low emission-line equivalent widths \citep{Kennicutt2008}. 
We adapt the IRAF package kindly provided by Salzer 
to conduct continuum subtraction, etc. 
A set of exposures for the same galaxy, including three H$\alpha$ frames and two/three {\it R} frames, are first shifted to register 
to an {\it R}-band image taken on a photometric night (called {\it R}-reference frame). 
Given the rough central coordinates and pixel scales, astrometry is obtained by shifting and rotating. 
Then we average the FWHM of over seven stars near the HIghMass galaxies and find out a frame with the worst seeing. 
All the other frames are convolved with a gaussian kernel to match this worst PSF. 
The same set of stars are used to determine the scaling factors subsequently. 

All H$\alpha$ images are normalized to the one taken on a photometric night and combined as the final `ON' frame; 
similarly combined {\it R} image yields the final `OFF' frame. 
Net H$\alpha$ images are obtained by subtracting a scaled `OFF' frame from the `ON' frame. 
This scaling factor is determined by the ratio of transmissivity of two filters, 
the sky transparency if under non-photometric conditions, and the difference in exposure time, 
but we estimate it by the average of ratios between foreground star counts in the two frames, forcing 
the residual fluxes of stars to reach a minimum in the net H$\alpha$ images. 
Our method implicitly assumes that field stars have no significant H$\alpha$ emission on average, 
which is most likely to be valid, and it allows for sky transparency changes. 
For some large programs, e.g., \citet{Kennicutt2008}, a uniform scaling factor can be derived by averaging 
the values from all galaxies observed in the same H$\alpha$ filter. 
In our case, 5/8 H$\alpha$ filters were used for only one or two galaxies thereby we skip this process.  

\subsection{Surface Photometry}

The reduction steps outlined above produced three combined images per galaxy 
that will be of further use: a net H$\alpha$, an unscaled {\it R}-band, and a scaled {\it R}-band. 
Considering the future studies, e.g., a comparison between the gas and SF surface density, 
both the integrated fluxes and one-dimensional profiles are of interest. 
We therefore brief surface photometry process as follows.  

\subsubsection{Isophotal Fitting}
We make use of the GALPHOT package, a collection of IRAF scripts 
modified to accommodate the KPNO images, 
which performs sky background subtraction and contamination cleaning 
\citep{Huang2012a}. 
Starting from an initial guess, elliptical surface brightness contours were fitted 
to these cleaned images, 
outwards to the radius at which the fitting fails to converge and inwards to the seeing limit. 
We adopt a logarithmic scale in radius increment, with $r_{\rm outer}$ = 1.25 $r_{\rm inner}$.  
The center coordinates, ellipticities, and position angles are all allowed to vary.
Final fits were visually inspected for robustness. 
This process is applied to all three combined images in order to refill the masked regions by interpolation. 
However, for consistency, only the set of ellipses as a result of fit to the unscaled {\it R}-band image 
are retained and used for aperture photometry on H$\alpha$ and scaled {\it R}-band images, 
yielding the azimuthally-averaged surface brightness profiles comparable in all bands. 
The disk portion of the surface brightness profile,  with the most linear appearance, is selected; 
within this semi-major-axis range of the disk, averages of ellipse centroids, position angles ($\theta$) 
and ellipticities ($\epsilon\equiv 1-b/a$) are measured to characterize the global properties of a galaxy 
(tabulated in Table~\ref{tab:KPNO}). 
This region is then fit to an exponential profile, yielding the disk scale length $r_{\rm d}$ and 
surface brightness interpolating to center, $\mu_0$. 

\subsubsection{Global Magnitudes}
Although we record several sets of global magnitudes, including ones at fixed isophotal levels, 
partial magnitudes integrated to a certain number of disk scale length (e.g., mag$_8$ to $8r_{\rm d}$), 
and asymptotic magnitudes extrapolated to infinity from the exponential outer disk fit, 
we report hereafter the Petrosian magnitudes for the majority of the HIghMass galaxies. 
We follow the SDSS convention, but calculate fluxes in elliptical annuli rather than in circular rings. 
The Petrosian ratio at a galacticentric radius $r$, $R_P(r)$, is calculated inside an elliptical annuli with an inner 
semi-major-axis of  $0.8r$ and outer of $1.25r$. 
The local surface brightness in the lower left panels of Fig.~\ref{fig:6168} -- \ref{fig:9234} also represents 
the mean H$\alpha$ surface brightness inside these annuli. 
The Petrosian radius ($r_P$) is defined as the semi-major axis at which Petrosian ratio $R_P(r)=0.2$, 
i.e., the radius where local surface brightness drops to 0.2 of the enclosed surface brightness. 
Our Petrosian magnitude denotes the flux in an elliptical aperture with $r=2r_P$, and 
centroid, $\theta$, and $\epsilon$ as given by the global measurements.  
We also calculate $r_{\rm petro,50}$ and $r_{\rm petro,90}$ as the semi-major axes of ellipses 
that contain 50\% and 90\% of the Petrosian flux, respectively. 

The errors given on magnitudes in this paper include the contributions from the 
photometric zero point error for a given night and the uncertainty in determining the sky level 
(set to 0.15$\sigma$ of the sky values in the sky boxes), in addition to the formal errors. 
In most cases, the Petrosian magnitudes agree with mag$_8$ 
(similar to the model magnitudes of disk galaxies in SDSS), within the magnitude errors 
(see below for the exception). 
However, in four galaxies, AGC~190796, AGC~203522, UGC~8089, and AGC~726428, 
the LSB H$\alpha$ emission drops so slowly outwards that the outermost ellipse of 
confidential surface brightness measurement is reached before the Petrosian radius can be determined. 
We report mag$_8$ instead of Petrosian magnitudes for them in Table~\ref{tab:KPNO}. 

If a significant change in slope exists in the light profile, we mark the inner and outer disk regions and 
fit two exponential functions to each portion individually. 
The inner fit is used to determine the general disk properties, 
e.g., $r_{\rm d}$, $\theta$, and $\epsilon$ as presented in Table~\ref{tab:KPNO} and \ref{tab:SDSS}, 
whereas the outer fit is used to extrapolate the surface brightness profile beyond the outermost measured isophote 
in the calculation of mag$_8$, and to determine the Petrosian aperture. 

\subsubsection{Absolute Flux Calibration}
The instrumental magnitudes $m_{\rm inst} \equiv -2.5\log{\rm CR}$ (CR being the observed count rate in unites of counts s$^{-1}$), 
are calibrated to Johnson-Cousins and AB magnitude systems, respectively, for {\it R}-band and H$\alpha$ measurements. 
For {\it R}-band calibration, we solve for the color coefficient ($cc$), 
extinction coefficient ($\kappa$), and zero point magnitudes (ZPs). 
in this equation with all Landolt standards from the same night: 
\begin{equation}
\label{eqa:bb}
	m = m_{\rm inst} + cc(R-I)-\kappa~{\rm AM}+{\rm ZP}, 
\end{equation}
where $m$ is the calibrated magnitudes available in \citet{Landolt1992}, $R-I$ is the color of stars we measured, and AM is airmass. 
We get $\kappa = 0.091\pm0.002$, given multiple exposures for the same star on the first photometric night. 
The ZPs are determined to be 23.968$\pm$0.006, 24.034$\pm$0.003, and 23.982$\pm$0.008 for the three photometric nights, 
respectively, with high photometric quality. 

For calibrations of H$\alpha$ filters, because of the narrow bandpasses, the color term is dropped and we adopt the 
standard KPNO $\kappa$ values. 
The width of H$\alpha$ filters we used translates to a ZP offset, and we only need to solve for ZP in this equation: 
\begin{equation}
\label{eqa:nb}
	m_\nu - 2.5\log {\rm FWHM} = m_{\rm inst} - \kappa~{\rm AM}+{\rm ZP}, 
\end{equation}
where $m_\nu$ is the monochromatic magnitudes of spectrophotometric standards available in \citet{Oke1983} 
and FWHM is the width of H$\alpha$ filters in units of {\rm \AA}. 
The monochromatic magnitudes ($m_\nu$) can be converted to flux density 
($f_\nu$ in units of erg~s$^{-1}$~cm$^{-2}$~$\rm \AA$) 
via $m_\nu = -2.5\log f_\nu-48.60$ \citep{Oke1983}, so that 
	$m_\nu = -2.5 \log (\lambda^2 f_\lambda)-2.408$, 
where wavelength $\lambda$ is in units of $\rm \AA$. 
Therefore, given the calibrated H$\alpha$ magnitude $m_{\rm cal}$, 
the line flux in units of erg~s$^{-1}$~cm$^{-2}$ is derived as 
\begin{equation}
\label{eqa:cal}
	f_{\rm H \alpha+[NII]} = 10^{-0.4(m_{\rm cal}+2.408+5\log\lambda)}. 
\end{equation}

It must be taken into account that the {\it R}-band image used is contaminated by H$\alpha$ flux, 
so that the continuum is over-subtracted. 
We correct for this effect following a similar approach as in \citet{Kennicutt2008}. 
Because $cz$ of our targets match very well with the central wavelengths of the H$\alpha$ filters, 
the normalized H$\alpha$ filter transmission is close to unity. 
The final correction factor is thus 
`$1/(1-\frac{T_R}{T_{\rm H \alpha}} \frac{t_R}{t_{\rm H \alpha}} \frac{1}{F})$', 
where $T_{\rm filter}$ is filter transmission at wavelength $\lambda$, $t_{\rm filter}$ is exposure time 
(3~min in {\it R} and 15~min in H$\alpha$), and the 
{\it R} frame is divided by factor $F$ so as to scale to H$\alpha$ frame for the purpose of continuum subtraction. 
This corresponds to an increment of the final H$\alpha$ flux by $\sim 4\%$ for most of the HIghMass galaxies. 

Examples of the variation with semi-major axis of our KPNO measurements are shown by filled symbols 
in Fig.~\ref{fig:6168} -- \ref{fig:9234} for three galaxies, UGC~6168, UGC~6692, and UGC~9234, respectively 
(see more discussions of these profiles in text). 
Black upwards triangles with error bars denote the {\it R}-band results in all of these plots; 
black filled circles with error bars are data points for H$\alpha$. 
The top two panels show position angles and ellipticities of a series of elliptical apertures as a result 
of the isophotal fitting. 
Black solid horizontal lines mark the global values as averages in the disk region 
(in between the two vertical dash-dotted lines); dotted horizontal lines illustrate the uncertainty range. 
The middle panels are surface brightness profiles and accumulated magnitudes, 
converted to AB magnitude system to be consistent with the SDSS results (see Appendix \ref{app:SDSS}). 
Solid horizontal lines mark the Petrosian magnitudes; dashed horizontal lines mark the mag$_8$. 
The bottom panels are the same quantities for H$\alpha$ in physical units, corrected for continuum over-subtraction. 
Definitions of solid and dashed horizontal lines are the same as in the middle panels; 
dotted horizontal lines illustrate the uncertainty range. 

It is demonstrated that the Petrosian magnitudes recover essentially all of the flux in our disk galaxies, 
and they generally agree with mag$_8$ within the uncertainty 
(in Fig.~\ref{fig:6168} and \ref{fig:6692}). 
The only exception is UGC~9234 with a compact central bulge but a shallow outer disk (see the middle left panel of 
Fig.~\ref{fig:9234}). 
The strong break in surface brightness profile of this galaxy causes a large deviation between the two magnitudes, 
in a sense that the Petrosian magnitude severely undercounts the extended flux (see the middle and bottom 
right panels). 

\subsubsection{H$\alpha$ Equivalent Widths}

Finally, we calculated the H$\alpha$ equivalent widths (EWs) from net H$\alpha$ and scaled {\it R}-band images. 
By definition, it compares the line flux against underlying continuum level, so that EW can be derived without a calibration 
of absolute flux. It characterizes the strength of current SF relative to the past. 
The EW in units of $\rm \AA$, corrected for continuum over-subtraction is: 
\begin{equation}
\label{eqa:EW}
	{\rm EW_{H\alpha+[NII]}} = {\rm FWHM} \frac{t_{\rm H \alpha}}{t_R} \frac{1}{10^{0.4(m_{\rm H \alpha} - m_{R^{\prime}})} - T_R/T_{\rm H \alpha}}, 
\end{equation}
where FWHM is the width of H$\alpha$ filters in units of {\rm \AA} 
and $m_{\rm filter}$s are instrumental magnitudes 
in net H$\alpha$ and scaled {\it R}-band images, respectively. 
We present the EW profiles of all 29 HIghMass galaxies with our KPNO measurements in Fig.~\ref{fig:ew}, 
in order of ascending galaxy $M_*$. 
The EW is plotted as a function of semi-major axis in the second column; 
disk region with an exponential surface brightness profile is marked in between the 
vertical dash-dotted lines. 
Cleaned {\it R}-band and H$\alpha$ images are shown in the third and forth columns, color scales inverted. 
Apertures within which the Petrosian magnitudes are measured are overlaid on these images, 
being absent if Petrosian radius is undefined. 

\subsection{H$\alpha$ Photometry External Check}

UGC~7686 in the GOLDmine dataset is not re-observed. 
All the H$\alpha$ images presented here are new narrowband observations, 
and there are few literature measurements to compare for the purpose of external photometry quality check. 
We only find that UGC~8573 (NGC~5230) has been observed spectroscopically by \citet{Jansen2000b}. 
\citet{Jansen2000b} obtained integrated and nuclear spectra for 196 nearby galaxies (Nearby Field Galaxy Survey), 
in order to measure the current SFRs and metallicities of these galaxies. 
For UGC~8573, 77\% of the total {\it B}-filter light is sampled in the integrated spectra, 
resulting in the spectroscopic induce measurements of EW$\rm _{[NII]\lambda\lambda6548}$~=~2.1, 
EW$\rm _{H\alpha}$~=~20.0, and EW$\rm _{[NII]\lambda\lambda6584}$~=~7.0~$\rm \AA$ for the integrated spectra. 
The corresponds numbers for the nuclear spectra are 1.2, 8.0, and 3.4 $\rm \AA$, respectively. 
Our global EW$_{\rm H\alpha+[NII]}$ is determined to be 34.51$\pm$2.84~$\rm \AA$, given the flux inside the Petrosian aperture. 
Taking into account of the different apertures and the variation of EW in the galaxy, this agreement is satisfactory. 
Although only the emission-line fluxes relative to H$\beta$ are presented in \citet{Jansen2000b}, 
\citet{Kewley2002} calibrate these integrated fluxes to absolute fluxes by comparison with the {\it B}-band photometry. 
Accounting for the extinction inferred from the IR data, 
they derived an H$\alpha$ SFR for NGC~5230 to be log SFR = 0.93$~M_\odot~{\rm yr^{-1}}$, 
in agreement with our H$\alpha$ result, being log SFR(H$\alpha$) = 0.96$\pm$0.15$~M_\odot~{\rm yr^{-1}}$, 
and our SED fitting result, being log SFR(SED) = 0.89$\pm$0.23$~M_\odot~{\rm yr^{-1}}$ (see Section \ref{Cal}). 

We perform an independent check of the calibration of our H$\alpha$ measurements by 
comparing the fluxes in the 3~arcsec nuclear apertures in our H$\alpha$ images with the flux in the 
H$\alpha$+[NII] lines in the SDSS DR8 spectroscopic data \citep{Brinchmann2004} obtained in 3~arcsec fibers 
(UGC~12506 is outside of the SDSS spectroscopic footprint). 
To reduce the impact of aperture effect, we focus on 18 galaxies with better smoothed PSFs. 
Additional flux uncertainties are introduced into this comparison as a result of our astrometry uncertainty, 
as well as the offset from center in case of the SDSS fiber positions (e.g., UGC~8573, UGC~9334). 
We characterize this by the difference in fluxes we obtained from our net H$\alpha$ image when 
placing the 3~arcsec aperture (1) at the coordinate of SDSS-reported fiber position and 
(2) at the {\it R}-band global centroid we derived in Appendix \ref{app:KPNO}. 
Note that the SDSS line fluxes are corrected for Galactic extinction.  
We assume $A({\rm H\alpha}) = 2.6E(B-V)$ following \citet{Kennicutt2008}, and $E(B-V)$ is given by the 
DIRBE measurements of diffuse IR emission \citep{Schlegel1998}. 
Based on the HI systemic velocity, we find that [NII]$\lambda\lambda$6584 is outside of the H$\alpha$ filter 
for AGC~203522, UGC~8408, UGC~9023, UGC~9234, and AGC~188749, so that only the SDSS line fluxes of 
H$\alpha$ and [NII]$\lambda\lambda$6548 are added up. 
The final comparison is shown in Fig.~\ref{fig:Ha}. 
Considering that the fluxes from our survey  and the SDSS dataset have been measured using entirely different techniques 
(imaging {\it vs.}~spectroscopic), the consistency of the respective fluxes is excellent. 
The only outlier, UGC~6066, has large uncertainties in SDSS measurements as a result of the low S/N spectra, 
making our flux more reliable. 

\section{Reprocessed SDSS Photometry}
\label{app:SDSS}

\subsection{Surface and Integral Photometry}

Because many gas rich galaxies appear to be patchy due to star forming regions throughout the disk, 
the standard SDSS pipeline will easily suffer from shredding \citep{Huang2012a}. 
Furthermore, the automatic sky background subtraction is optimized for more distant galaxies 
with small angular sizes, but will lead to over subtraction for large or LSB galaxies. 
As a result, we independently recomputed photometric quantities for all 34 HIghMass galaxies, 
making use of an IRAF pipeline adapted from `GALPHOT'. 

Header-supplemented corrected frames are retrieved from SDSS DR7 in five bands, {\it u, g, r, i, z}. 
Multiple adjacent frames are combined if our target is on image edge. 
Following the treatment to H$\alpha$ and {\it R}-band images in Appendix \ref{app:KPNO}, 
all the other bands are first shifted to align with the {\it r}-band image and smoothed to a common PSF. 
Elliptical isophotal fitting is applied to all images in order to interpolate over the contaminating sources for all bands, 
but we use the same set of apertures given by the {\it r}-band fitting on all SDSS images for consistency. 
We obtain one-dimensional profiles, as a function of semi-major axis, 
of the centroid, $\theta$, $\epsilon$, and surface brightness of all elliptical isophots. 

Among various global magnitudes we calculated, the Petrosian magnitudes are preferred. 
Alternatively, we adopt mag$_8$ in all bands if Petrosian radius is undetermined in any band 
to ensure the reliability of colors that are obtained from comparable apertures. 
To convert counts to AB magnitudes, we follow the convention of SDSS asinh magnitudes, 
where the photometric zeropoint, 
extinction coefficient, 
and softening parameter 
are found in image headers. 
However, we confirm that all of the HIghMass galaxies are detected in SDSS bands with high S/N 
so that the difference between asinh magnitude and conventional magnitude is less than 1\%. 

The SDSS measurements of three galaxies are presented by colored open symbols along with the 
KPNO results by black filled symbols in the top four panels of Fig.~\ref{fig:6168} -- \ref{fig:9234}; 
{\it u}-band data are denoted by blue circles, {\it g} by cyan upward triangles, {\it r} by green downward triangles, 
{\it i} by yellow diamonds, and {\it z} by red squares. 
Definitions of the lines are the same as those of the KPNO data. 
Although only three representative cases are shown, 
it is generally true that the elliptical isophot fitting to {\it r}-band image performed independently from 
the {\it R}-band fitting generates similar sets of apertures possessing similar variations of the 
$\theta$ and $\epsilon$ with semi-major axis: 
the fitting result is robust and insensitive to the initial assumptions of aperture orientation. 
The global $\theta$ and $\epsilon$ measurements from {\it r}- and {\it R}-band images also agree 
with each other within the uncertainties. 
Because {\it R} and H$\alpha$ surface photometries are performed in the same set of apertures, 
we can derive EW$_{\rm H\alpha}$ profiles. Similarly, surface brightness is measured in the same 
set of apertures for all SDSS bands, and thus we can obtain the azimuthally-averaged broadband 
color profiles, which are demonstrated in the first column of Fig.~\ref{fig:ew} 
for the $u-r$ color. In between the vertical dash-dotted lines are the disk regions in {\it u} (blue lines) and 
{\it r} (green lines) bands, respectively. 

\subsection{Broadband Photometry Quality Check}

\subsubsection{Internal Comparison}

We have checked the internal consistency of broadband photometry by comparing 
the KPNO {\it R}-band magnitudes with the same quantities inferred by the combined SDSS magnitudes, 
using the transformation, $R = r - 0.2936(r - i) - 0.1439;~\sigma = 0.0072$. 
The KPNO magnitudes on the Johnson-Cousins standard photometric system are placed onto the AB system 
by correcting for the color zero-point differences between these systems: $R_{\rm Johnson} = R_{\rm AB} - 0.055$.
The most extreme outlier from the one-to-one relation is AGC~190796. Its KPNO {\it R}-band magnitude is measured inside 
the Petrosian aperture, whereas its SDSS combined magnitude comes from mag$_8$ because the Petrosian 
radius is available only on the {\it r}-band image. This Malin-like galaxy has a relatively steeper inner disk but 
the surface brightness drops off slowly throughout the outer disk. The disk scale length is thus underestimated 
from the inner fit. In all, we demonstrate that our SDSS-derived {\it R}-band magnitudes agree with the KPNO 
measurements without any systematic offset. 
This gives further confidence that our KPNO flux calibration is reliable. 
Plus, the two separate ellipse fittings give consistent $\theta$, $\epsilon$, and surface brightness profiles. 

\subsubsection{External Comparison}

We perform external checks with the {\it I}-band catalog from SFI++ Tully-Fisher Survey \citep{Springob2007}, 
as well as the SDSS DR8 pipeline measurements. {\it I}-band magnitudes are derived from our SDSS magnitudes 
via $I = i - 0.3780(i - z)  -0.3974;~\sigma = 0.0063$. We note that the SFI++ program was led by our earlier group 
members and a similar photometry procedure was followed. It is not surprising to see the excellent agreement 
between our results (Petrosian magnitudes preferred) and the SFI++ results (mag$_8$) for the 12 galaxies in common. 
We mention that UGC~8573 has been observed by \citet{Jansen2000a} and their {\it B}-band magnitude also agree with ours. 

In addition we retrieved the SDSS images from DR7 database but we compare our results with the DR8 pipeline measurements, 
given the fact that DR8 is superior to DR7 mainly due to the changes in image processing rather than re-observing. 
Although the sky background subtraction is improved in DR8 \citep{Blanton2011}, the problem of shredding still remains. 
We identify the HIghMass galaxies with problematic DR8 photometry in at least one band and they are represented by 
open circles in Fig.~\ref{fig:scheck}. The rest of galaxies as filled circles in the same figure can be cross correlated to 
a single photometric object that dominates in flux. The top panels illustrate comparisons between our mag$_8$ 
and the DR8 model magnitudes in five bands, respectively. 
Obviously, fluxes are under represented by the pipeline magnitudes, especially in bluer bands and among the galaxies 
suffering from shredding. It suggests an inconsistency of flux redistribution among multiple children in different bands, 
in a sense that the bluer star-forming regions are more likely to be identified as separate photometric objects. 
Therefore, the pipeline magnitudes result in redder colors than our measurements, and subsequently lower SSFRs. 
The lower panels show comparisons of more quantities, e.g., axial ratio, $r_{\rm petro, 50}$, $r_{\rm petro, 90}$, and 
disk scale length $r_{\rm d}$. Both axial ratio measurements have been corrected for the seeing effect and no 
systematic deviation exists. However, our Petrosian magnitudes are derived in elliptical apertures, so that 
our $r_{\rm petro, 50}$ and $r_{\rm petro, 90}$ are systematically larger than the same pipeline values derived in circular 
apertures. Large scatters are seen in the last two panels; the most extreme outliers with much smaller 
pipeline $r_{\rm d}$s suffer from shredding (being open circles). 
These indicate that the pipeline axial ratio can be a good proxy to infer the circular velocity from HI line width by 
an inclination correction, whereas the derived quantities depending on $r_{\rm d}$ are more vulnerable to large scatter, 
e.g., the empirical $\lambda$ values obtained in \citet{Huang2012b}. 

\end{appendices}


\begin{deluxetable}{rrrrrrrrrrrrr}
\rotate
\tablecolumns{13}
\tablewidth{0pt}
\tabletypesize{\scriptsize}
\tablecaption{Basic properties and selected observing programs of the HIghMass sample \label{tab:HI}}
\tablehead{
\colhead{AGC}	&\colhead{Other}		&\colhead{R.A.}	&\colhead{Dec.}		&\colhead{morph}	&
\colhead{$W_{50}$}	&\colhead{$V_{\rm rot}$}	&
\colhead{$cz$}	&
\colhead{$D$}	&\colhead{$\log M_{\rm HI}$}	&
\colhead{\it GALEX}	&\colhead{H$\alpha$}	&\colhead{HI}	\\
\colhead{}	&\colhead{}	&\colhead{[$^\circ$]}	&\colhead{[$^\circ$]}	&\colhead{}	&
\colhead{[km~s$^{-1}$]} 	&\colhead{[km~s$^{-1}$]} 	&
\colhead{[km~s$^{-1}$]}		&
\colhead{[Mpc]}	&\colhead{[M$_\odot$]}	&
\colhead{}	&\colhead{}	&\colhead{}	\\
\colhead{(1)} & \colhead{(2)} & \colhead{(3)} & \colhead{(4)} &
\colhead{(5)} & \colhead{(6)} & \colhead{(7)} & \colhead{(8)} &
\colhead{(9)} & \colhead{(10)} & \colhead{(11)} & \colhead{(12)} &
\colhead{(13)}
}
\startdata
188749	&SDSS J080649.98+120341.7	&121.7079	&12.062	&ExtendedSrc	&136	&105	&10972	&160	&10.14(0.03)	&AIS	&K	&GMRT11	\\
4599		&UGC 04599	&131.9238	&13.419	&(R)S0	&148	&195	&2071	& 32	&10.04(0.15)	&AIS		&K	&...	\\
190796	&LSBC F634-V02	&137.7754	&13.122	&S/Malin-like	&123	&158	&8895	&131	&10.14(0.02)	&AIS	&K	&...	\\
721391	&CGCG 121-063	&140.3658	&25.062	&...	&293	&152	&7537	&111	&10.07(0.02)	&AIS	&K	&...	\\
190277	&CGCG 062-031	&141.3088	&12.156	&Sdm(flat)	&286	&144	&8658	&128	&10.22(0.02)	&MIS	&K	&...	\\
5543	&UGC 05543	&154.0850	& 4.822	&Sc	&547	&340	&13737	&201	&10.73(0.03)	&MIS	&K	&JVLA11	\\
5648	&UGC 05648	&156.5362	& 4.372	&S	&322	&169	&6862	&103	&10.14(0.02)	&MIS	&K	&...	\\
5711	&NGC 3270	&157.8750	&24.869	&S(r)b:	&522	&271	&6253	& 93	&10.51(0.01)	&AIS	&K	&JVLA11	\\
203522	&SDSS J103304.79+073453.2	&158.2700	& 7.581	&...	&190	&108	&10886	&160	&10.21(0.04)	&AIS	&K	&GMRT11	\\
6043	&UGC 06043	&164.0642	&15.223	&Scd:	&276	&150	&8148	&121	&10.06(0.03)	&AIS	&K	&...	\\
6066	&UGC 06066	&164.7471	& 6.522	&Sab:edge-on	&667	&342	&11807	&173	&10.71(0.02)	&...	&K	&JVLA11	\\
6168	&UGC 06168	&166.7679	& 7.804	&(R')S(r)bc:	&363	&193	&8073	&120	&10.35(0.02)	&AIS	&K	&JVLA11	\\
6536	&NGC 3728	&173.3158	&24.447	&Sb	&430	&269	&6962	&103	&10.58(0.01)	&AIS	&K	&WSRT11	\\
6692	&NGC 3833	&175.8704	&10.161	&Sc	&390	&219	&6060	& 91	&10.37(0.01)	&AIS	&K	&... \\
213964	&SDSS J114637.62+061017.0	&176.6567	& 6.171	&...	& 88	&57	&9876	&146	&10.08(0.03)	&AIS	&K	&GMRT11	\\
6895	&NGC 3968	&178.8696	&11.968	&S(rs)bc	&490	&324	&6389	& 96	&10.65(0.01)	&AIS	&K	&GMRT09	\\
6967	&NGC 4017	&179.6908	&27.452	&Sbc	&253	&242	&3453	& 51	&10.23(0.09)	&GII	&K	&...	\\
7220	&IC 3046	&183.2829	&12.918	&S?/Sc(s)I	&443	&225	&8102	&120	&10.52(0.01)	&GII	&K	&JVLA11	\\
7686	&IC 3467	&188.1025	&11.787	&Scd:	&274	&140	&7510	&112	&10.22(0.02)	&DIS	&G	&GMRT11	\\
7899	&IC 3704	&190.9400	&10.770	&Sbc/Sc	&433	&221	&8696	&128	&10.42(0.02)	&GII	&K	&JVLA11	\\
8089	&UGC 08089	&194.6254	& 9.544	&SBdm:	&117	&105	&7142	&106	&10.06(0.01)	&...	&K	&GMRT11	\\
8408	&NGC 5115	&200.7512	&13.951	&SBcd:	&290	&171	&7303	&108	&10.26(0.02)	&AIS	&K	&...	\\
8475	&NGC 5162	&202.3575	&11.008	&Scd:	&591	&329	&6835	&101	&10.66(0.01)	&GII	&K	&GMRT09	\\
8573	&NGC 5230	&203.8829	&13.676	&S(s)c	&152	&154	&6857	&101	&10.53(0.01)	&AIS	&K	&GMRT09	\\
8797	&UGC 08797	&208.2600	&24.560	&S?	&559	&305	&17104	&247	&10.70(0.04)	&AIS	&...	&JVLA11	\\
9023	&UGC 09023	&211.7129	& 9.321	&Scd:	&231	&128	&7203	&106	&10.05(0.02)	&AIS	&K	&...	\\
248881	&KUG 1405+151	&211.8550	&14.919	&Spiral	&192	&115	&7660	&112	&10.10(0.02)	&AIS	&K	&...	\\
9037	&UGC 09037	&212.1213	& 7.058	&Scd:	&294	&174	&5939	& 88	&10.33(0.01)	&MIS	&K	&JVLA11	\\
726428	&2MASX J14220675+2659490	&215.5283	&26.997	&...	&208	&142	&8892	&129	&10.02(0.03)	&AIS	&K	&...	\\
9234	&UGC 09234	&216.1942	&26.139	&S?	&589	&433	&10890	&158	&10.68(0.02)	&GII	&K	&WSRT11	\\
9334	&NGC 5650	&217.7542	& 5.978	&Sbc	&413	&259	&7493	&110	&10.57(0.01)	&MIS	&K	&JVLA11	\\
714145	&2MASX J14342549+0835510	&218.6063	& 8.597	&...	&283	&158	&8262	&121	&10.18(0.03)	&AIS	&K	&...	\\
9410	&UGC 09410	&219.3079	& 8.646	&Sbc	&317	&170	&8421	&123	&10.19(0.03)	&AIS	&K	&...	\\
12506	&UGC 12506	&349.8771	&16.074	&Scd:	&457	&230	&7237	& 98	&10.53(0.01)	&AIS	&K*	&JVLA11	\\
\hline
\enddata
\end{deluxetable}

\begin{deluxetable}{rrrrrrrrrrrrrr}
\rotate
\tablecolumns{14}
\tablewidth{0pt}
\tabletypesize{\scriptsize}
\tablecaption{KPNO measurements and derived values of the HIghMass sample \label{tab:KPNO}}
\tablehead{
\colhead{AGC}	&\colhead{H$\alpha$ filter}		&\colhead{$\theta$}	&\colhead{$\epsilon$}		&\colhead{$\mu_0$}	&
\colhead{$r_{\rm d}$}	&
\colhead{$r_{\rm d, out}$}	&\colhead{$r_{\rm petro, 50}$}	&
\colhead{$r_{\rm petro, 90}$}	&\colhead{$d_{25}$}	&\colhead{{\it R}}	&\colhead{$\log F_{\rm H\alpha+[NII]}$}	&\colhead{EW$_{\rm H\alpha+[NII]}$}	&
\colhead{$\log {\rm SFR (H\alpha)}$}	\\
\colhead{}	&\colhead{}	&\colhead{[deg]}	&\colhead{}	&\colhead{[mag arcsec$^{-2}$]}	&
\colhead{[arcsec]}		&
\colhead{[arcsec]}	&\colhead{[arcsec]}	&
\colhead{[arcsec]}	&\colhead{[arcsec]}	&\colhead{[mag]}	&\colhead{[erg s$^{-1}$ cm$^{-2}$]}	&\colhead{[\rm \AA]}	&
\colhead{[$M_\odot$~yr$^{-1}$]}\\
\colhead{(1)} & \colhead{(2)} & \colhead{(3)} & \colhead{(4)} &
\colhead{(5)} & \colhead{(6)} & \colhead{(7)} & \colhead{(8)} &
\colhead{(9)} & \colhead{(10)} & \colhead{(11)} & \colhead{(12)} &
\colhead{(13)} & \colhead{(14)}
}
\startdata
188749&	1497&	-17(28)&	0.27(0.14)&	20.87&	 4.2&	...&	 8.5&	18.5&	 32.2&	16.00(0.03)&	-13.86(0.13)&	17.37(5.56)&	-0.39(0.18)\\
190796&	1496&	 85(43)&	0.09(0.05)&	21.92&	 8.8&	...&	17.9&	57.9&	 49.8&	14.96(0.19)&	-13.47(0.12)\tablenotemark{*}&	19.54(5.68)&	-0.22(0.19)\\
721391&	1496&	 78(1)&	0.69(0.05)&	20.22&	 9.9&	 5.7&	12.4&	27.8&	 87.0&	14.81(0.03)&	-13.16(0.04)&	28.82(2.95)&	 0.32(0.28)\\
5543&	1517&	-11(6)&	0.39(0.06)&	19.83&	10.1&	...&	15.2&	37.7&	 96.5&	13.30(0.02)&	-12.53(0.02)&	30.63(1.41)&	 1.25(0.24)\\
5648&	1566&	 18(1)&	0.63(0.05)&	19.46&	 6.7&	...&	10.6&	27.0&	 68.5&	14.40(0.01)&	-12.98(0.02)&	34.26(1.83)&	 0.34(0.24)\\
5711&	1566&	 12(1)&	0.65(0.06)&	20.16&	23.4&	...&	29.9&	89.7&	208.5&	12.22(0.01)&	-12.45(0.03)&	13.98(0.88)&	 1.03(0.36)\\
203522&	1497&	-84(5)&	0.43(0.06)&	21.13&	 4.3&	 5.1&	...&	...&	 31.0&	15.98(0.02)\tablenotemark{*}&	-13.14(0.01)\tablenotemark{*}&	90.66(2.85)&	 0.47(0.17)\\
6043&	1496&	 62(6)&	0.56(0.04)&	21.07&	10.7&	34.7&	17.9&	37.0&	 77.6&	14.75(0.01)&	-13.21(0.06)&	24.41(3.90)&	 0.22(0.25)\\
6066&	1498&	 40(2)&	0.68(0.08)&	20.45&	14.5&	...&	10.9&	49.2&	121.7&	13.62(0.02)&	-13.17(0.06)&	 9.36(1.29)&	 0.88(0.34)\\
6168&	1496&	-56(2)&	0.58(0.07)&	19.92&	 8.7&	...&	13.1&	36.4&	 81.7&	14.05(0.01)&	-12.89(0.02)&	27.25(1.50)&	 0.57(0.26)\\
6536&	1566&	 21(5)&	0.38(0.05)&	20.88&	18.2&	...&	11.7&	48.3&	138.1&	12.68(0.01)&	-12.95(0.01)&	 6.72(0.25)&	 0.13(0.20)\\
6692&	1495&	 36(6)&	0.49(0.06)&	20.27&	17.6&	 6.8&	24.3&	38.6&	153.6&	13.10(0.01)&	-12.56(0.04)&	23.97(2.45)&	 0.49(0.20)\\
213964&	1497&	 35(46)&	0.32(0.10)&	21.18&	 5.7&	...&	10.9&	29.3&	 40.0&	15.68(0.02)&	-13.26(0.03)&	51.19(3.91)&	 0.08(0.15)\\
6895&	1566&	 20(13)&	0.31(0.04)&	19.78&	19.8&	...&	29.0&	74.6&	190.0&	11.62(0.02)&	-12.11(0.04)&	16.54(1.42)&	 0.96(0.24)\\
6967&	1565&	-29(51)&	0.36(0.17)&	19.41&	12.4&	...&	28.6&	47.5&	127.8&	12.60(0.02)&	-12.03(0.01)&	39.78(1.32)&	 0.31(0.11)\\
7220&	1496&	-49(1)&	0.70(0.09)&	20.26&	14.6&	11.7&	14.7&	43.2&	127.3&	13.84(0.01)&	-12.88(0.03)&	22.95(1.61)&	 0.90(0.35)\\
7899&	1496&	 41(2)&	0.70(0.06)&	19.53&	11.7&	 5.0&	17.3&	37.5&	118.0&	13.63(0.01)&	-12.45(0.01)&	53.87(1.14)&	 1.20(0.30)\\
8089&	1566&	-19(44)&	0.29(0.19)&	21.17&	 7.0&	...&	...&	...&	 49.7&	15.23(0.03)\tablenotemark{*}&	-13.50(0.04)\tablenotemark{*}&	21.42(2.22)&	-0.47(0.14)\\
8408&	1566&	 99(3)&	0.42(0.11)&	20.08&	10.6&	...&	16.9&	39.1&	 95.8&	13.52(0.02)&	-12.83(0.04)&	19.51(2.04)&	 0.41(0.19)\\
8475&	1566&	-15(3)&	0.52(0.05)&	19.88&	23.6&	...&	37.9&	86.8&	222.4&	11.73(0.01)&	-12.10(0.01)&	19.46(0.62)&	 1.33(0.30)\\
8573&	1566&	 75(46)&	0.15(0.06)&	19.84&	15.5&	...&	25.9&	52.0&	147.0&	12.01(0.02)&	-11.97(0.03)&	34.51(2.84)&	 0.96(0.15)\\
9023&	1566&	138(5)&	0.51(0.09)&	20.96&	10.1&	 6.0&	14.4&	32.2&	 75.2&	14.88(0.06)&	-13.19(0.12)&	30.68(8.99)&	 0.04(0.21)\\
248881&	1496&	-64(14)&	0.43(0.12)&	19.82&	 5.0&	12.5&	10.5&	25.3&	 47.3&	14.85(0.02)&	-12.90(0.01)&	58.02(2.44)&	 0.38(0.22)\\
9037&	1495&	-14(6)&	0.43(0.07)&	20.20&	12.2&	...&	17.2&	45.9&	108.0&	13.29(0.02)&	-12.43(0.04)&	39.95(3.62)&	 0.56(0.20)\\
726428&	1496&	 -4(7)&	0.28(0.05)&	20.93&	 5.8&	...&	 9.2&	25.8&	 43.3&	15.42(0.05)&	-13.34(0.02)\tablenotemark{*}&	37.06(1.65)&	-0.13(0.12)\\
9234&	1497&	-42(25)&	0.27(0.13)&	23.32&	33.3&	...&	 6.2&	19.2&	103.2&	14.36(0.02)&	-13.55(0.04)&	 6.70(0.67)&	-0.02(0.19)\\
9334&	1496&	-73(5)&	0.38(0.09)&	19.71&	13.1&	...&	22.8&	52.1&	127.7&	12.53(0.02)&	-12.16(0.03)&	35.34(2.54)&	 1.02(0.21)\\
714145&	1496&	-12(5)&	0.52(0.05)&	20.56&	 6.7&	...&	 9.9&	25.2&	 54.5&	15.23(0.04)&	-13.43(0.07)&	22.64(3.80)&	-0.07(0.22)\\
9410&	1496&	-50(2)&	0.58(0.04)&	20.21&	 8.7&	...&	12.6&	33.3&	 76.6&	14.45(0.01)&	-12.99(0.02)&	31.17(1.58)&	 0.40(0.22)\\
\hline
\enddata
\\
\begin{flushleft}
\footnotetext{*}{Values with superscripts are measured by extrapolating the radial 
profiles to eight times the disk scale length, in substitute for the 
Petrosian magnitudes. See Appendix \ref{app:KPNO} for the definitions of the two types 
of global magnitudes and validation of their consistency.}
\end{flushleft}
\end{deluxetable}

\clearpage
\begin{deluxetable}{rrrrrrrrrrrrrr}
\rotate
\tablecolumns{14}
\tablewidth{0pt}
\tabletypesize{\scriptsize}
\tablecaption{SDSS measurements and derived values of the HIghMass sample \label{tab:SDSS}}
\tablehead{
\colhead{AGC}	&\colhead{\it u}		&\colhead{\it g}	&\colhead{\it r}		&\colhead{\it i}	&
\colhead{\it z}	&\colhead{$\cos i$}	&\colhead{$\mu_e(r)$}	&
\colhead{$A_r$}	&
\colhead{$M_r$}	&\colhead{$\log M_*$}	&\colhead{$\log \mu_*$}	&
\colhead{$\log {\rm SFR(SED)}$}	&\colhead{$\log {b}$}\\
\colhead{}	&\colhead{[mag]}	&\colhead{[mag]}	&\colhead{[mag]}	&\colhead{[mag]}	&
\colhead{[mag]}		&\colhead{}	&\colhead{[mag arcsec$^{-2}$]}	&
\colhead{[mag]}	&\colhead{[mag]}	&\colhead{[$M_\odot$]}	&\colhead{[$M_\odot$ kpc$^{-2}$]}	&
\colhead{[$M_\odot$~yr$^{-1}$]}	&\colhead{}\\
\colhead{(1)} & \colhead{(2)} & \colhead{(3)} & \colhead{(4)} &
\colhead{(5)} & \colhead{(6)} & \colhead{(7)} & \colhead{(8)} &
\colhead{(9)} & \colhead{(10)} & \colhead{(11)} & \colhead{(12)} & \colhead{(13)}& \colhead{(14)}
}
\startdata
188749	&18.00(0.14)	&16.68(0.02)	&16.31(0.02)	&16.01(0.02)	&16.05(0.09)	&0.76	&23.5	&0.18(0.14)	&-19.96	& 9.50(0.08)	&7.32	&-1.28(0.98)	&-1.23(0.73)\\
4599	&15.71(0.05)	&14.18(0.01)	&13.54(0.01)	&13.14(0.01)	&12.95(0.02)	&0.93	&21.6	&0.16(0.13)	&-19.23	& 9.62(0.17)	&8.52	&-2.05(0.93)	&-1.89(0.70)\\
190796	&17.23(0.07)\tablenotemark{*}	&15.94(0.01)\tablenotemark{*}	&15.55(0.01)\tablenotemark{*}	&15.18(0.01)\tablenotemark{*}	&15.12(0.03)\tablenotemark{*}	&0.92	&24.0	&0.20(0.16)	&-20.34\tablenotemark{*}	& 9.76(0.11)	&...	&-0.61(0.88)	&-0.90(0.74)\\
721391	&16.57(0.05)	&15.45(0.01)	&15.02(0.01)	&14.71(0.01)	&14.54(0.03)	&0.26	&23.3	&0.62(0.30)	&-20.93	& 9.84(0.08)	&7.89	& 0.41(0.25)	&-0.22(0.20)\\
190277	&17.25(0.06)	&16.11(0.01)	&15.62(0.01)	&15.27(0.01)	&15.28(0.04)	&0.13	&24.6	&0.50(0.22)	&-20.54	& 9.70(0.07)	&7.76	& 0.07(0.23)	&-0.33(0.18)\\
5543	&15.44(0.05)	&14.16(0.01)	&13.60(0.01)	&13.23(0.01)	&13.07(0.02)	&0.59	&22.2	&0.47(0.26)	&-23.48	&11.07(0.09)	&8.22	& 1.08(0.31)	&-0.58(0.26)\\
5648	&16.23(0.05)	&15.12(0.01)	&14.66(0.01)	&14.40(0.01)	&14.19(0.02)	&0.30	&22.7	&0.54(0.27)	&-21.02	& 9.92(0.08)	&8.21	& 0.38(0.23)	&-0.28(0.16)\\
5711	&14.76(0.07)	&13.31(0.01)	&12.61(0.01)	&12.23(0.01)	&11.92(0.03)	&0.26	&23.2	&0.90(0.39)	&-23.19	&11.05(0.10)	&8.72	& 0.83(0.58)	&-0.74(0.45)\\
203522	&17.39(0.15)	&16.43(0.02)	&16.23(0.02)	&16.04(0.03)	&16.12(0.13)	&0.48	&24.3	&0.33(0.19)	&-20.18	& 9.25(0.08)	&6.91	& 0.12(0.51)	&-0.15(0.46)\\
6043	&16.63(0.11)	&15.45(0.01)	&15.05(0.02)	&14.87(0.02)	&14.61(0.06)	&0.40	&23.9	&0.50(0.27)	&-20.93	& 9.80(0.10)	&7.33	& 0.18(0.61)	&-0.39(0.47)\\
6066	&16.38(0.07)	&14.70(0.01)	&13.93(0.01)	&13.47(0.01)	&13.22(0.02)	&0.23	&23.2	&0.91(0.37)	&-23.26	&11.13(0.10)	&9.11	&-0.01(1.03)	&-1.48(0.79)\\
6168	&16.08(0.06)	&14.91(0.01)	&14.35(0.01)	&14.00(0.01)	&13.76(0.03)	&0.34	&23.2	&0.55(0.29)	&-21.69	&10.37(0.08)	&8.38	& 0.57(0.24)	&-0.39(0.19)\\
6536	&15.08(0.05)	&13.63(0.01)	&12.96(0.01)	&12.61(0.01)	&12.35(0.02)	&0.60	&21.8	&0.34(0.22)	&-22.54	&10.87(0.10)	&8.84	&-0.06(0.90)	&-1.30(0.70)\\
6692	&15.39(0.07)	&14.08(0.01)	&13.39(0.01)	&13.06(0.01)	&12.83(0.03)	&0.46	&22.5	&0.35(0.21)	&-21.93	&10.56(0.09)	&8.10	& 0.20(0.46)	&-0.84(0.38)\\
213964	&17.14(0.05)\tablenotemark{*}	&16.21(0.01)\tablenotemark{*}	&15.88(0.01)\tablenotemark{*}	&15.75(0.01)\tablenotemark{*}	&15.71(0.03)\tablenotemark{*}	&0.64	&23.9	&0.21(0.16)	&-20.20\tablenotemark{*}	& 9.39(0.06)	&6.97	& 0.06(0.18)	&-0.21(0.18)\\
6895	&13.67(0.06)	&12.49(0.01)	&11.88(0.01)	&11.49(0.01)	&11.27(0.02)	&0.65	&22.1	&0.41(0.26)	&-23.51	&11.24(0.09)	&8.45	& 1.15(0.24)	&-0.58(0.21)\\
6967	&14.07(0.05)	&13.04(0.01)	&12.60(0.01)	&12.41(0.01)	&12.33(0.03)	&0.85	&21.8	&0.16(0.12)	&-21.19	&10.00(0.12)	&7.67	& 0.19(0.23)	&-0.47(0.20)\\
7220	&16.00(0.06)	&14.73(0.01)	&14.12(0.01)	&13.76(0.01)	&13.40(0.03)	&0.18	&23.5	&0.92(0.38)	&-22.29	&10.59(0.08)	&8.48	& 0.88(0.27)	&-0.32(0.18)\\
7686	&16.47(0.07)	&15.40(0.01)	&15.01(0.01)	&14.75(0.01)	&14.65(0.04)	&0.22	&23.7	&0.56(0.27)	&-20.90	& 9.71(0.07)	&7.75	& 0.39(0.23)	&-0.19(0.20)\\
7899	&15.53(0.05)	&14.37(0.01)	&13.90(0.01)	&13.57(0.01)	&13.39(0.02)	&0.21	&23.1	&0.75(0.33)	&-22.47	&10.49(0.08)	&8.32	& 0.99(0.25)	&-0.24(0.20)\\
8089	&16.53(0.06)\tablenotemark{*}	&15.63(0.01)\tablenotemark{*}	&15.33(0.02)\tablenotemark{*}	&15.19(0.01)\tablenotemark{*}	&15.05(0.04)\tablenotemark{*}	&0.83	&24.0	&0.18(0.15)	&-20.05\tablenotemark{*}	& 9.37(0.07)	&6.93	& 0.04(0.17)	&-0.20(0.18)\\
8408	&15.46(0.05)	&14.34(0.01)	&13.88(0.01)	&13.61(0.01)	&13.49(0.03)	&0.53	&22.6	&0.32(0.20)	&-21.69	&10.24(0.08)	&7.97	& 0.50(0.22)	&-0.42(0.20)\\
8475	&14.09(0.06)	&12.79(0.01)	&12.10(0.01)	&11.74(0.01)	&11.41(0.02)	&0.44	&22.7	&0.74(0.33)	&-23.80	&11.28(0.09)	&8.33	& 1.31(0.27)	&-0.50(0.22)\\
8573	&13.90(0.05)	&12.77(0.01)	&12.29(0.01)	&11.98(0.01)	&11.86(0.03)	&0.87	&21.9	&0.20(0.16)	&-23.02	&10.89(0.09)	&8.00	& 0.89(0.23)	&-0.58(0.22)\\
8797	&16.62(0.06)	&15.26(0.01)	&14.59(0.01)	&14.21(0.01)	&14.00(0.03)	&0.40	&22.8	&0.62(0.31)	&-23.05	&10.98(0.09)	&8.49	& 0.79(0.42)	&-0.72(0.33)\\
9023	&16.83(0.14)	&15.57(0.02)	&15.17(0.02)	&14.96(0.02)	&14.88(0.08)	&0.44	&23.6	&0.31(0.19)	&-20.36	& 9.59(0.08)	&7.50	&-0.81(1.01)	&-0.96(0.75)\\
248881	&16.27(0.05)	&15.32(0.01)	&15.08(0.01)	&14.90(0.01)	&14.68(0.03)	&0.55	&22.8	&0.41(0.24)	&-20.64	& 9.56(0.11)	&7.46	& 0.43(0.23)	&-0.10(0.27)\\
9037	&14.99(0.06)	&13.97(0.01)	&13.60(0.01)	&13.37(0.01)	&13.19(0.04)	&0.53	&22.5	&0.36(0.22)	&-21.57	&10.09(0.08)	&7.89	& 0.63(0.21)	&-0.25(0.18)\\
726428	&17.10(0.05)\tablenotemark{*}	&16.07(0.01)\tablenotemark{*}	&15.66(0.01)\tablenotemark{*}	&15.33(0.01)\tablenotemark{*}	&15.43(0.03)\tablenotemark{*}	&0.68	&23.3	&0.18(0.13)	&-20.13\tablenotemark{*}	& 9.58(0.08)	&7.62	&-0.21(0.19)	&-0.46(0.20)\\
9234	&17.09(0.06)	&15.34(0.01)	&14.54(0.01)	&14.14(0.01)	&13.89(0.02)	&0.73	&21.8	&0.28(0.20)	&-21.77	&10.75(0.09)	&8.77	&-1.21(0.95)	&-2.12(0.72)\\
9334	&14.33(0.05)	&13.24(0.01)	&12.78(0.01)	&12.53(0.01)	&12.29(0.03)	&0.60	&22.3	&0.40(0.23)	&-22.92	&10.73(0.08)	&8.05	& 1.11(0.21)	&-0.31(0.16)\\
714145	&17.04(0.11)	&15.90(0.02)	&15.49(0.02)	&15.23(0.02)	&15.11(0.06)	&0.45	&23.3	&0.40(0.23)	&-20.40	& 9.64(0.08)	&7.71	& 0.01(0.30)	&-0.37(0.22)\\
9410	&16.39(0.09)	&15.25(0.01)	&14.73(0.01)	&14.44(0.01)	&14.28(0.04)	&0.36	&23.1	&0.44(0.24)	&-21.26	&10.10(0.08)	&8.07	& 0.35(0.24)	&-0.40(0.19)\\
12506	&16.13(0.09)	&14.65(0.01)	&13.96(0.01)	&13.71(0.01)	&13.22(0.03)	&0.13	&24.2	&0.82(0.38)	&-21.98	&10.46(0.11)	&8.37	& 0.40(0.64)	&-0.64(0.52)\\
\hline
\enddata
\\
\begin{flushleft}
\footnotetext{*}{Values with superscripts are measured by extrapolating the radial 
profiles to eight times the disk scale length, in substitute for the 
Petrosian magnitudes. See Appendix \ref{app:KPNO} for the definitions of the two types 
of global magnitudes and validation of their consistency.}
\end{flushleft}
\end{deluxetable}

\eject

\begin{figure*}
\center{
\includegraphics[scale=0.8]{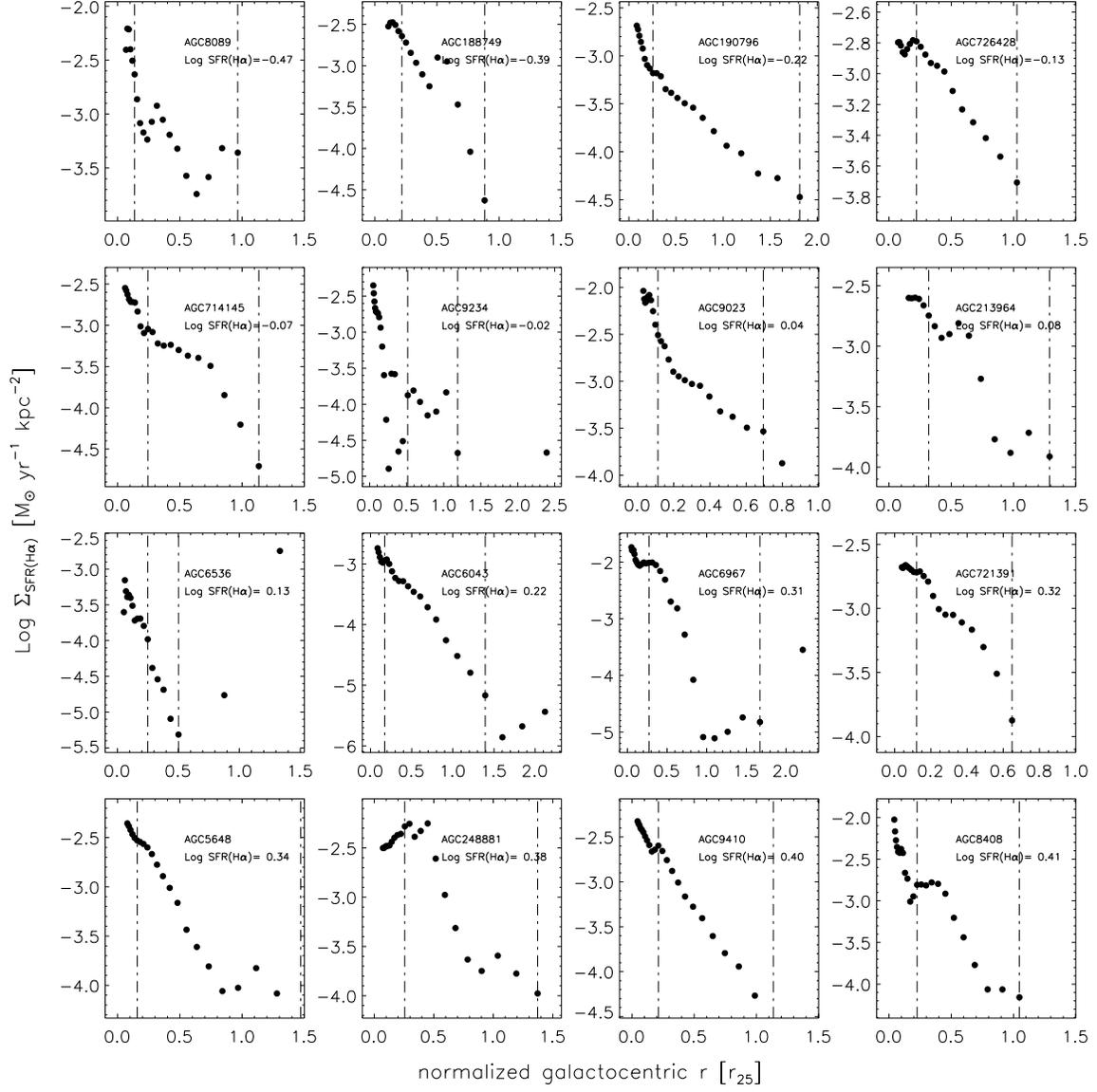}
}
\caption[]{
The $\Sigma_{\rm SFR}$ radial profiles of 29 HIghMass galaxies, in order of increasing integrated SFR(H$\alpha$). 
Values of $\Sigma_{\rm SFR}$ are deprojected by a factor of $\cos i$, where $i$ is the inclination of the disk.
The galactocentric radius is normalized by $r_{25}$ on the $x$-axis. 
The marked inner disk region is denoted by vertical dash-dotted lines (see Appendix).
}
\label{fig:spf}
\end{figure*}

\addtocounter{figure}{-1}

\begin{figure*}
\center{
\includegraphics[scale=0.8]{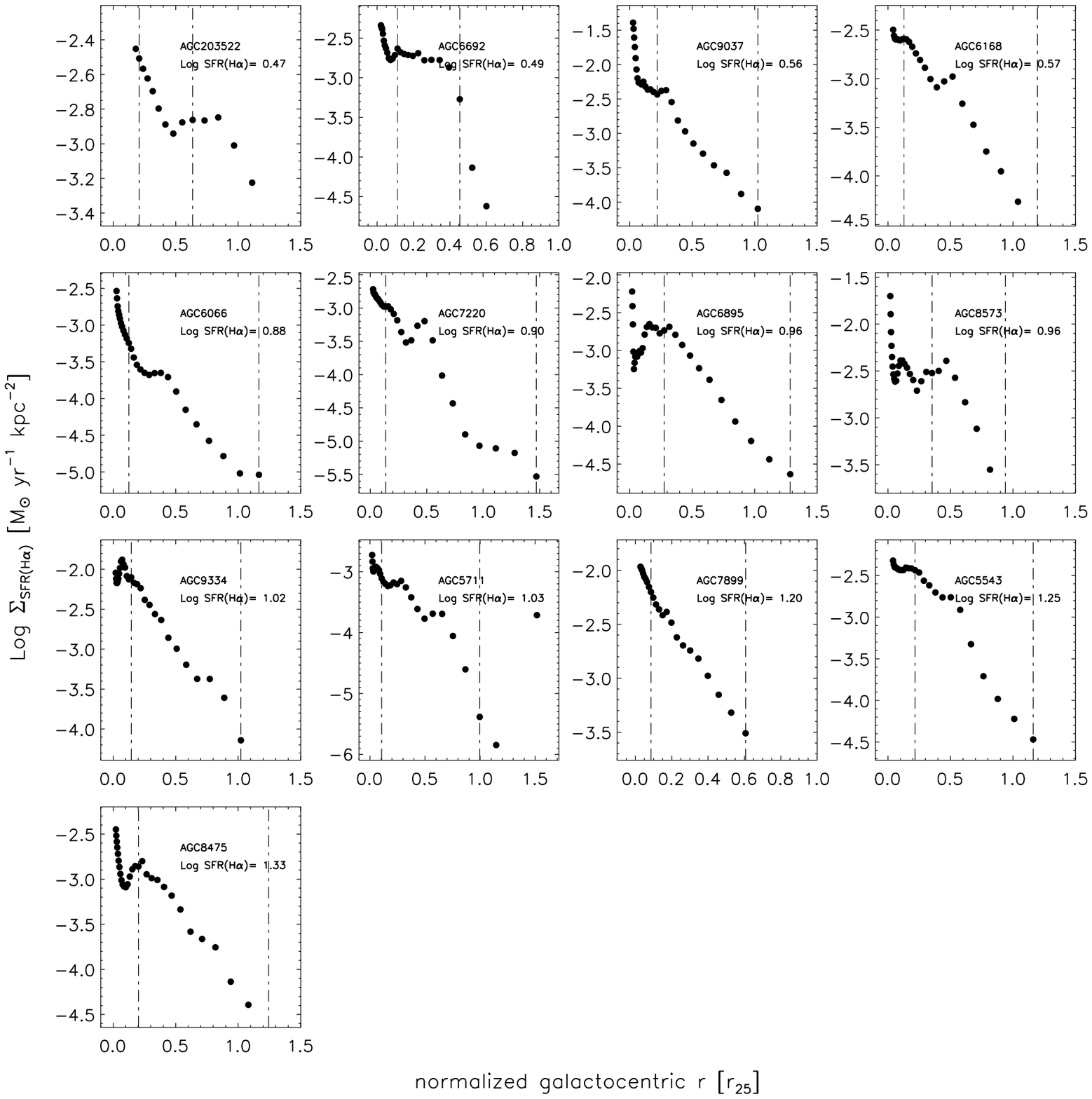}
}
\caption[]{
\it Continued.}
\end{figure*}

\begin{figure*}
\center{
\includegraphics[scale=0.8]{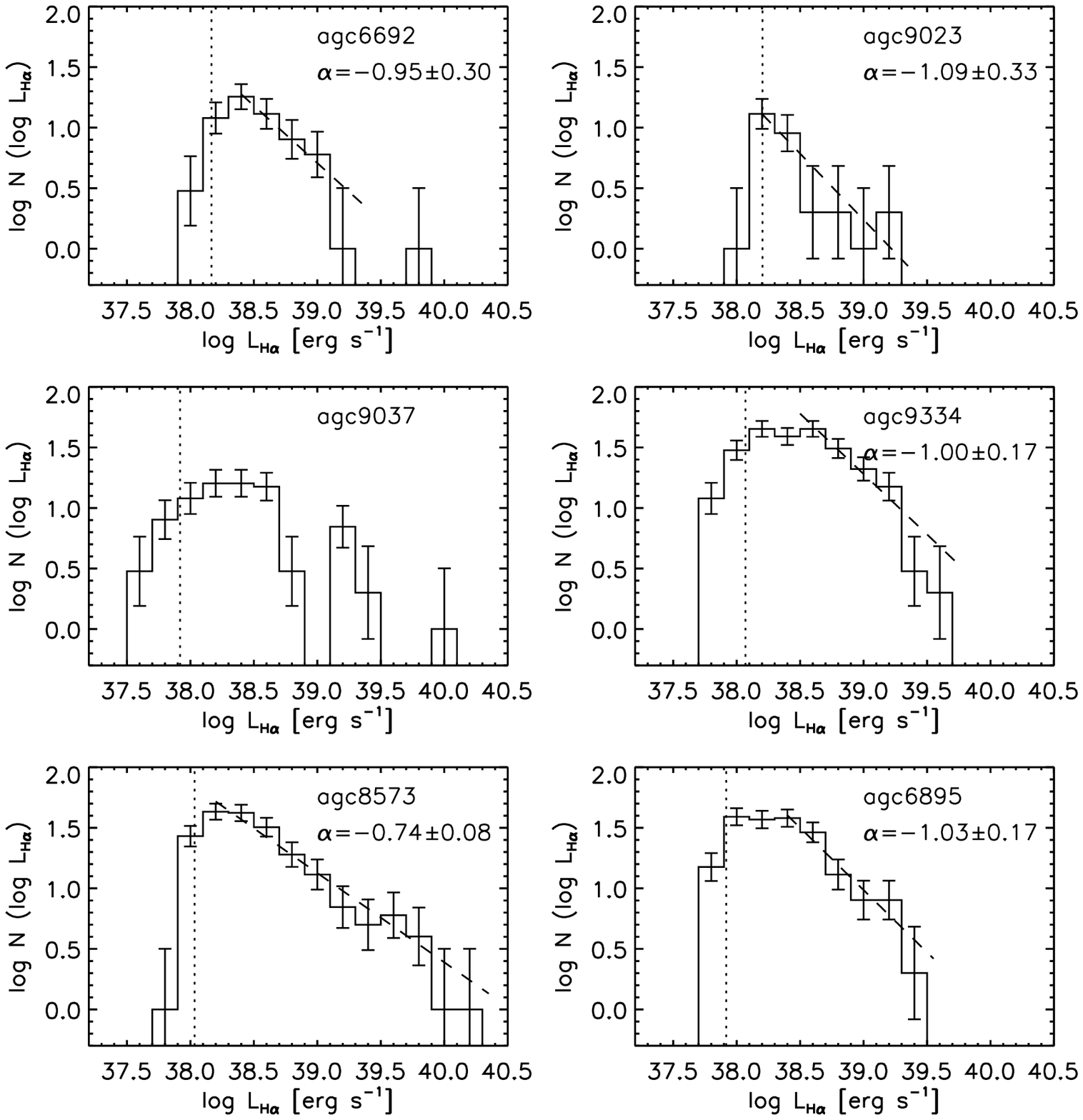}
}
\caption[]{
HII region luminosity functions for the six HIghMass galaxies with the best linear resolutions. 
The vertical dotted lines are 5$\sigma$ detection limit. The dashed lines are power law fits, 
$dN/d \log L_{\rm H\alpha} \propto L_{\rm H\alpha}^\alpha$, to the upper luminosity functions only (see text). 
Most of the slopes are consistent with $dN/dL_{\rm H\alpha} \propto L_{\rm H\alpha}^{\alpha-1} \propto L_{\rm H\alpha}^{-2}$.}
\label{fig:LF}
\end{figure*}

\begin{figure*}
\center{
\includegraphics[scale=0.85]{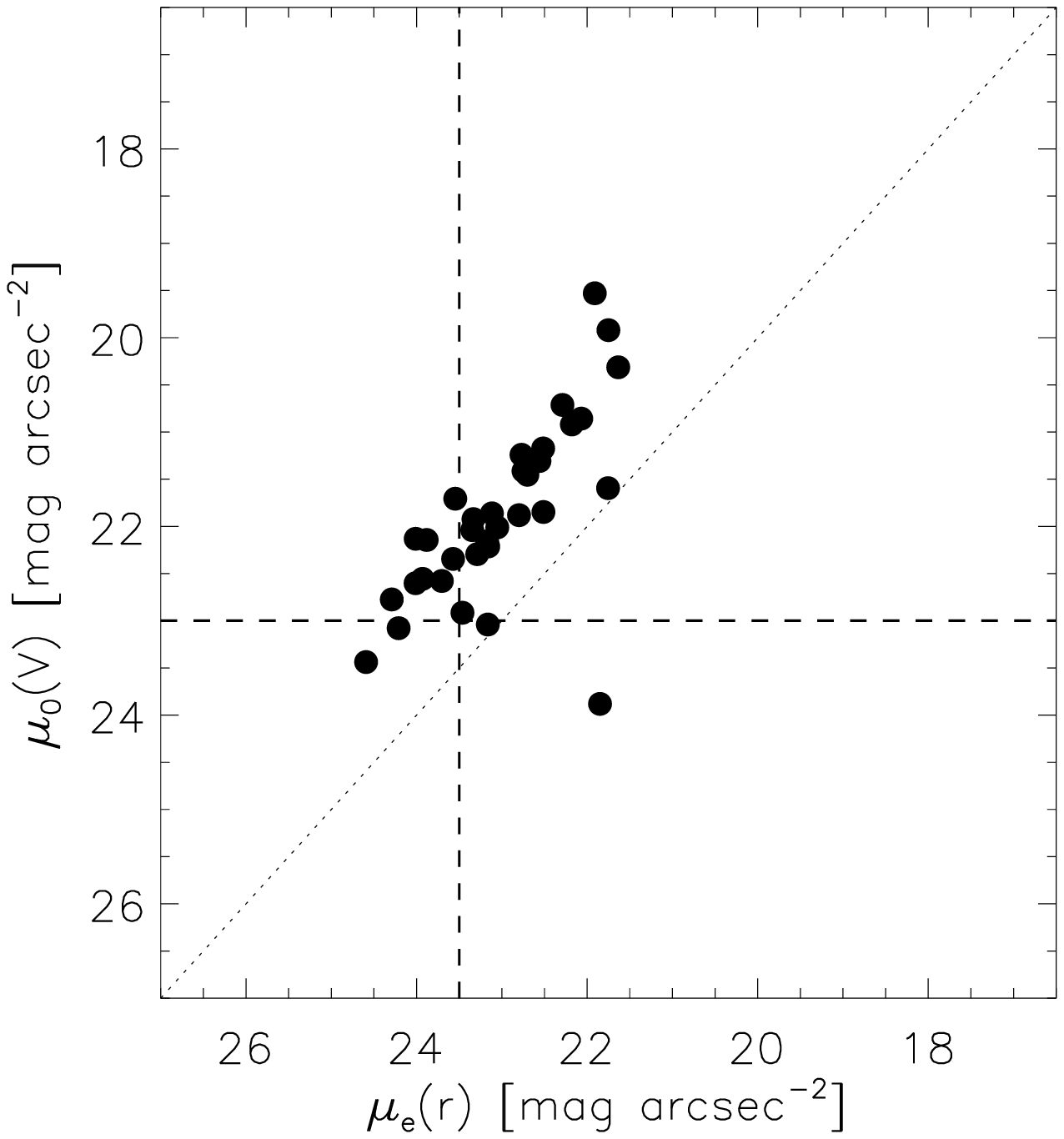}
}
\caption[]{
The {\it r}-band effective surface brightness on the $x$-axis and 
{\it V}-band disk surface brightness interpolated to the center on the $y$-axis, 
both after inclination correction. The dotted diagonal line illustrates the one-to-one relation. 
The criteria of LSB galaxies used in previous works are marked by dashed lines in this plot. 
}
\label{fig:SB}
\end{figure*}

\begin{figure*}
\center{
\includegraphics[scale=0.7]{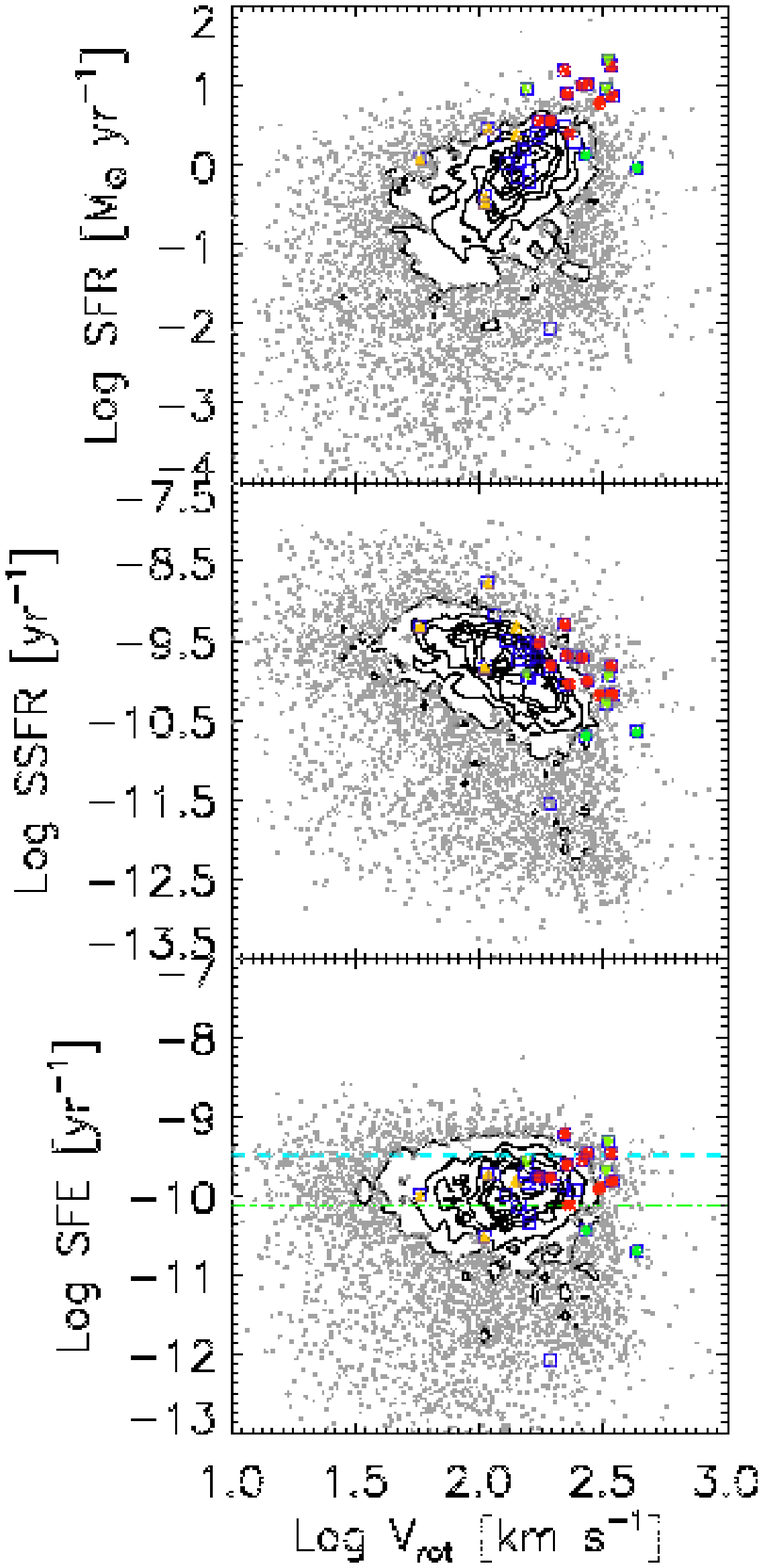}
}
\caption[]{
The star formation properties of the HIghMass galaxies in comparison with the parent $\alpha$.40--SDSS (DR8) sample. 
Definition of the colored symbols, contours, and points are the same as those in Fig.~\ref{fig:CMD}. 
The vertical layout shows the run of SFR, SSFR, and SFE, all examined as a function of HI rotation velocity, $V_{\rm rot}$. 
In the last panel, the horizontal dash-dotted line in green corresponds to an HI depletion timescale of a Hubble time. 
Cyan dashed line marks the average SFE derived from the GASS as an optically-selected sample. 
}
\label{fig:SFS}
\end{figure*}

\begin{figure*}
\center{
\includegraphics[scale=0.7]{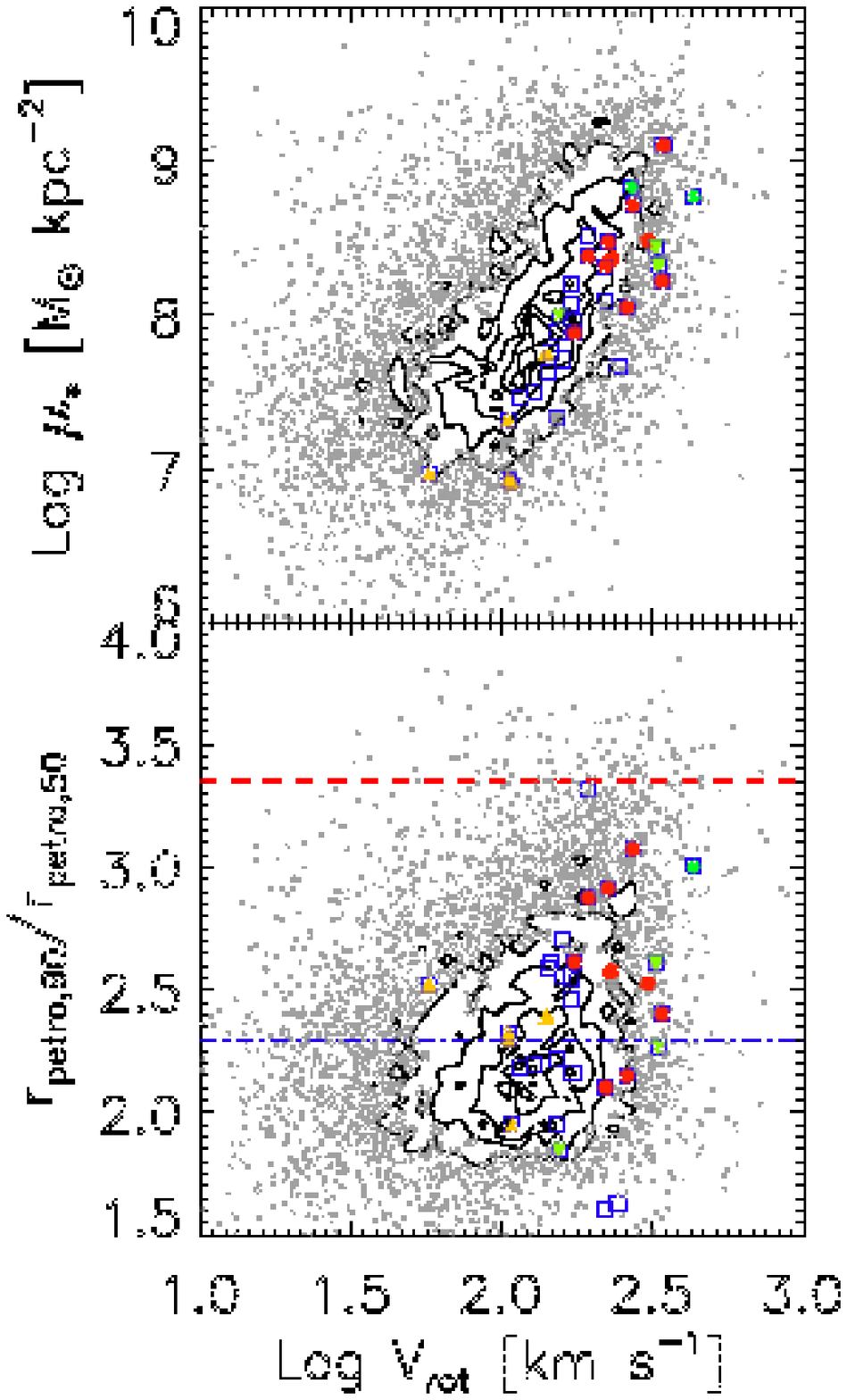}
}
\caption[]{
The stellar disk properties of the HIghMass galaxies in comparison with the parent $\alpha$.40--SDSS (DR8) sample. 
Definition of the colored symbols, contours, and points are the same as those in Fig.~\ref{fig:CMD}. 
Stellar mass surface density $\mu_*$ and concentration index $r_{\rm petro,90}/r_{\rm petro,50}$ are inspected 
with respect to HI rotation velocity in the upper and lower panels, respectively. 
In the lower panel, the red dashed line on top denotes the $r_{\rm petro,90}/r_{\rm petro,50}$ value for a 
de Vaucouleurs model; the blue dash-dotted line associates with an exponential disk. 
}
\label{fig:muext}
\end{figure*}

\begin{figure*}
\center{
\includegraphics[scale=0.7]{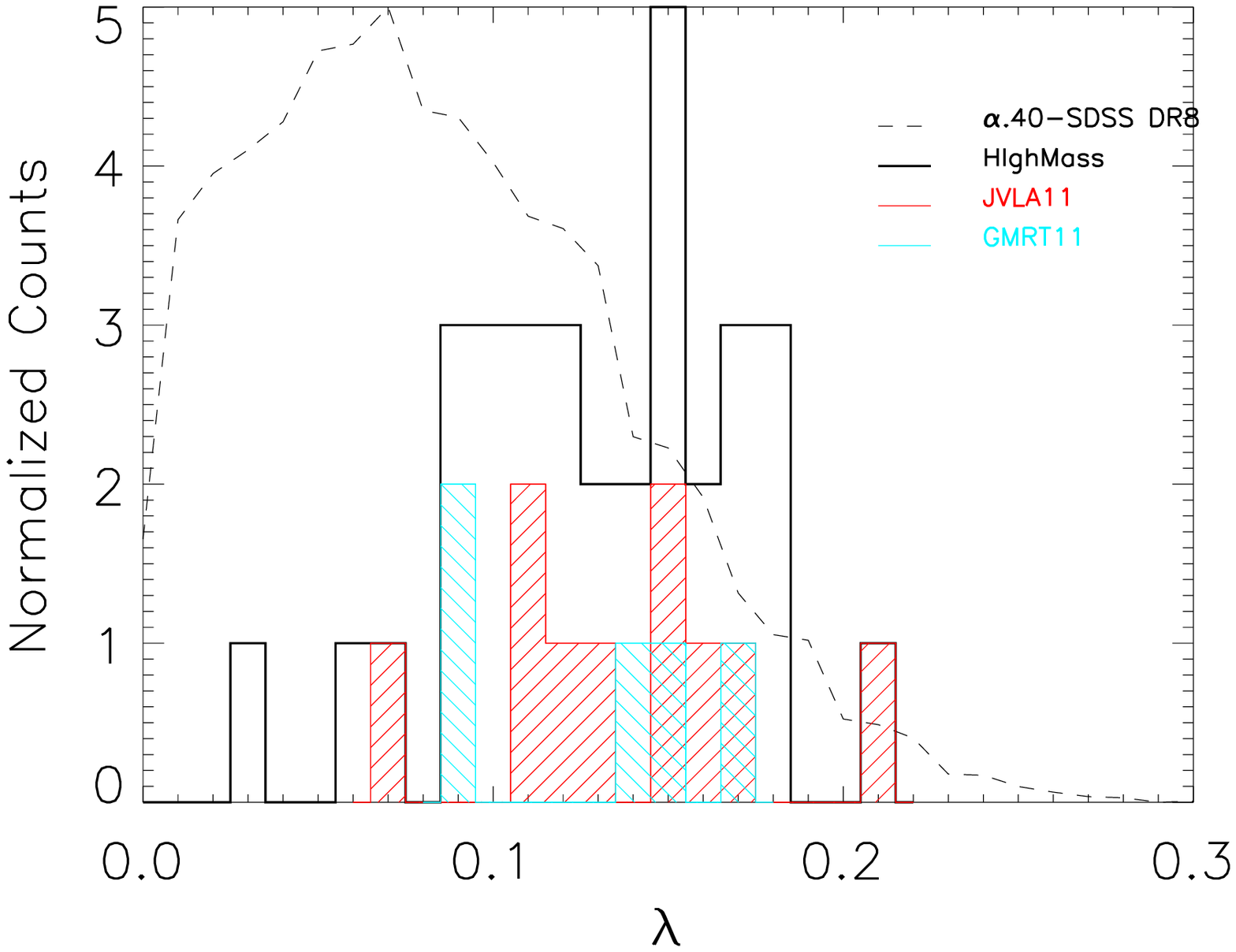}
}
\caption[]{Empirical distributions of the spin parameter ($\lambda$) of host halo. We follow the same approach as in \citet{Huang2012b} 
to estimate $\lambda$ for both the HIghMass and $\alpha$.40--SDSS parent sample. 
Being extremely HI rich, the HighMass galaxies have on average higher $\lambda$ values.}
\label{fig:spin}
\end{figure*}

\begin{figure*}
\center{
\includegraphics[scale=0.6]{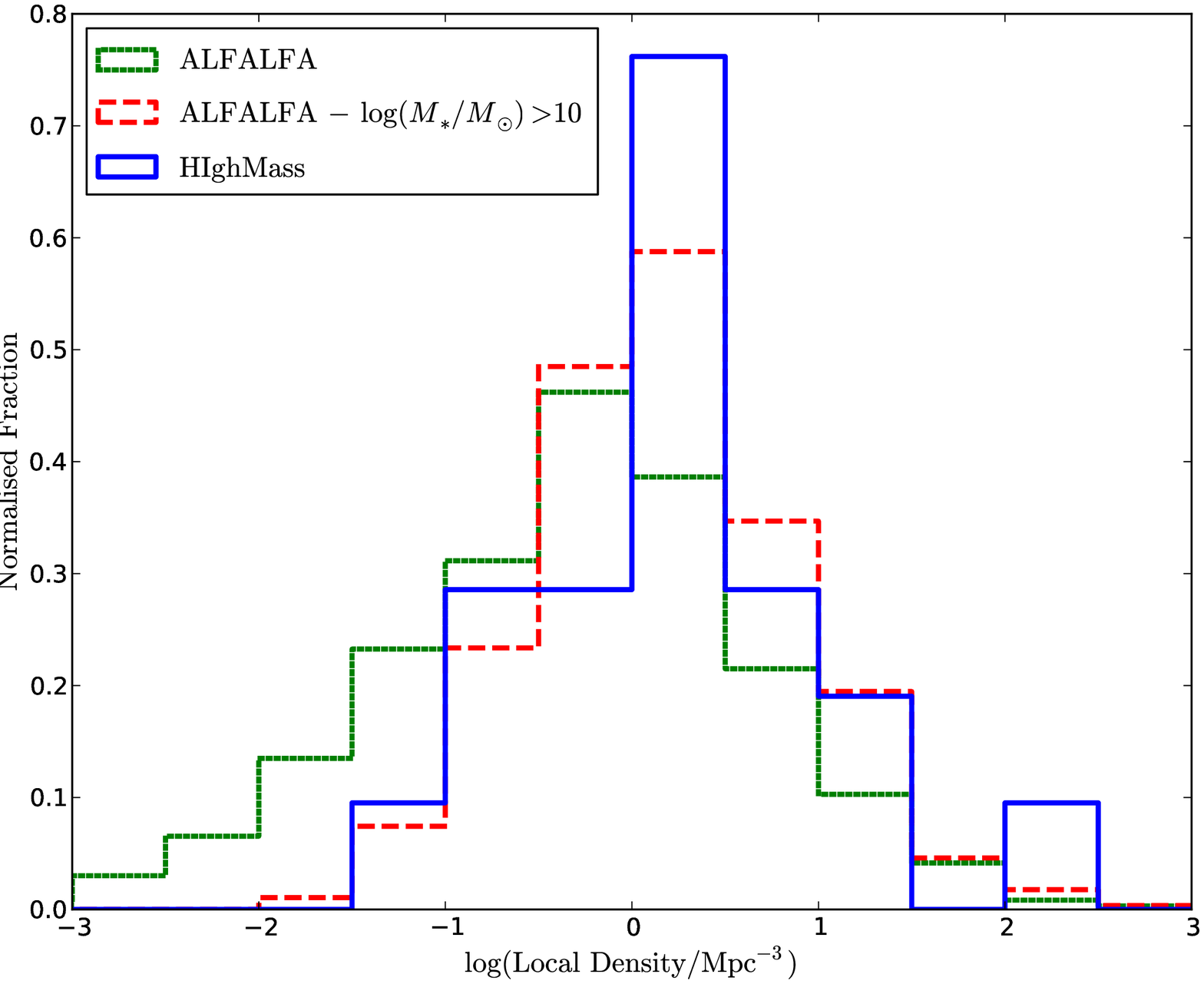}
}
\caption[]{
Local galaxy density estimated by averaging the logarithm of the 1st, 2nd, and 3rd nearest neighbor densities, 
plotted for the HIghMass (solid line), ALFALFA overall (dotted line, normalized to the HIghMass distribution), 
and ALFALFA galaxies with $M_* > 10^{10}~M_\odot$ (dashed line, normalized) respectively. 
The HIghMass and ALFALFA high stellar mass samples appear to be found in similar environments, 
but both avoid the lowest density environments, compared to the parent ALFALFA sample. 
}
\label{fig:env}
\end{figure*}

\begin{figure*}
\center{
\includegraphics[scale=1]{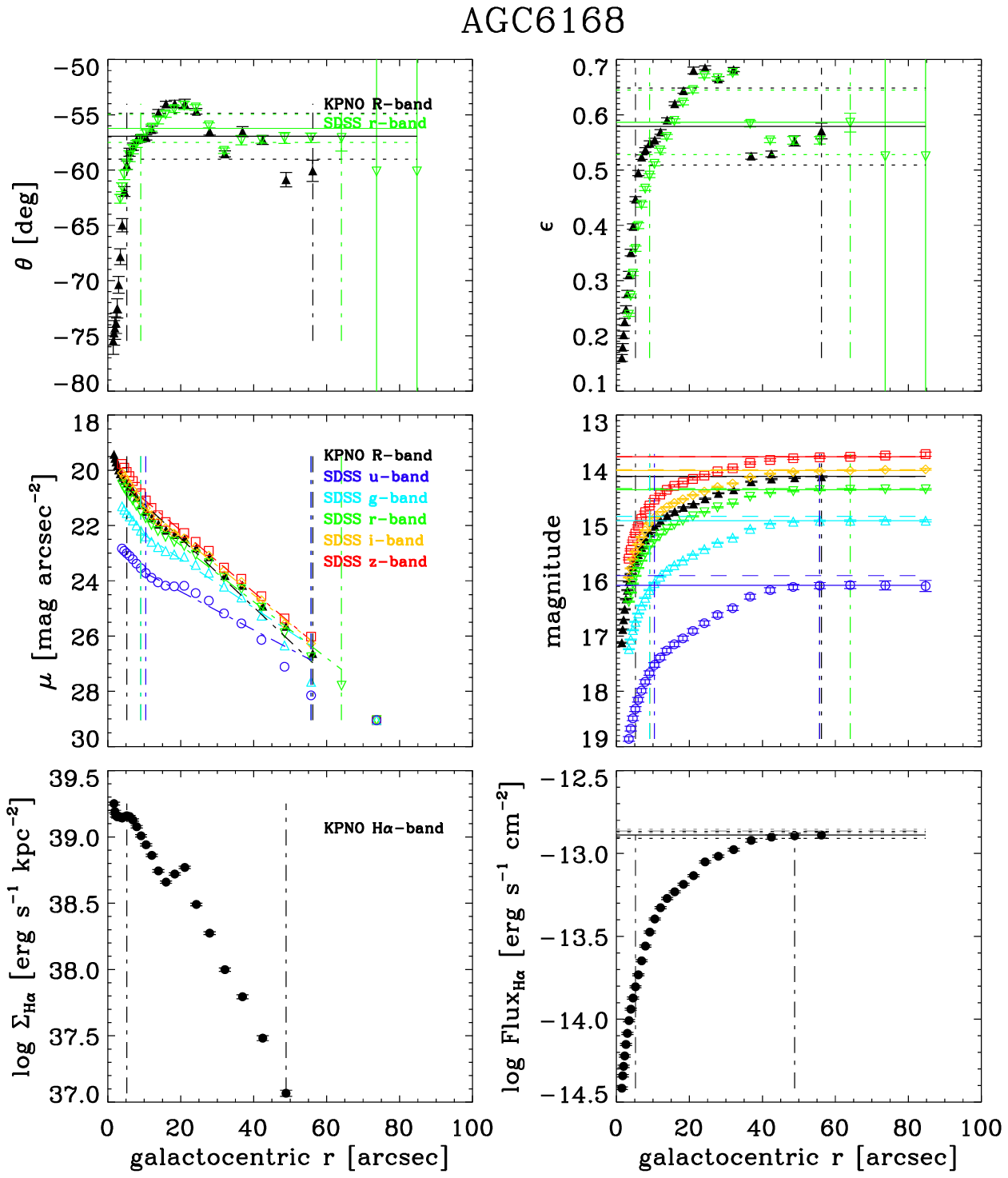}
}
\caption[]{
Example of the optical isophotal fitting for UGC~6168, a typical HIghMass galaxy. 
The variation with semi-major axis of our KPNO measurements are shown by black filled symbols, 
and that of our SDSS measurements are shown by colored open symbols. 
All panels in order from left to right and from top to bottom are position angle ($\theta$), 
ellipticity ($\epsilon\equiv 1-b/a$), surface brightness ($\mu$, not corrected for inclination), enclosed magnitude (AB system), 
H$\alpha$ surface brightness ($\Sigma_{\rm H\alpha+[NII]}$, not corrected for inclination), and enclosed H$\alpha$+[NII] flux. 
The marked inner disk region is denoted by vertical dash-dotted lines. 
Horizontal lines represent the magnitudes within 8$r_{\rm d}$ (mag$_8$, dashed) and 
Petrosian magnitudes (solid) in the magnitude or flux plot, 
and final values determined from the mean of data points in the inner disk region in the $\theta$ and $\epsilon$ plots. 
In between the horizontal dotted lines is the uncertainty range. 
Dashed lines in the $\mu$ plots show the linear fit to the light profiles.
}
\label{fig:6168}
\end{figure*}

\begin{figure*}
\center{
\includegraphics[scale=1]{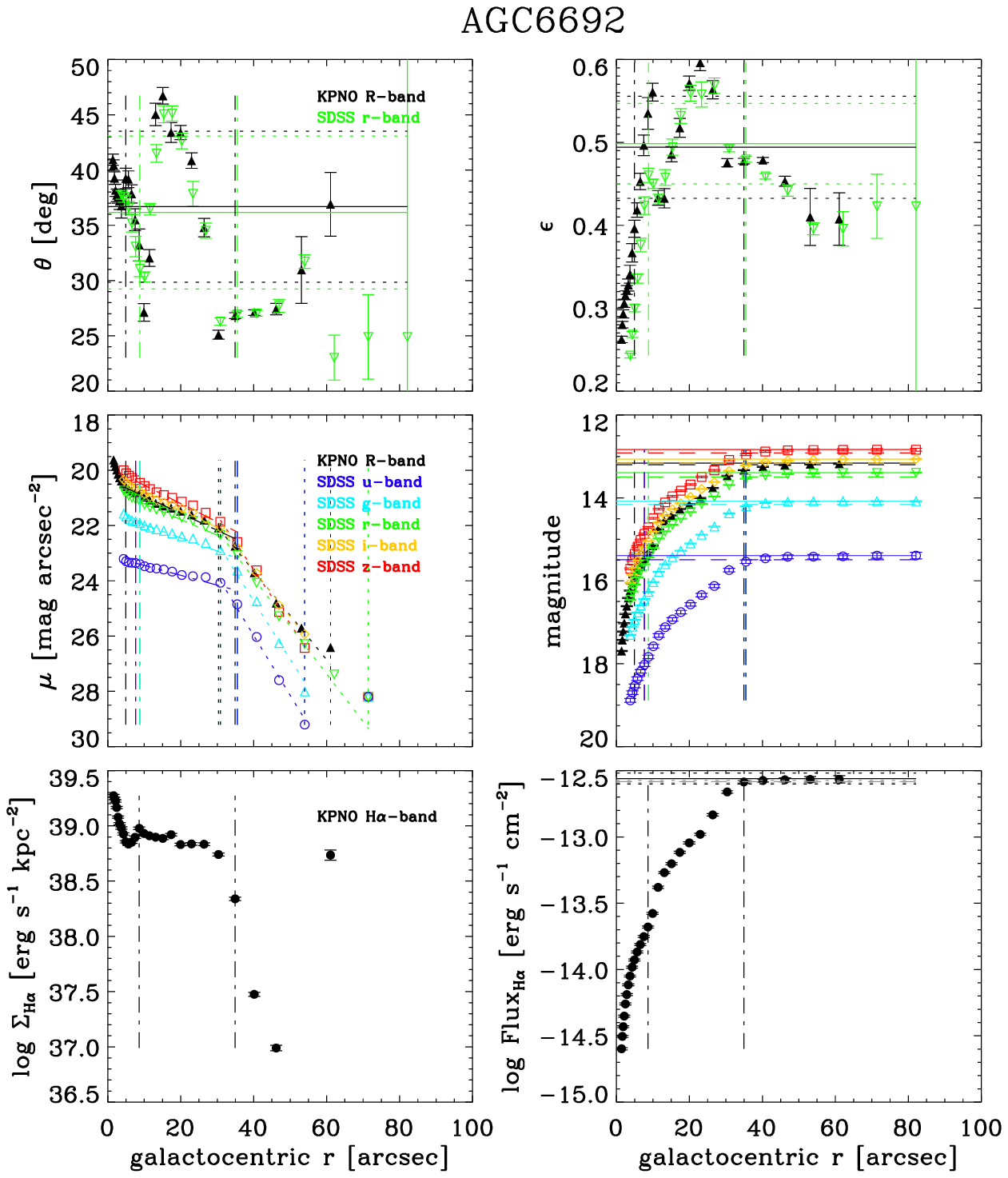}
}
\caption[]{
Example of the isophotal fitting for UGC~6692. The symbol definition is the same as that in Fig.~\ref{fig:6168}. 
This galaxy exhibits a broken exponential disk feature (downward break) and the outer disk 
is in between the vertical dotted lines. 
Similar results are observed in a total of six HIghMass galaxies, implying a threshold in SF (see text).
}
\label{fig:6692}
\end{figure*}

\begin{figure*}
\center{
\includegraphics[scale=1]{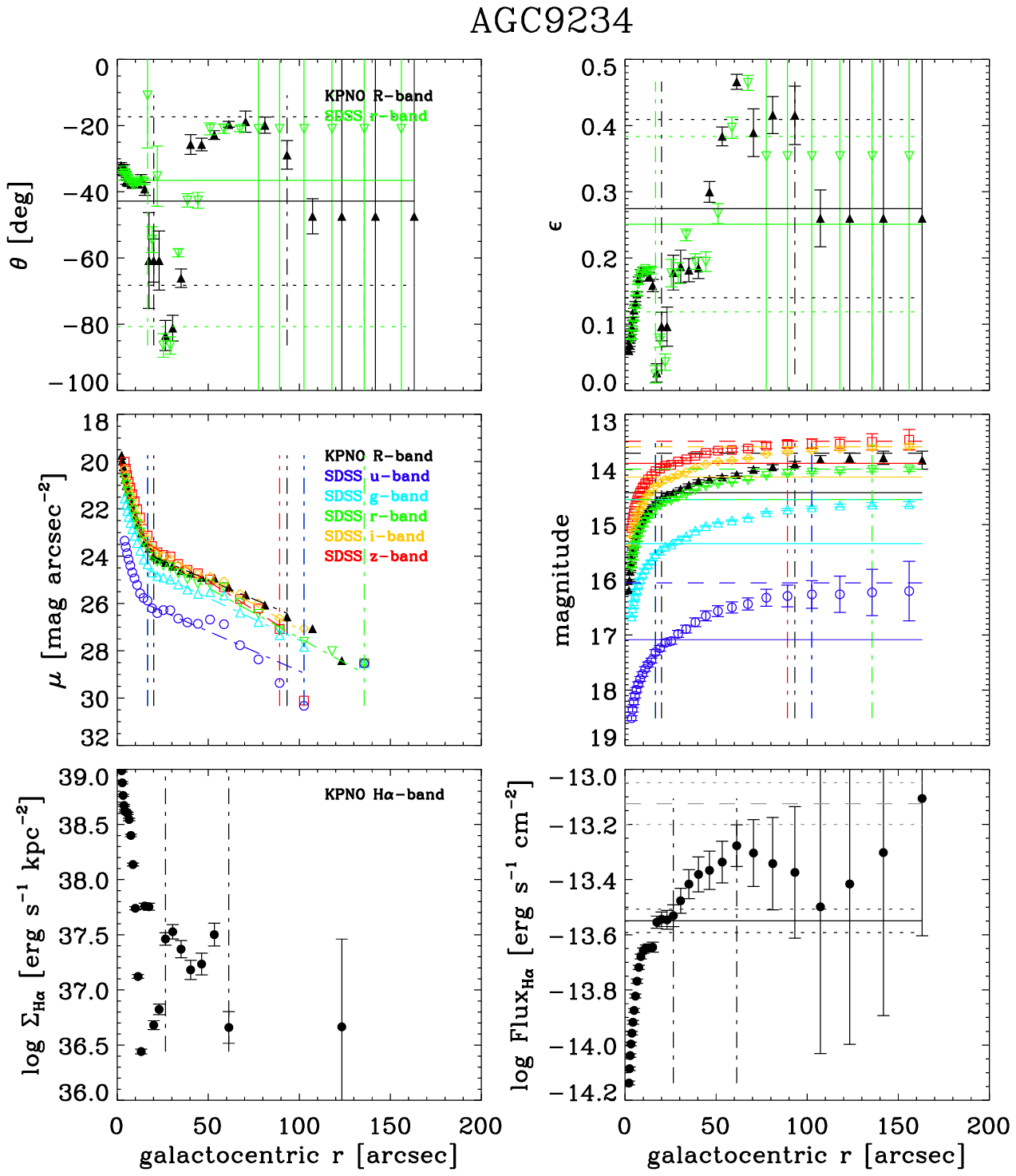}
}
\caption[]{
Example of the isophotal fitting for UGC~9234. The symbol definition is the same as that in Fig.~\ref{fig:6168}. 
This galaxy has a compact central bulge but a LSB shallow outer disk (bright in UV). 
As a result, significant discrepancies between the mag$_8$ and Petrosian magnitudes exist.
}
\label{fig:9234}
\end{figure*}

\clearpage
\begin{figure*}
\center{
\includegraphics[scale=0.8]{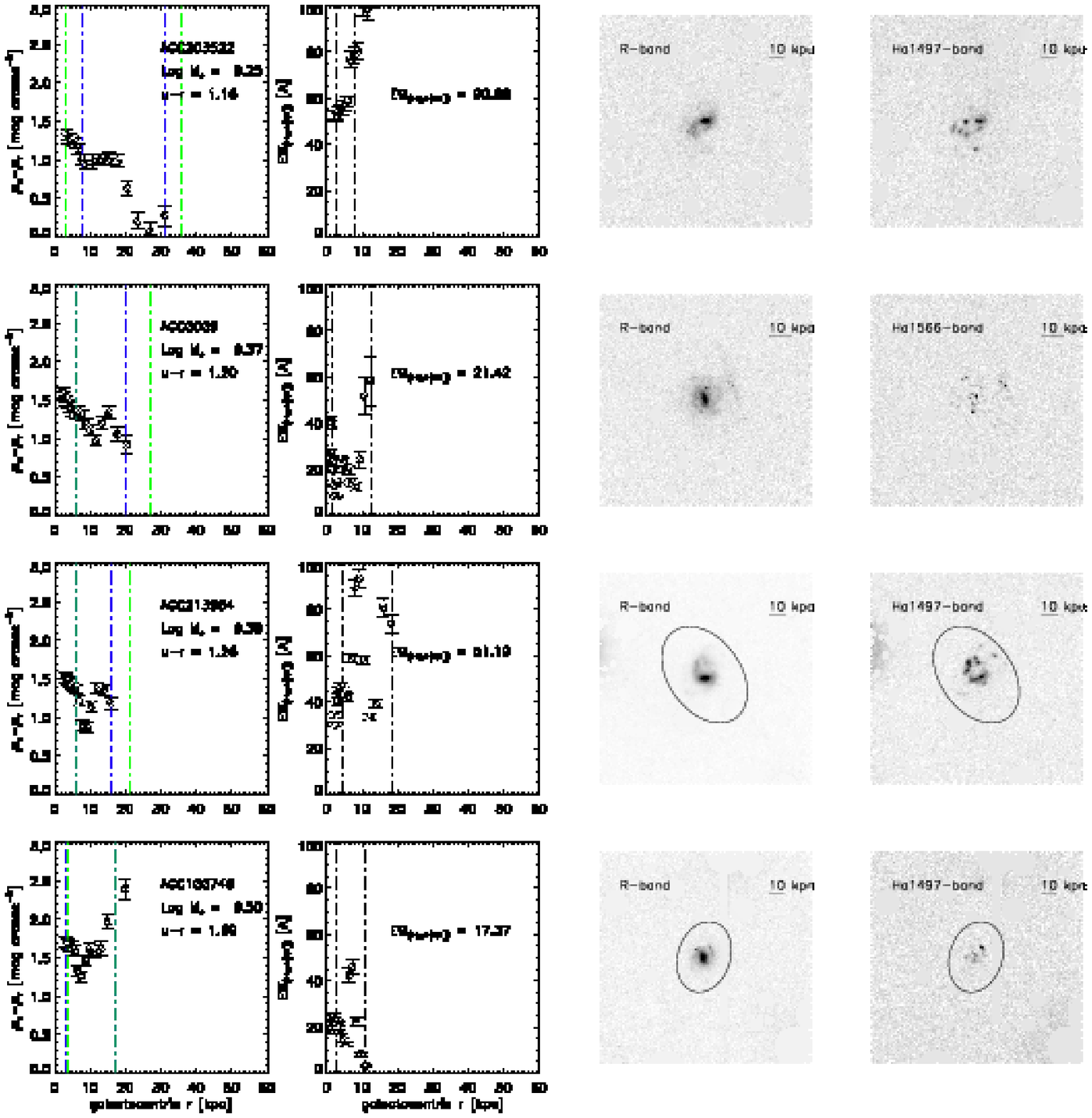}
}
\caption[]{
Profile measurements along with the inverted {\it R}-band and continuum-subtracted H$\alpha$ images (cleaned) for all 29 KPNO targets 
with photometry, in order of increasing $M_*$. 
Plotted in the first and second columns are the $u-r$ color and EW$_{\rm H\alpha+[NII]}$ radial profiles, respectively. 
The vertical line definition is the same as that in Fig.~\ref{fig:6168}: 
the inner disk region is in between the vertical dash-dotted lines (blue for {\it u} and green for {\it r} band, 
only green lines visible if disk edges in both bands overlap; black for H$\alpha$). 
The global color and EW values are given in corresponding panels. 
Petrosian magnitudes represent the flux within the elliptical apertures overlaid on the images, 
The mag$_8$ are adopted alternatively if the Petrosian radius are undetermined, 
in which cases the elliptical apertures are absent in the images.
}
\label{fig:ew}
\end{figure*}

\addtocounter{figure}{-1}

\begin{figure*}
\center{
\includegraphics[scale=0.8]{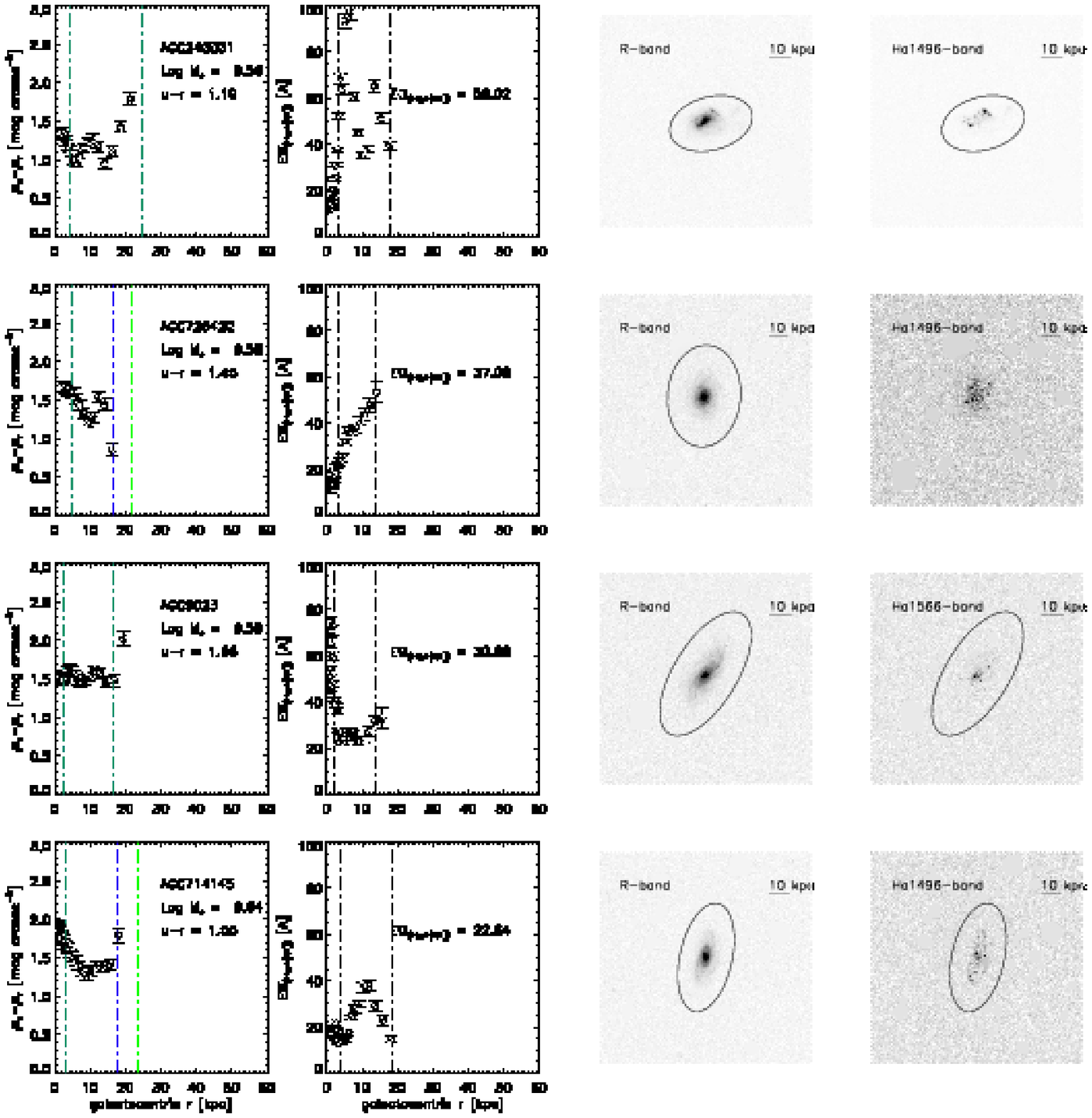}
}
\caption[]{
\it Continued.}
\end{figure*}

\addtocounter{figure}{-1}

\begin{figure*}
\center{
\includegraphics[scale=0.8]{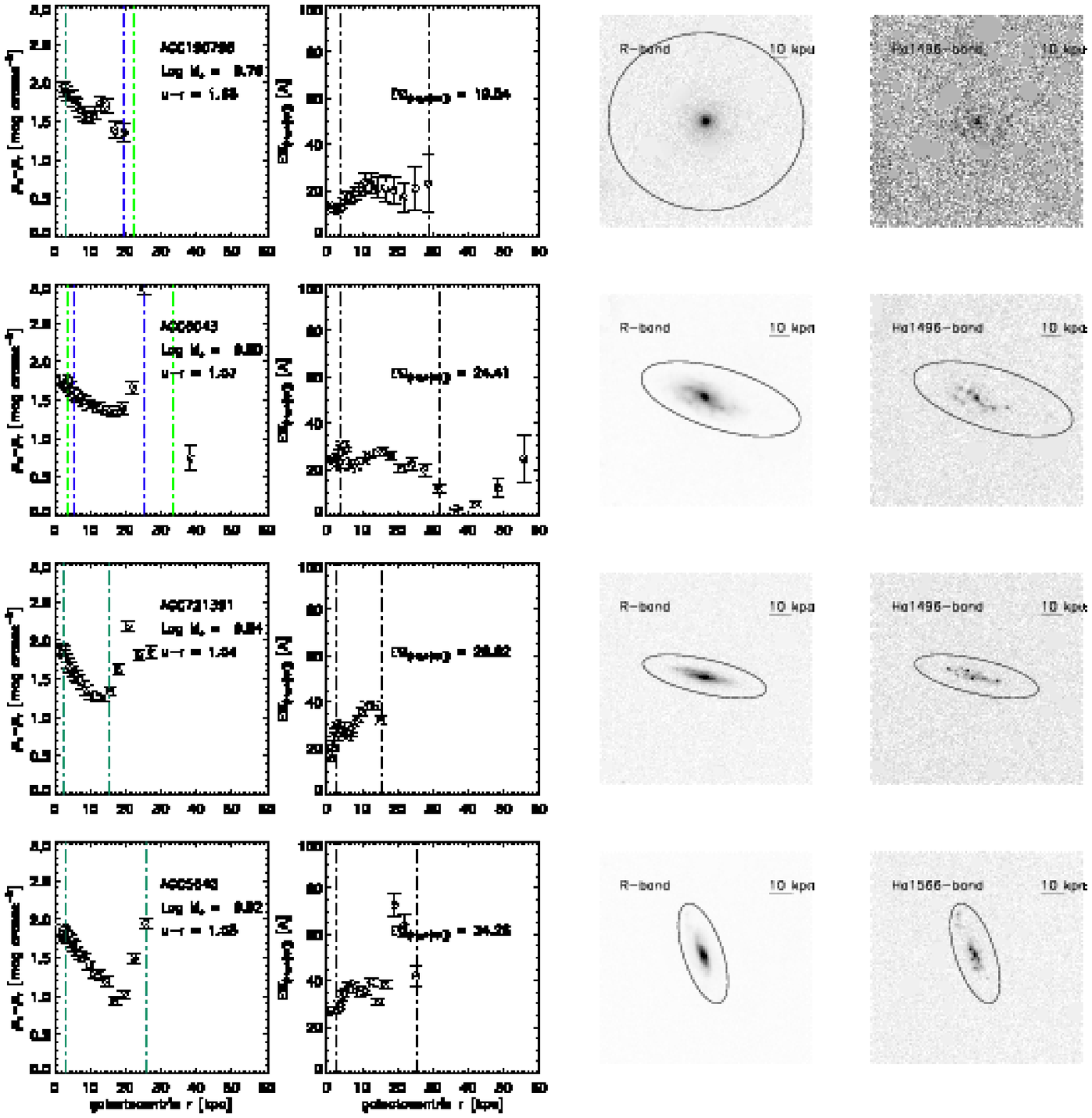}
}
\caption[]{
\it Continued.}
\end{figure*}

\addtocounter{figure}{-1}

\begin{figure*}
\center{
\includegraphics[scale=0.8]{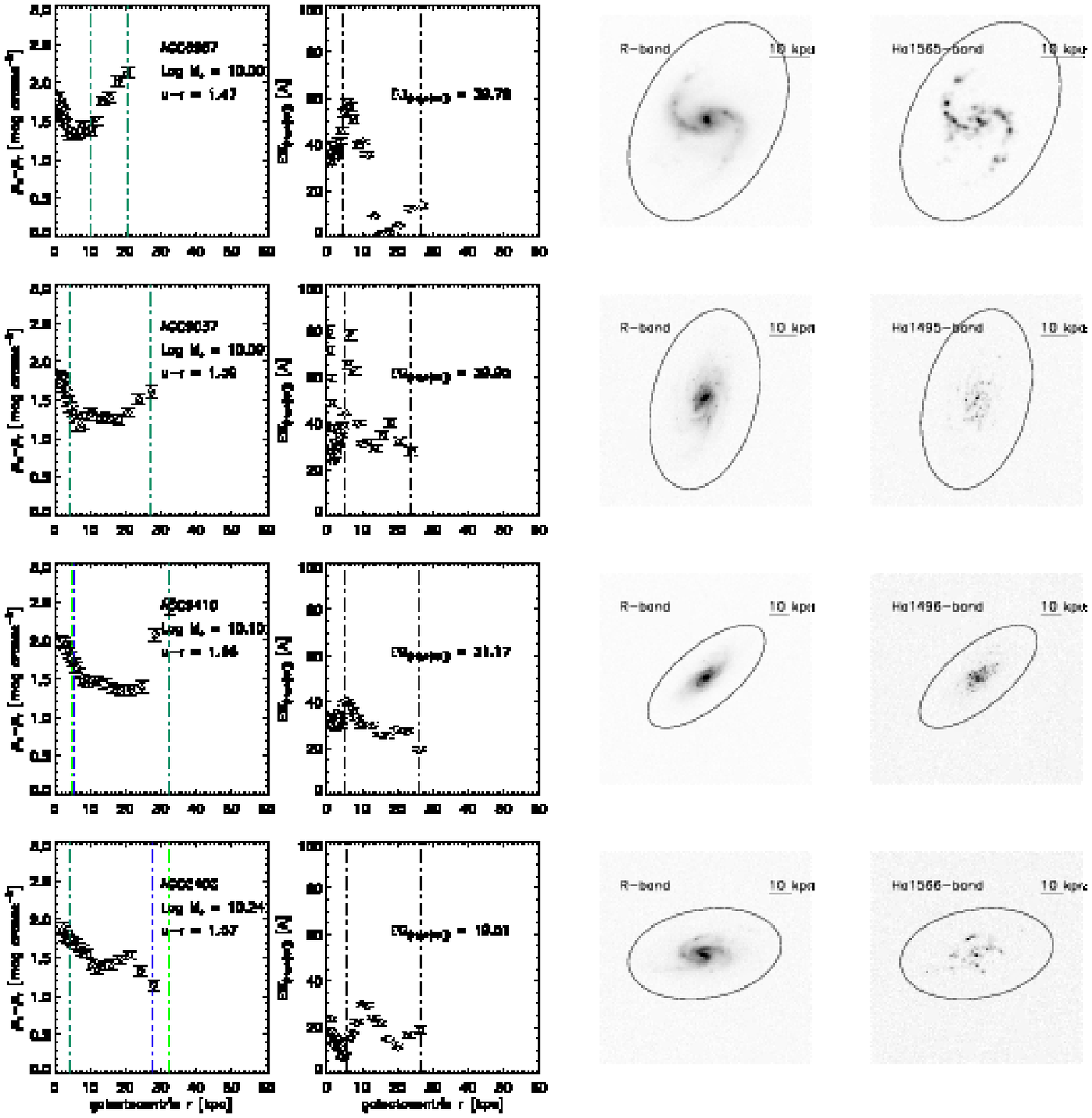}
}
\caption[]{
\it Continued.}
\end{figure*}

\addtocounter{figure}{-1}

\begin{figure*}
\center{
\includegraphics[scale=0.8]{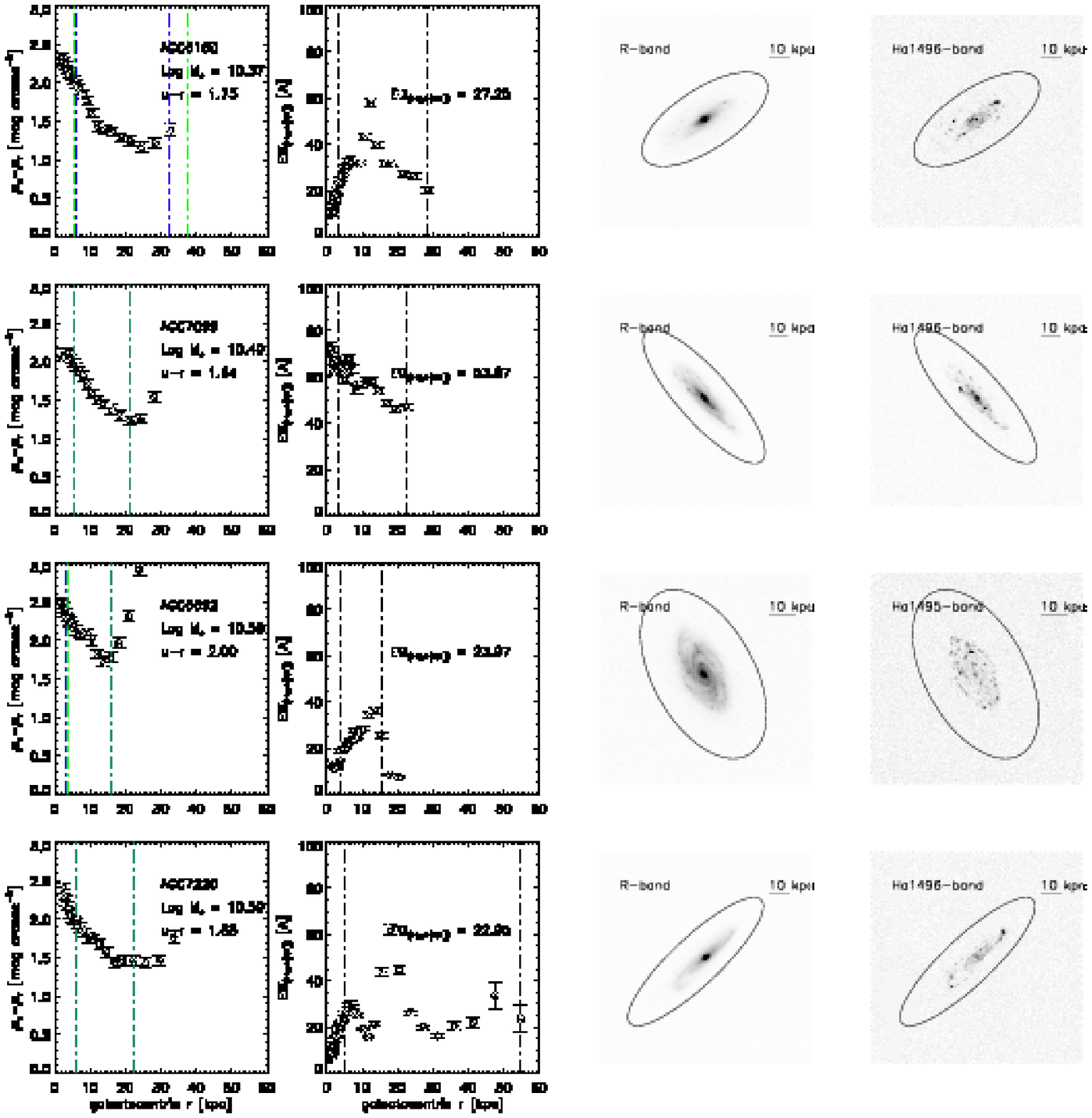}
}
\caption[]{
\it Continued.}
\end{figure*}

\addtocounter{figure}{-1}

\begin{figure*}
\center{
\includegraphics[scale=0.8]{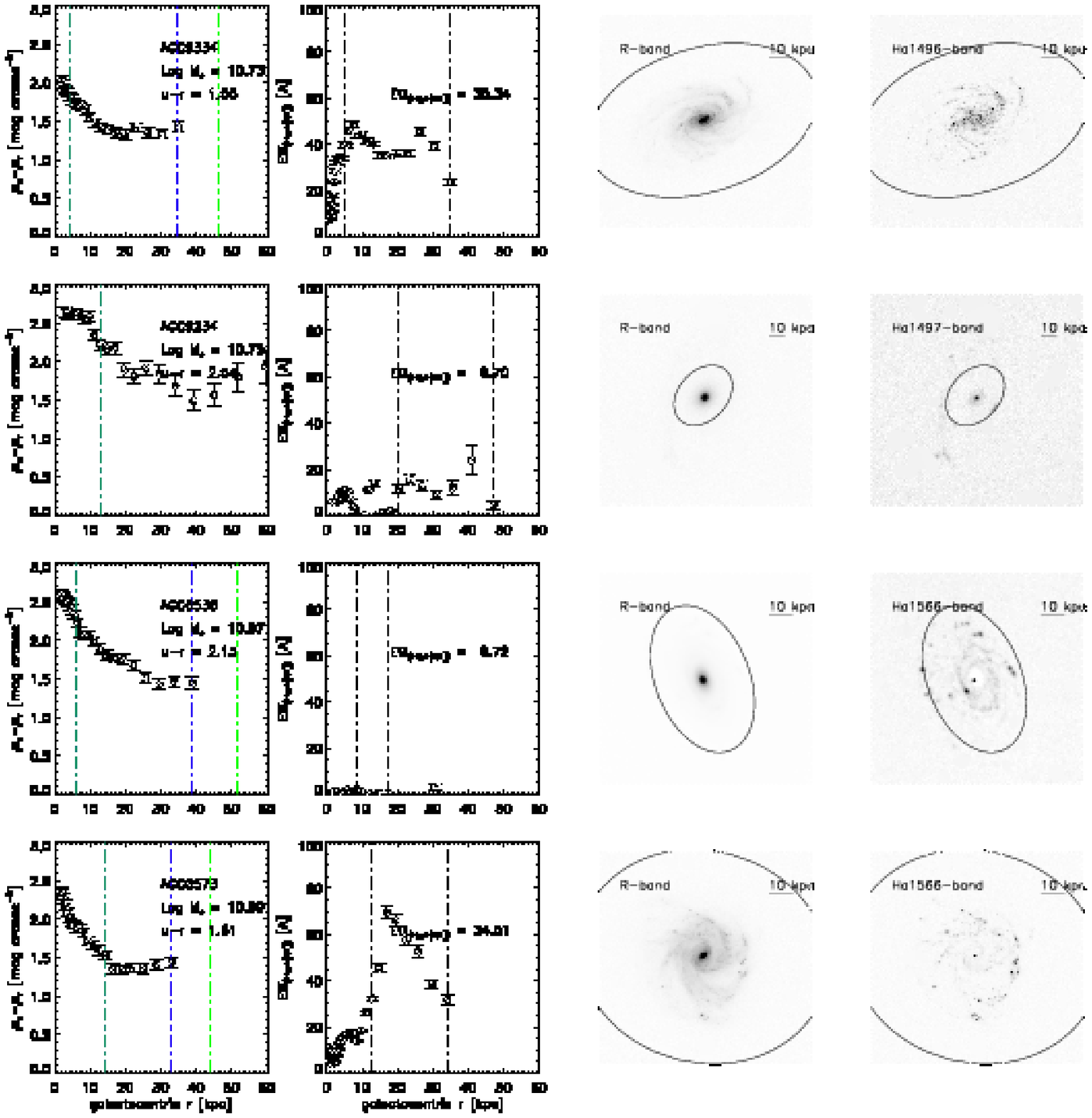}
}
\caption[]{
\it Continued.}
\end{figure*}

\addtocounter{figure}{-1}

\begin{figure*}
\center{
\includegraphics[scale=0.8]{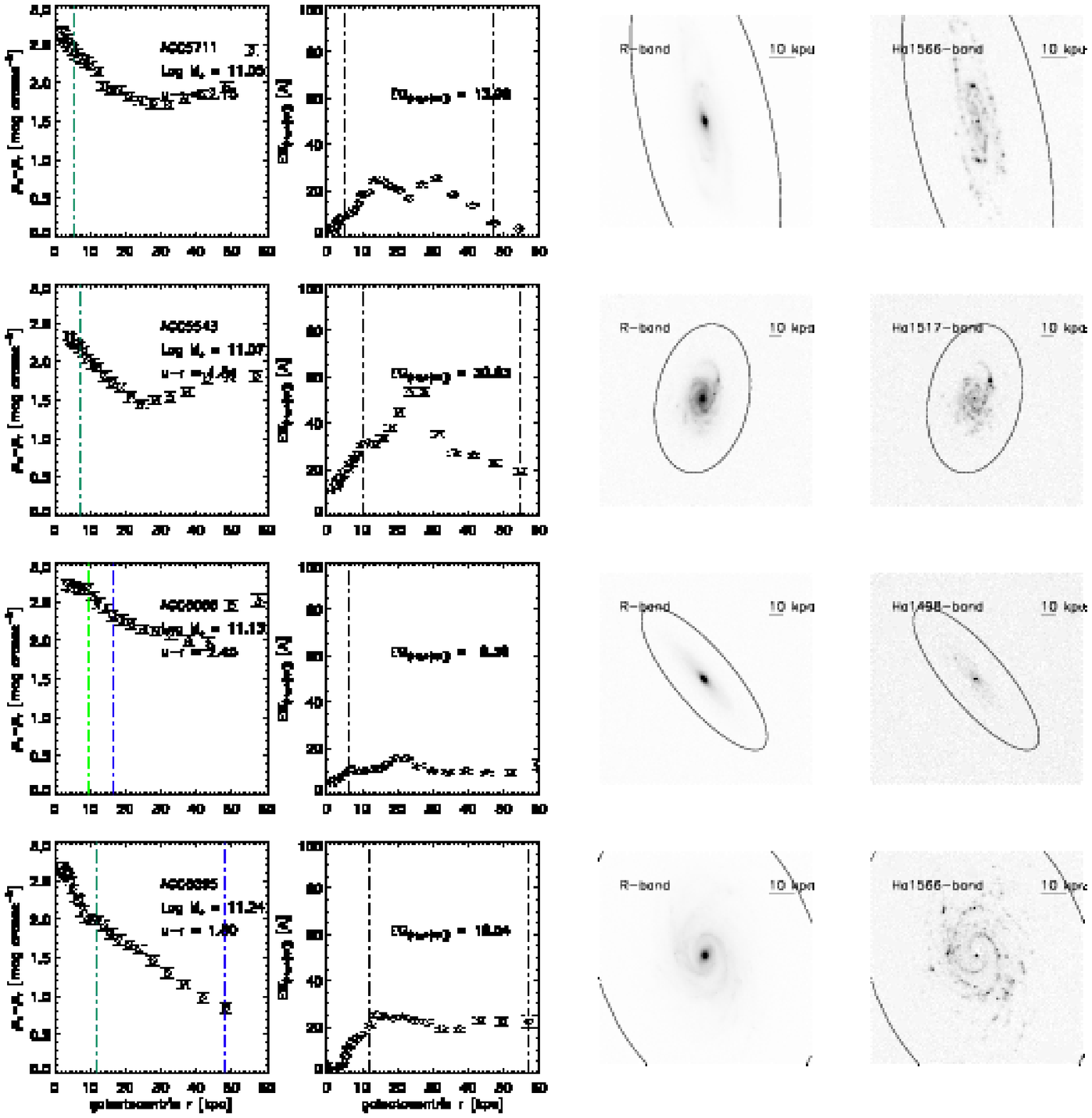}
}
\caption[]{
\it Continued.}
\end{figure*}

\addtocounter{figure}{-1}

\begin{figure*}
\center{
\includegraphics[scale=0.8]{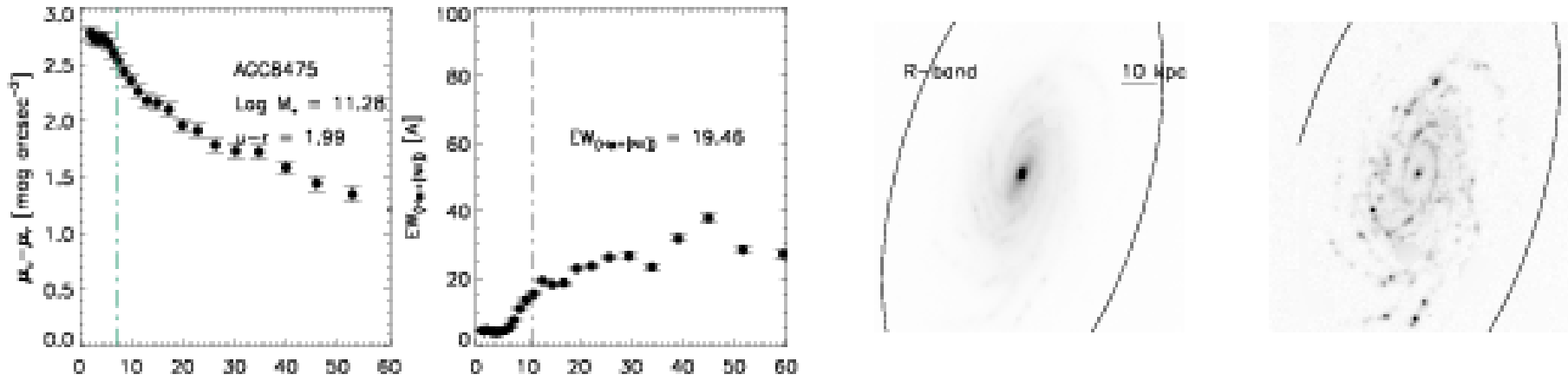}
}
\caption[]{
\it Continued.}
\end{figure*}

\begin{figure*}
\center{
\includegraphics[scale=0.85]{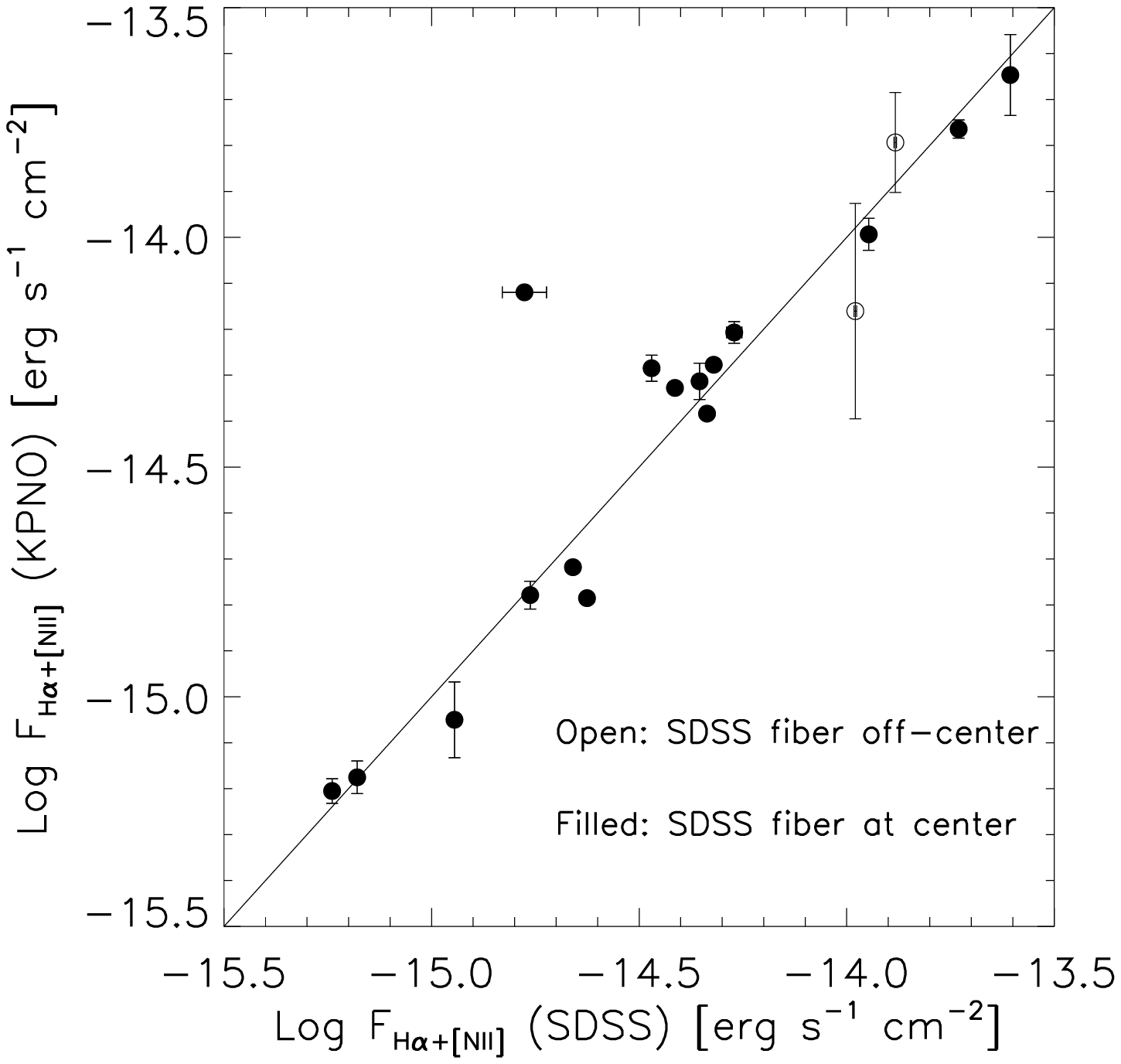}
}
\caption[]{
H$\alpha$ photometry external check by comparing the fluxes in the 
nuclear apertures in our continuum-subtracted H$\alpha$ images with the fluxes in the H$\alpha$+[NII] lines 
according to the SDSS DR8 spectroscopic measurements. 
The open symbols represent the galaxies with the SDSS fiber position off-center, 
in opposite to the filled symbols at center. The solid diagonal line marks one-to-one relation.
}
\label{fig:Ha}
\end{figure*}



\clearpage
\begin{figure*}
\center{
\includegraphics[scale=0.65]{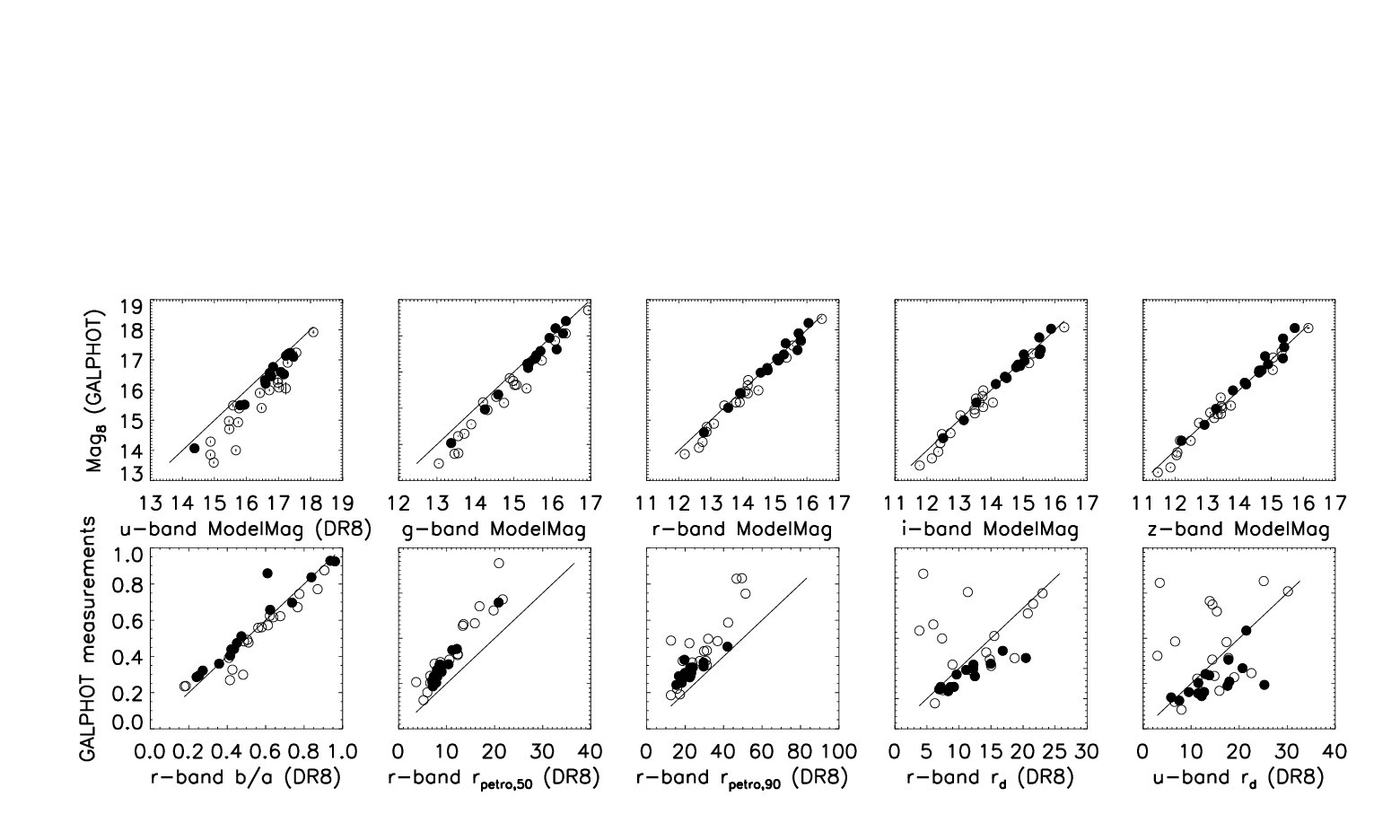}
}
\caption[]{
GALPHOT measurements in comparison with the SDSS DR8 pipeline results. 
The solid diagonal line marks one-to-one relation in all panels. 
GALPHOT-derived mag$_8$ are plotted against the SDSS model magnitudes 
for all five bands in the first row. 
The shredded pipeline measurements confirmed by visual inspections 
are represented by open symbols, being filled if no evident shredding. 
Additional comparisons are presented in the second row for the {\it r}-band axial ratio, 
$r_{\rm petro,50}$, $r_{\rm petro,90}$, and disk scale length, as well as the {\it u}-band disk scale length.
}
\label{fig:scheck}
\end{figure*}


\newpage
\begin{thebibliography}{}

\bibitem[Abazajian et al.(2009)]{Abazajian2009} Abazajian, K.~N., Adelman-McCarthy, J.~K., Ag{\"u}eros, M.~A., et al.\ 2009, \apjs, 182, 543
\bibitem[Baldry et al.(2004)]{Baldry2004} Baldry, I.~K., Glazebrook, K., Brinkmann, J., et al.\ 2004, \apj, 600, 681
\bibitem[Baldry et al.(2006)]{Baldry2006} Baldry, I.~K., Balogh, M.~L., Bower, R.~G., et al.\ 2006, \mnras, 373, 469
\bibitem[Bell et al.(2003)]{Bell2003} Bell, E.~F., McIntosh, D.~H., Katz, N., \& Weinberg, M.~D.\ 2003, \apjs, 149, 289
\bibitem[Bigiel et al.(2008)]{Bigiel2008} Bigiel, F., Leroy, A., Walter, F., et al.\ 2008, \aj, 136, 2846
\bibitem[Bigiel et al.(2010)]{Bigiel2010} Bigiel, F., Leroy, A., Walter, F., et al.\ 2010, \aj, 140, 1194
\bibitem[Bigiel et al.(2011)]{Bigiel2011} Bigiel, F., Leroy, A.~K., Walter, F., et al.\ 2011, \apjl, 730, L13
\bibitem[Blanton et al.(2005)]{Blanton2005} Blanton, M.~R., Lupton, R.~H., Schlegel, D.~J., et al.\ 2005, \apj, 631, 208
\bibitem[Blanton et al.(2011)]{Blanton2011} Blanton, M.~R., Kazin, E., Muna, D., Weaver, B.~A., \& Price-Whelan, A.\ 2011, \aj, 142, 31
\bibitem[Boissier \& Prantzos(2000)]{Boissier2000} Boissier, S., \& Prantzos, N.\ 2000, \mnras, 312, 398
\bibitem[Boissier et al.(2007)]{Boissier2007} Boissier, S., Gil de Paz, A., Boselli, A., et al.\ 2007, \apjs, 173, 524
\bibitem[Bothun(1985)]{Bothun1985a} Bothun, G.~D.\ 1985, \aj, 90, 1982
\bibitem[Bothun et al.(1985)]{Bothun1985b} Bothun, G.~D., Beers, T.~C., Mould, J.~R., \& Huchra, J.~P.\ 1985, \aj, 90, 2487
\bibitem[Bothun et al.(1987)]{Bothun1987} Bothun, G.~D., Impey, C.~D., Malin, D.~F., \& Mould, J.~R.\ 1987, \aj, 94, 23
\bibitem[Bothun et al.(1990)]{Bothun1990} Bothun, G.~D., Schombert, J.~M., Impey, C.~D., \& Schneider, S.~E.\ 1990, \apj, 360, 427
\bibitem[Bouch{\'e} et al.(2010)]{Bouche2010} Bouch{\'e}, N., Dekel, A., Genzel, R., et al.\ 2010, \apj, 718, 1001
\bibitem[Brinchmann et al.(2004)]{Brinchmann2004} Brinchmann, J., Charlot, S., White, S.~D.~M., et al.\ 2004, \mnras, 351, 1151
\bibitem[Bruzual \& Charlot(2003)]{Bruzual2003} Bruzual, G., \& Charlot, S.\ 2003, \mnras, 344, 1000
\bibitem[Caldwell et al.(1991)]{Caldwell1991} Caldwell, N., Kennicutt, R., Phillips, A.~C., \& Schommer, R.~A.\ 1991, \apj, 370, 526
\bibitem[Calzetti et al.(2000)]{Calzetti2000} Calzetti, D., Armus, L., Bohlin, R.~C., et al.\ 2000, \apj, 533, 682
\bibitem[Catinella et al.(2008)]{Catinella2008} Catinella, B., Haynes, M.~P., Giovanelli, R., Gardner, J.~P., \& Connolly, A.~J.\ 2008, \apjl, 685, L13
\bibitem[Catinella et al.(2010)]{Catinella2010} Catinella, B., Schiminovich, D., Kauffmann, G., et al.\ 2010, \mnras, 403, 683
\bibitem[Catinella et al.(2013)]{Catinella2013} Catinella, B., Schiminovich, D., Cortese, L., et al.\ 2013, \mnras, 436, 34
\bibitem[Chabrier(2003)]{Chabrier2003} Chabrier, G.\ 2003, \pasp, 115, 763
\bibitem[Cortese et al.(2012)]{Cortese2012} Cortese, L., Boissier, S., Boselli, A., et al.\ 2012, \aap, 544, A101
\bibitem[Cowie et al.(1996)]{Cowie1996} Cowie, L.~L., Songaila, A., Hu, E.~M., \& Cohen, J.~G.\ 1996, \aj, 112, 839
\bibitem[Croton et al.(2006)]{Croton2006} Croton, D.~J., Springel, V., White, S.~D.~M., et al.\ 2006, \mnras, 365, 11
\bibitem[de Vaucouleurs et al.(1991)]{deV1991} de Vaucouleurs, G., de Vaucouleurs, A., Corwin, H.~G., Jr., et al.\ 1991, 
	Third Reference Catalogue of Bright Galaxies
\bibitem[Disney \& Phillipps(1987)]{Disney1987} Disney, M., \& Phillipps, S.\ 1987, \nat, 329, 203
\bibitem[Duffy et al.(2012)]{Duffy2012} Duffy, A.~R., Meyer, M.~J., Staveley-Smith, L., et al.\ 2012, \mnras, 426, 3385
\bibitem[Finkelman \& Brosch(2011)]{Finkelman2011} Finkelman, I. \& Brosch, N. \ 2011, \mnras, 413, 2621
\bibitem[Fraternali \& Tomassetti(2012)]{Fraternali2012} Fraternali, F., \& Tomassetti, M.\ 2012, \mnras, 426, 2166
\bibitem[Freeman(1970)]{Freeman1970} Freeman, K.~C.\ 1970, \apj, 160, 811
\bibitem[Fu et al.(2010)]{Fu2010} Fu, J., Guo, Q., Kauffmann, G., \& Krumholz, M.~R.\ 2010, \mnras, 409, 515
\bibitem[Gavazzi et al.(2003)]{Gavazzi2003} Gavazzi, G., Boselli, A., Donati, A., Franzetti, P., \& Scodeggio, M.\ 2003, \aap, 400, 451
\bibitem[Gavazzi et al.(2012)]{Gavazzi2012} Gavazzi, G., Fumagalli, M., Galardo, V., et al.\ 2012, \aap, 545, A16
\bibitem[Giovanelli \& Haynes(1985)]{Giovanelli1985} Giovanelli, R., \& Haynes, M.~P.\ 1985, \apj, 292, 404
\bibitem[Giovanelli et al.(1994)]{Giovanelli1994} Giovanelli, R., Haynes, M.~P., Salzer, J.~J., et al.\ 1994, \aj, 107, 2036
\bibitem[Giovanelli et al.(1997)]{Giovanelli1997} Giovanelli, R., Haynes, M.~P., Herter, T., et al.\ 1997, \aj, 113, 22
\bibitem[Giovanelli et al.(2005)]{Giovanelli2005} Giovanelli, R., Haynes, M.~P., Kent, B.~R., et al.\ 2005, \aj, 130, 2598
\bibitem[Hallenbeck et al.(2014)]{Hallenbeck2014} Hallenbeck, G., Huang, S., Haynes, M.P. et al.\ 2014, (submitted)
\bibitem[Haynes et al.(2011)]{Haynes2011} Haynes, M.~P., Giovanelli, R., Martin, A.~M., et al.\ 2011, \aj, 142, 170
\bibitem[Helmboldt et al.(2005)]{Helmboldt2005} Helmboldt, J.~F., Walterbos, R.~A.~M., Bothun, G.~D., \& O'Neil, K.\ 2005, \apj, 630, 824
\bibitem[Hernandez \& Cervantes-Sodi(2006)]{Hernandez2006} Hernandez, X., \& Cervantes-Sodi, B.\ 2006, \mnras, 368, 351
\bibitem[Hernandez et al.(2007)]{Hernandez2007} Hernandez, X., Park, C., Cervantes-Sodi, B., \& Choi, Y.-Y.\ 2007, \mnras, 375, 163
\bibitem[Hopkins et al.(2003)]{Hopkins2003} Hopkins, A.~M., Miller, C.~J., Nichol, R.~C., et al.\ 2003, \apj, 599, 971
\bibitem[Huang et al.(2012a)]{Huang2012a} Huang, S., Haynes, M.~P., Giovanelli, R., et al.\ 2012, \aj, 143, 133
\bibitem[Huang et al.(2012b)]{Huang2012b} Huang, S., Haynes, M.~P., Giovanelli, R., \& Brinchmann, J.\ 2012, \apj, 756, 113
\bibitem[Hunter \& Elmegreen(2004)]{Hunter2004} Hunter, D.~A., \& Elmegreen, B.~G.\ 2004, \aj, 128, 2170
\bibitem[Hunter et al.(2010)]{Hunter2010} Hunter, D.~A., Elmegreen, B.~G., \& Ludka, B.~C.\ 2010, \aj, 139, 447
\bibitem[Impey \& Bothun(1997)]{Impey1997} Impey, C., \& Bothun, G.\ 1997, \araa, 35, 267
\bibitem[James et al.(2004)]{James2004} James, P.~A., Shane, N.~S., Beckman, J.~E., et al.\ 2004, \aap, 414, 23
\bibitem[Jansen et al.(2000)]{Jansen2000a} Jansen, R.~A., Franx, M., Fabricant, D., \& Caldwell, N.\ 2000, \apjs, 126, 271
\bibitem[Jansen et al.(2000)]{Jansen2000b} Jansen, R.~A., Fabricant, D., Franx, M., \& Caldwell, N.\ 2000, \apjs, 126, 331
\bibitem[Karachentsev \& Kaisin(2010)]{Karachentsev2010} Karachentsev, I.~D., \& Kaisin, S.~S.\ 2010, \aj, 140, 1241
\bibitem[Karachentseva(1973)]{Karachentseva1973} Karachentseva, V.~E.\ 1973, Astrofizicheskie Issledovaniia Izvestiya Spetsial'noj 
Astrofizicheskoj Observatorii, 8, 3
\bibitem[Kennicutt(1983)]{Kennicutt1983} Kennicutt, R.~C., Jr.\ 1983, \apj, 272, 54
\bibitem[Kennicutt(1988)]{Kennicutt1988} Kennicutt, R.~C., Jr.\ 1988, \apj, 334, 144
\bibitem[Kennicutt et al.(1989)]{Kennicutt1989a} Kennicutt, R.~C., Jr., Edgar, B.~K., \& Hodge, P.~W.\ 1989, \apj, 337, 761
\bibitem[Kennicutt et al.(1989)]{Kennicutt1989b} Kennicutt, R.~C., Jr., Keel, W.~C., \& Blaha, C.~A.\ 1989, \aj, 97, 1022
\bibitem[Kennicutt(1989)]{Kennicutt1989c} Kennicutt, R.~C., Jr.\ 1989, \apj, 344, 685
\bibitem[Kennicutt(1998)]{Kennicutt1998} Kennicutt, R.~C., Jr.\ 1998, \apj, 498, 541
\bibitem[Kennicutt et al.(2003)]{Kennicutt2003} Kennicutt, R.~C., Jr., Armus, L., Bendo, G., et al.\ 2003, \pasp, 115, 928
\bibitem[Kennicutt et al.(2008)]{Kennicutt2008} Kennicutt, R.~C., Jr., Lee, J.~C., Funes, S.~J., Jos{\'e} G., Sakai, S., \& Akiyama, S.\ 2008, \apjs, 178, 247
\bibitem[Kennicutt \& Evans(2012)]{Kennicutt2012} Kennicutt, R.~C., Jr, \& Evans, N.~J., II 2012, arXiv:1204.3552
\bibitem[Kere{\v s} et al.(2005)]{Keres2005} Kere{\v s}, D., Katz, N., Weinberg, D.~H., \& Dav{\'e}, R.\ 2005, \mnras, 363, 2
\bibitem[Kewley et al.(2002)]{Kewley2002} Kewley, L.~J., Geller, M.~J., Jansen, R.~A., \& Dopita, M.~A.\ 2002, \aj, 124, 3135
\bibitem[Kewley et al.(2010)]{Kewley2010} Kewley, L.~J., Rupke, 
D., Zahid, H.~J., Geller, M.~J., \& Barton, E.~J.\ 2010, \apjl, 721, L48
\bibitem[Kim \& Lee(2013)]{Kim2013} Kim, J.-h., \& Lee, J.\ 2013, \mnras, 432, 1701
\bibitem[Kravtsov(2013)]{Kravtsov2013} Kravtsov, A.~V.\ 2013, \apjl, 764, L31
\bibitem[Krumholz(2013)]{Krumholz2013} Krumholz, M.~R.\ 2013, \mnras, 2443
\bibitem[Landolt(1992)]{Landolt1992} Landolt, A.~U.\ 1992, \aj, 104, 340
\bibitem[Lee et al.(2007)]{Lee2007} Lee, J.~C., Kennicutt, R.~C., Funes, S.~J., Jos{\'e} G., Sakai, S., \& Akiyama, S.\ 2007, \apjl, 671, L113
\bibitem[Lelli et al.(2010)]{Lelli2010} Lelli, F., Fraternali, F., \& Sancisi, R.\ 2010, \aap, 516, A11 
\bibitem[Lemonias et al.(2011)]{Lemonias2011} Lemonias, J.~J., Schiminovich, D., Thilker, D., et al.\ 2011, \apj, 733, 74 
\bibitem[Leroy et al.(2008)]{Leroy2008} Leroy, A.~K., Walter, F., Brinks, E., et al.\ 2008, \aj, 136, 2782
\bibitem[Li et al.(2012)]{Li2012} Li, C., Kauffmann, G., Fu, J., et al.\ 2012, \mnras, 424, 1471
\bibitem[Liu et al.(2013)]{Liu2013} Liu, G., Calzetti, D., Kennicutt, R.~C., Jr., et al.\ 2013, \apj, 772, 27
\bibitem[Martin \& Roy(1994)]{Martin1994} Martin, P., \& Roy, J.-R.\ 1994, \apj, 424, 599
\bibitem[Martin et al.(2012)]{Martin2012} Martin, A.~M., Giovanelli, R., Haynes, M.~P., \& Guzzo, L.\ 2012, \apj, 750, 38
\bibitem[Mart{\'{\i}}n-Navarro et al.(2012)]{MN2012} Mart{\'{\i}}n-Navarro, I., Bakos, J., Trujillo, I., et al.\ 2012, \mnras, 427, 1102
\bibitem[Masters et al.(2012)]{Masters2012} Masters, K.~L., Nichol, R.~C., Haynes, M.~P., et al.\ 2012, \mnras, 424, 2180
\bibitem[Meurer et al.(2006)]{Meurer2006} Meurer, G.~R., Hanish, D.~J., Ferguson, H.~C., et al.\ 2006, \apjs, 165, 307
\bibitem[Meyer et al.(2004)]{Meyer2004} Meyer, M.~J., Zwaan, M.~A., Webster, R.~L., et al.\ 2004, \mnras, 350, 1195
\bibitem[Meyer(2009)]{Meyer2009} Meyer, M.\ 2009, Panoramic Radio Astronomy: Wide-field 1-2 GHz Research on Galaxy Evolution, 
\bibitem[Mo et al.(1994)]{Mo1994} Mo, H.~J., McGaugh, S.~S., \& Bothun, G.~D.\ 1994, \mnras, 267, 129
\bibitem[Mo et al.(1998)]{Mo1998} Mo, H.~J., Mao, S., \& White, S.~D.~M.\ 1998, \mnras, 295, 319
\bibitem[Moran et al.(2010)]{Moran2010} Moran, S.~M., Kauffmann, G., Heckman, T.~M., et al.\ 2010, \apj, 720, 1126
\bibitem[Mu{\~n}oz-Mateos et al.(2011)]{Munoz-Mateos2011} 
	Mu{\~n}oz-Mateos, J.~C., Boissier, S., Gil de Paz, A., et al.\ 2011, \apj, 731, 10
\bibitem[Oey \& Clarke(1998)]{Oey1998} Oey, M.~S., \& Clarke, C.~J.\ 1998, \aj, 115, 1543
\bibitem[Oke \& Gunn(1983)]{Oke1983} Oke, J.~B., \& Gunn, J.~E.\ 1983, \apj, 266, 713
\bibitem[Papastergis et al.(2011)]{Papastergis2011} Papastergis, E., Martin, A.~M., Giovanelli, R., \& Haynes, M.~P.\ 2011, \apj, 739, 38
\bibitem[Pickering et al.(1997)]{Pickering1997} Pickering, T.~E., Impey, C.~D., van Gorkom, J.~H., \& Bothun, G.~D.\ 1997, \aj, 114, 1858
\bibitem[Pohlen \& Trujillo(2006)]{Pohlen2006} Pohlen, M., \& Trujillo, I.\ 2006, \aap, 454, 759
\bibitem[Portas et al.(2010)]{Portas2010} Portas, A.~M., Brinks, E., Filho, M.~E., et al.\ 2010, \mnras, 407, 1674
\bibitem[Ro{\v s}kar et al.(2008)]{Roskar2008} Ro{\v s}kar, R., Debattista, V.~P., Stinson, G.~S., et al.\ 2008, \apjl, 675, L65
\bibitem[Saintonge et al.(2011)]{Saintonge2011} Saintonge, A., Kauffmann, G., Kramer, C., et al.\ 2011, \mnras, 415, 32
\bibitem[Salim et al.(2007)]{Salim2007} Salim, S., Rich, R.~M., Charlot, S., et al.\ 2007, \apjs, 173, 267
\bibitem[S{\'a}nchez-Gallego et al.(2012)]{SG2012} S{\'a}nchez-Gallego, J.~R., Knapen, J.~H., Wilson, C.~D., et al.\ 2012, \mnras, 422, 3208
\bibitem[Sancisi et al.(2008)]{Sancisi2008} Sancisi, R., Fraternali, F., Oosterloo, T., \& van der Hulst, T.\ 2008, \aapr, 15, 189
\bibitem[Schiminovich et al.(2010)]{Schiminovich2010} Schiminovich, D., Catinella, B., Kauffmann, G., et al.\ 2010, \mnras, 408, 919
\bibitem[Schlegel et al.(1998)]{Schlegel1998} Schlegel, D.~J., Finkbeiner, D.~P., \& Davis, M.\ 1998, \apj, 500, 525
\bibitem[Schombert et al.(1992)]{Schombert1992} Schombert, J.~M., Bothun, G.~D., Schneider, S.~E., \& McGaugh, S.~S.\ 1992, \aj, 103, 1107
\bibitem[Schombert et al.(2011)]{Schombert2011} Schombert, J., Maciel, T., \& McGaugh, S.\ 2011, Advances in Astronomy, 2011,
\bibitem[Schruba et al.(2011)]{Schruba2011} Schruba, A., Leroy, A.~K., Walter, F., et al.\ 2011, \aj, 142, 37
\bibitem[Sprayberry et al.(1995)]{Sprayberry1995} Sprayberry, D., Impey, C.~D., Bothun, G.~D., \& Irwin, M.~J.\ 1995, \aj, 109, 558
\bibitem[Springel et al.(2006)]{Springel2006} Springel, V., Frenk, C.~S., \& White, S.~D.~M.\ 2006, \nat, 440, 1137
\bibitem[Springob et al.(2007)]{Springob2007} Springob, C.~M., Masters, K.~L., Haynes, M.~P., Giovanelli, R., \& Marinoni, C.\ 2007, \apjs, 172, 599
\bibitem[Stewart et al.(2013)]{Stewart2013} Stewart, K.~R., Brooks, A.~M., Bullock, J.~S., et al.\ 2013, arXiv:1301.3143
\bibitem[Taylor et al.(2011)]{Taylor2011} Taylor, E.~N., Hopkins, A.~M., Baldry, I.~K., et al.\ 2011, \mnras, 418, 1587
\bibitem[Thilker et al.(2000)]{Thilker2000} Thilker, D.~A., Braun, R., \& Walterbos, R.~A.~M.\ 2000, \aj, 120, 3070
\bibitem[Thilker et al.(2007)]{Thilker2007} Thilker, D.~A., Bianchi, L., Meurer, G., et al.\ 2007, \apjs, 173, 538
\bibitem[Tully et al.(1996)]{Tully1996} Tully, R.~B., Verheijen, M.~A.~W., Pierce, M.~J., Huang, J.-S., \& Wainscoat, R.~J.\ 1996, \aj, 112, 2471
\bibitem[Verdes-Montenegro et al.(2005)]{VM2005} Verdes-Montenegro, L., Sulentic, J., Lisenfeld, U., et al.\ 2005, \aap, 436, 443
\bibitem[Walter et al.(2008)]{Walter2008} Walter, F., Brinks, E., de Blok, W.~J.~G., et al.\ 2008, \aj, 136, 2563
\bibitem[Wang et al.(2013)]{Wang2013} Wang, J., Kauffmann, G., J{\'o}zsa, G.~I.~G., et al.\ 2013, \mnras, 433, 270
\bibitem[Willett et al.(2013)]{Willett2013} Willett, K.~W., Lintott, C.~J., Bamford, S.~P., et al.\ 2013, \mnras, 435, 2835
\bibitem[Zwaan et al.(2005)]{Zwaan2005} Zwaan, M.~A., Meyer, M.~J., Staveley-Smith, L., \& Webster, R.~L.\ 2005, \mnras, 359, L30
\end{thebibliography}
\end{document}